\documentclass[floatfix, prd, twocolumn, showpacs, showkeys, preprintnumbers, superscriptaddress,nofootinbib, 10pt]{revtex4-1}
\usepackage[usenames,dvipsnames,svgnames,table]{xcolor} 
\usepackage[sort&compress]{natbib}
\usepackage{array,mathtools,amssymb,booktabs,multirow,diagbox,hhline}
\pdfoutput=1
\usepackage{tikz-feynman,contour}
\usepackage{ulem}
\usepackage{graphicx}
\usepackage{subcaption}
\usepackage{graphicx,graphics,color}
\usepackage{dcolumn}
\usepackage{bm}
\usepackage{bbold}
\usepackage{hyperref}
\usepackage{cancel}
\hypersetup{ 
    pdfnewwindow=true,      
    colorlinks=true,       
    allcolors=SeaGreen
} 

\DeclareMathOperator*{\SumInt}{%
\mathchoice%
  {\ooalign{$\displaystyle\sum$\cr\hidewidth$\displaystyle\int$\hidewidth\cr}}
  {\ooalign{\raisebox{.14\height}{\scalebox{.7}{$\textstyle\sum$}}\cr\hidewidth$\textstyle\int$\hidewidth\cr}}
  {\ooalign{\raisebox{.2\height}{\scalebox{.6}{$\scriptstyle\sum$}}\cr$\scriptstyle\int$\cr}}
  {\ooalign{\raisebox{.2\height}{\scalebox{.6}{$\scriptstyle\sum$}}\cr$\scriptstyle\int$\cr}}
}
\usepackage{amsmath}
\usepackage{xcolor}
\usepackage{tikz}
\usepackage{tikz-feynman}
\usetikzlibrary{positioning}
\usepackage{feynmp}
\usepackage{simpler-wick}
\newcommand{\im}{\mbox{Im}\,}
\newcommand{\re}{\mbox{Re}\,}
\newcommand{\Od}{{\mathcal O}}

\newcommand{\I}{{\Bbb 1}}
\newcommand{\IZ}{{\Bbb Z}}

\newcommand{\modq}{\vert \vec{q} \,\vert}
\newcommand{\modqzero}{\vert \vec{q}_0 \vert}

\newcommand{\pp}{\vec{p}\,'}
\newcommand{\pq}{\hat{p}\cdot\hat{q}}
\newcommand{\qpp}{\hat{q}\cdot\hat{p}\,'}
\newcommand{\vppp}{\vec{p}\cdot\vec{p}\,'}
\newcommand{\qp}{\vec{q}\cdot\vec{p}}

\newcommand{\ppp}{\vec{p}\cdot\vec{p}\,'}

\newcommand{\T}{\mathcal{T}}

\newcommand{\J}{\mathcal{J}}

\newcommand{\V}{\mathcal{V}}

\newcolumntype{P}[1]{>{\centering\arraybackslash}p{#1}}
\begin{document}

\title{Pion scattering in finite volume within the Inverse Amplitude Method}
\author{Angel G\'omez Nicola}
\email{gomez@ucm.es}
\affiliation{Universidad Complutense de Madrid, Facultad de Ciencias F\'isicas, Departamento de F\'isica Te\'orica and
IPARCOS, Plaza de las Ciencias 1, 28040 Madrid, Spain}
\author{Raquel Molina}
\email{Raquel.Molina@ific.uv.es}
\affiliation{Departamento de F\'isica Te\'orica and IFIC, Centro Mixto Universidad de Valencia-CSIC, Parc Cient\'ific UV, C/ Catedr\'atico Jos\'e Beltr\'an, 2, 46980 Paterna, Spain}
\author{Juli\'an A S\'anchez}
\email{julian.sanchez@ific.uv.es}
\affiliation{Departamento de F\'isica Te\'orica and IFIC, Centro Mixto Universidad de Valencia-CSIC, Parc Cient\'ific UV, C/ Catedr\'atico Jos\'e Beltr\'an, 2, 46980 Paterna, Spain}

\begin{abstract}

We study the effect of a finite volume for pion-pion scattering within Chiral Perturbation Theory (ChPT) and the Inverse Amplitude Method (IAM) in a $L^3$ box (rest frame). Our full ChPT calculation takes  into account the discretization not only in the $s$-channel loops but also in the $t,u$- channels and tadpole contributions. Hence, not only the unitarity right-hand cut but also the continuum contributions to the left-hand cut are calculated in the finite volume. A proper extension of the standard Veltman-Passarino identities is needed, as well as a suitable projection on the internal space spanned by the irreducible representations (irreps) of the octahedral group, based on either a finite set of cubic harmonics or the matrices which represent the irreps properly. From the ChPT we construct the IAM in the internal space, which provides the full volume dependence of the interacting energy levels of two pions scattering in the finite volume. Our results for various sets of low-energy constants show sizable corrections with respect to previous analyses in the literature for $ m_\pi L  \lesssim 2$, being compatible with lattice data on energy levels.
We expect that our analysis and results will help to optimize  the process of determining energy levels and phase-shifts with higher accuracy.

\end{abstract}


\maketitle

\section{Introduction}

In recent years, there has been a considerable increase in the application of QCD lattice techniques to extract information related to hadronic spectra 
\cite{Briceno:2017max}. Thus, lattice results for the scattering of two \cite{Beane:2007xs,Feng:2009ij,NPLQCD:2011htk,Fu:2011bz,Dudek:2013yja,Dudek:2014qha,Culver:2019qtx,Mai:2019pqr,ExtendedTwistedMass:2019omo,Fischer:2020jzp} and three particles \cite{Fischer:2020jzp,Mai:2021nul}   have paved the way for  our understanding of hadron resonances in the near future. 
In this regard, what can be determined directly in the lattice are the energy levels of particles interacting in the finite volume  for a given irreducible representation (irrep) of the symmetry group corresponding to the chosen boundary conditions in the box.
In this context, a crucial advance  has been to determine the connection of those energy levels with the  scattering amplitude in the continuum limit, from which predictions e.g. on 
 scattering lengths or resonances can be obtained. Such connection came originally from the so-called Lüscher formalism  which estabished the Quantization Condition (QC) for the two particle scattering in a box~\cite{Luscher:1986pf,Luscher:1990ux}. There are already extensions of this formalism to the scattering of three particles~\cite{Polejaeva:2012ut,Briceno:2012rv,Hansen:2014eka,Hansen:2015zga,Mai:2017bge,Hammer:2017kms,Doring:2018xxx} although our main interest here will be the two-particle case. Thus, lattice simulations with different values of the volume size and/or different boundary conditions,  provide in principle a collection of energy values to which one can relate  the physical scattering amplitude. See some examples of applications in~\cite{Wilson:2015dqa,Guo:2016zos,Guo:2018zss,Sadasivan:2021emk,Mai:2019fba,Hansen:2020otl} and in the review~\cite{Briceno:2017max}.
 
On the other hand, a natural way of dealing with hadron scattering in the finite volume is to consider the discretization in the momentum or coordinate space in Effective Field Theories (EFT's). In this way, one can arrive also to a QC for energy levels, 
determined as the values where the scattering matrix has real poles. Indeed, scattering in a finite volume is  not well-defined physically. By that concept we  actually mean the calculation of the corresponding amputated correlation function with on-shell external legs. 
The QC obtained in that way gives rise to the same levels as 
 the Lüscher method  \cite{Doring:2011vk,Briceno:2017max} up to  corrections of exponential order $\exp (-M L)$ with $M$ a typical mass for the particles involved in the scattering process and $L$ the volume size.  The main advantage of this method   is that it allows to use all the power of fully relativistic effective theories, which are based on the relevant symmetries of QCD within a well-defined power-counting scheme,  dealing with amplitudes with the correct properties of analiticity and unitarity. In particular, one should be able to compute consistently exponentially suppressed corrections to the pure Lüshcer approach. Indeed, in some cases, and specifically when a long range interaction  takes place, these exponentially suppressed corrections can be significant~\cite{Sato:2007ms}.  Such method has been carried out for the light meson-meson scattering  sector using a Bethe-Salpeter (BS) unitarized amplitude in the scalar \cite{Doring:2011vk}
and vector \cite{Chen:2012rp} channels for a vanishing total three-momentum $\vec{P}$ for the scattering process (rest frame). In \cite{Doring:2012eu} the extension  to moving frames was extensively studied, following previous works on the Lüscher formalism and scattering at finite volume for nonzero momentum     \cite{Rummukainen:1995vs,Kim:2005gf,Bour:2011ef,Davoudi:2011md}. This method has been extensively applied in~\cite{Guo:2016zos,Guo:2018zss,Molina:2015uqp,Hu:2016shf,Zhuang:2024udv,Gil-Dominguez:2023huq,Gil-Dominguez:2024zmr} for the case of two-particle systems. See also the review~\cite{Mai:2021lwb} for the application in multi-particle systems.

In this work we will analyze pion scattering in the  finite volume within Chiral Perturbation Theory  (ChPT) and its unitarized version through the  Inverse Amplitude Method (IAM). We will consider   periodic boundary conditions on a cubic box of size $L^3$ and zero total three-momentum. The main motivations  and objectives of our analysis are the following:

First of all, ChPT, as the low-energy effective theory of QCD,  provides a scattering amplitude at infinite volume which has the correct analytical structure, including a unitary cut and  a left-hand cut in the complex  $s$-plane and is the most general amplitude at a given order $\Od(p^n)$ in the low-energy chiral expansion in terms of meson masses and momenta \cite{Weinberg:1978kz,Gasser:1983yg,Gasser:1984gg,Scherer:2002tk}. 
We remark that the importance of the left-hand cut has been highlighted in recent finite-volume works \cite{Raposo:2025dkb,Raposo:2023oru,Hansen:2024ffk}.
ChPT also has by definition the correct functional dependence on meson masses, which is particularly useful when performing chiral extrapolations, either to the chiral limit or to the masses used in the lattice \cite{Nebreda:2011di}. The ChPT amplitudes depend  on a given set of Low-Energy Constants (LECs) which are universal (e.g. independent of the process considered) and can be fixed with experimental data or within lattice analyses \cite{Gasser:1983yg,Gasser:1984gg,FlavourLatticeAveragingGroupFLAG:2021npn}. Actually, the extraction from lattice data of the  LECs involved in scattering is usually performed through the Lüscher method combined with ChPT  for the chiral extrapolations to physical masses, usually near zero momentum, i.e.  where  the scattering lengths are the leading order parameters within the effective range expansion (ERE) around  threshold \cite{Beane:2007xs,Feng:2009ij,NPLQCD:2011htk,Fu:2011bz,Culver:2019qtx,Mai:2019pqr}.

It is important to remark that the above  properties are fulfilled for the complete ChPT amplitude, i.e. including properly $t$- and $u$-channel diagrams, as well as tadpole corrections to the relevant chiral order (see below). Thus, we will provide here the full ChPT amplitude at finite volume and  its projections into the three  isospin channels $I=0,1,2$ as well as into the counterpart of partial waves for the cubic box. We emphasize that the exponentially suppressed corrections coming from the $t$-, $u$- channel  and tadpole diagrams that we are including here, are actually of the same chiral order as those to the L\"uscher formula coming from $s$-channel diagrams \cite{Doring:2012eu}.  Previous works on $\pi\pi$ scattering at finite volume in ChPT  had addressed those issues partially. Namely, in \cite{Bedaque:2006yi}, the $I=2$ amplitude was studied at threshold, while in \cite{Albaladejo:2012jr} ($I=0$) and 
 \cite{Albaladejo:2013bra} ($I=1$) the finite-volume corrections to the ChPT amplitude due to the Lorentz structure of loop sum-integrals and the cubic box projections were not accounted for.

Secondly, unitarization methods have been traditionally implemented to improve the applicability energy range of ChPT demanding exact unitarity of the partial waves \cite{Truong:1988zp,Dobado:1989qm,Dobado:1996ps,Oller:1997ti,Oller:1998hw,GomezNicola:2001as,GomezNicola:2007qj}. Thus, unitarized amplitudes also generate dynamically light resonances ($\rho$, $f_0(500)$, $K_0^* (700)$, ...)  in accordance with experimental data. 
 In fact, when extracting phase shifts in the continuum from lattice energy levels using Lüscher formalism, using unitarized amplitudes to fit the lattice points allows to extend the energy applicability range with respect to the low-energy ERE \cite{Culver:2019qtx,Mai:2019pqr}.
Here, we will rely on the so-called Inverse Amplitude Method (IAM) which ensures exact unitarity of the meson-meson partial waves while  matching ChPT at low energies, and can be justified through dispersion relations  \cite{Dobado:1989qm,Dobado:1996ps,GomezNicola:2001as}. In addition, the IAM can be improved to incorporate properly the  behaviour around the so-called Adler zeros of the ChPT scattering amplitude \cite{GomezNicola:2007qj}. Such IAM amplitude, once properly projected in the cubic box, will provide a natural way to extract energy levels including  all the relevant exponential corrections within a consistent formalism, allowing to  check the robustness of other methods used previously and eventually to compare with lattice results relying on the analiticity and pion mass dependence of the ChPT amplitudes. Our analysis is  meant to be particularly useful when approaching the small-$L$ regime in lattice simulations and takes into account correctly the $t(u)$-channel contribution, which is not present in the standard BS  approach. Actually, we would like to highlight the fact that, at present, the discretization of the $t(u)-$channel loops in the IAM method and the investigation of the volume dependence obtained from those has not been done yet properly, although there have been some attempts \cite{Albaladejo:2012jr,Albaladejo:2013bra}. However, this is a relevant topic when considering the extraction of phase shifts from LQCD energy levels, since the L\"uscher formula does not incorporate the left-hand-cut effect, as it has been noted in $\pi\pi$ scattering~\cite{Rodas:2023nec}. Also, in some channels the left-hand-cut contribution can influence phase shifts for the energy levels nearby, as it happens in the heavy quark sector for the $T_{cc}(3875)$~\cite{Meng:2023bmz,Hansen:2024ffk,Gil-Dominguez:2024zmr}. 
 Actually, in \cite{Meng:2023bmz} the  procedure    to determine energy levels from scattering amplitudes \cite{Meng:2021uhz} is similar to the one we describe in section \ref{sec:cubic}, generalizing the standard partial wave expansion. However, the source of the left-hand cut in those works comes from pion exchange diagrams, while here it is originated from loops in the $t$-, $u$- channels with two-pion exchange. These terms have not been considered before in the finite volume in the meson-meson scattering problem. Furthermore, here we also treat the discretization and volume dependence of all diagrams present at one-loop NLO, including tadpole diagrams. 

Our paper is structured as follows: In section \ref{sec:IAMinf} we will provide a short review about the main properties of the IAM in the continuum. In section \ref{sec:chptfv} we will obtain the full ChPT $\pi\pi$ scattering amplitude at finite volume, paying special attention to the modifications of the usual Veltman-Passarino relations between loop  sum-integrals coming from the loss of Lorentz covariance at finite volume.  In section \ref{sec:cubic} we discuss in detail the projection of the amplitude in the cubic box within two different formalisms suitable for the description of energy levels, while in section \ref{sec:IAMV} the IAM at finite volume and the corresponding QC for the energy levels are derived. Our numerical results, comparing our approach with previous analyses,  are presented in section \ref{sec:results}. Various technical details are collected in the Appendices.

\section{The $\pi\pi$ ChPT and IAM infinite volume scattering amplitudes}
\label{sec:IAMinf}

\begin{figure}[h]
\centering
  \includegraphics[width=0.45\textwidth]{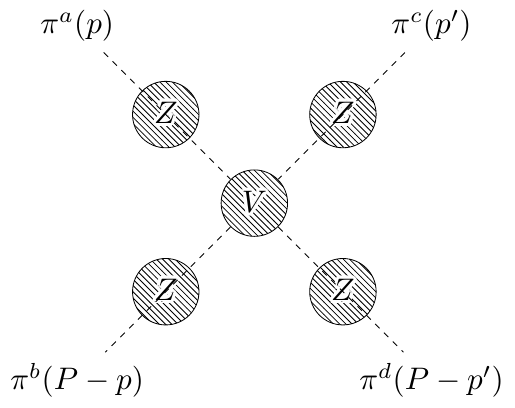}
 \caption{Generic diagram for elastic pion-pion scattering amplitude, where $V$ and $Z$ account respectively for the generic interaction vertex and the renormalization of the external legs.} 
 \label{fig:int}
 \end{figure}

The  elastic pion scattering amplitude $\pi^a \pi^b \rightarrow \pi^c \pi^d$, which we denote $T_{abcd}$ in the continuum,  can be symbolically depicted through the diagram in Fig.\ref{fig:int}, where we set our notation for the external momenta with total four-momentum $P$. The external legs are on the mass shell and their corresponding propagator must be properly renormalized.  In the infinite volume limit, the amplitude is a function  of the usual Mandelstam variables $s,t,u$ with $s+t+u=4m_\pi^2$, $m_\pi$ denoting the physical pion mass. Isospin invariance and crossing symmetry imply that only one function, $A (s,t,u)$, is needed to characterize any pion scattering amplitude, namely, 

\begin{eqnarray}
\label{crossinginf}
T_{abcd}(s,t,u)=A(s,t,u)\delta_{ab}\delta_{cd} + A(t,s,u)\delta_{ac}\delta_{bd} +\\
A(u,t,s)\delta_{ad}\delta_{bc}, \notag
\end{eqnarray}
where $A(s,t,u)$ is the $\pi^+\pi^-\rightarrow \pi^0\pi^0$ amplitude.

\begin{figure}[h]
  \centering
\includegraphics[width=0.5\textwidth]{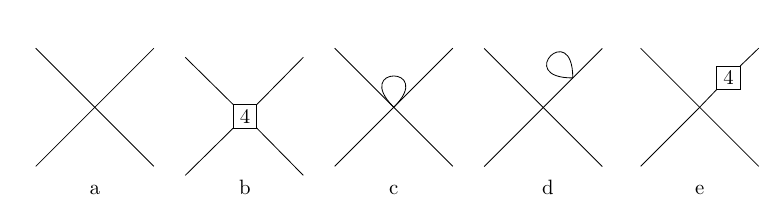}\\
\includegraphics[width=0.5\textwidth]{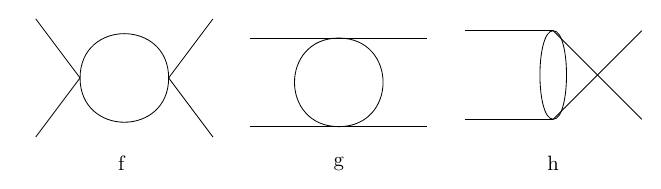}
 \caption{Diagrams contributing to pion scattering in ChPT up to $\Od(p^4)$. Vertices coming from the ${\cal L}_4$ lagrangian are explicitly indicated. The remaining vertices come from ${\cal L}_2$.}
 \label{fig:diag}
 \end{figure}

More specifically, if we consider ChPT, where Weinberg's power counting assigns a power $\Od(p^n)$ to a given diagram \cite{Weinberg:1978kz,Gasser:1983yg,Gasser:1984gg,Scherer:2002tk}, $p$ denoting 
a generic low-energy scale representing pion field derivatives, momenta or masses, the scattering amplitude up to fourth-order $\Od(p^4)$ is given by the diagrams showed in Fig. \ref{fig:diag}. Thus, we write $T=T_2+T_4+\dots$ with $T_n$ the $\Od(p^n)$ contribution, and we denote the effective lagrangian by ${\cal L}={\cal L}_2+{\cal L}_4+\dots$ with ${\cal L}_n$ the $\Od(p^n)$ lagrangian.  The explicit expressions for ${\cal L}_{2}$ and ${\cal L}_{4}$ for two-light flavors can be found in \cite{Gasser:1983yg,Scherer:2002tk}.
 Considering this counting order, the diagram a in Fig. \ref{fig:diag} is the only contribution to $T_2$. It represents the tree-level scattering amplitude with ${\cal L}_2$-type vertices, including derivative and non-derivative, the latter being proportional to the tree-level pion mass squared $m^2$. 
 Diagram b represents the tree-level scattering with vertices of the ${\cal L}_4$ type. Hence, this term is  proportional to the LECs $l_i, i=1-4$ in the $SU(2)$ ChPT formalism for two light flavours.  The third diagram, c, is the one-loop tadpole-like contribution coming from the six-pion ${\cal L}_2$-type vertices\footnote{Recall that every loop introduces an extra $\Od(p^2)$.}. Diagrams d,e  renormalize the propagator of the external legs, diagram e introducing additional LECs. At this order, such renormalization amounts, on the one hand,  to a shift in the pion mass from its tree level value $m^2$ to the one-loop one, which at this order can be identified with the physical mass $m_\pi^2$. On the other hand, there is a multiplicative residue coming from the wave function renormalization of the pion field $Z_\pi$. Thus, including diagrams d,e in the scattering amplitude with asymptotic pion states of squared mass $m^2$ is equivalent at this order to consider asymptotic states with mass $m_\pi^2$  and multiply by the $Z_\pi$ factor for the lowest order diagram a  in Fig. \ref{fig:diag}. Finally, diagrams f, g and h are  the $s$-, $t$- and $u$- channel one-loop diagrams with ${\cal L}_2$-type vertices, respectively. All the UV divergences coming from the loop diagrams in  Fig. {\ref{fig:diag} c, d, f-h,}  are absorbed in the LECs of diagrams b, e, rendering a finite and scale-independent amplitude.  

The finite part of the LECs  have to be fixed by experimental data or lattice results \cite{Hanhart:2008mx,Molina:2020qpw,FlavourLatticeAveragingGroupFLAG:2021npn}. Following the standard procedure \cite{Gasser:1983yg}, we denote by $l_i^r(\mu)$ the renormalized $SU(2)$ LEC, which depend on the dimensional regularization renormalization scale $\mu$ but are independent of the pion mass. Thus, the scattering tree-level amplitude to fourth order (diagram b in Fig.\ref{fig:diag}) depends on $l_1^r$ and $l_2^r$. In addition, we will express the scattering amplitude in terms of the physical mass $m_\pi\simeq$ 140 MeV and the physical pion decay constant $f_\pi\simeq$ 92.4 MeV, which introduces an additional LECs dependence on $l^r_{3,4}$, since they are related to the tree level values $m^2$ and $f_0$  by the  one-loop tadpole corrections 

\begin{equation}
    m_{\pi}^{2} = m^{2}  \left(1 + \frac{m^{2}}{32 \pi^{2} f_{0}^{2}} \bar{l}_{3}\right), \quad 
    f_{\pi} = f_0\left( 1 + \frac{m^{2}}{16 \pi^{2} f_{0}^{2} } \bar{l}_{4} \right)
\label{pionmassdecay}
\end{equation}
where the tree level pion mass is related to the quark masses through  $m^2=B(m_u+m_d)$ with $\langle \bar u u \rangle=\langle \bar d d \rangle=-Bf_0^2$ the tree-level quark condensate, and the $\bar l_i$  LECs are  related to the $l_i^r(\mu)$ as \cite{Gasser:1983yg}

\begin{equation}
    l_{i}^{r}(\mu) = \frac{\gamma_{i}}{32 \pi^{2}} \left( \bar{l}_{i} + \log{\frac{m^{2}}{\mu^{2}}} \right)
\end{equation}
where  $\gamma_{1} = \tfrac{1}{3}$, $\gamma_{2} = \tfrac{2}{3}$, $\gamma_{3} = -\tfrac{1}{2}$, and $\gamma_{4} = 2$. The $\bar l_i$ are independent of the scale $\mu$ but they depend on the pion mass.

Specifically, the infinite volume $\pi\pi$ scattering amplitude up to $\Od(p^4)$ in ChPT is given by \cite{Gasser:1983yg}

\begin{equation}
\label{Ainfgen}
A(s,t,u)=A_2(s,t,u)+ A_4^{pol}(s,t,u)+ A_4^{loop}(s,t,u)
\end{equation}
where $A_2$ is the $\Od(p^2)$ contribution coming from diagram a in Fig.\ref{fig:diag},

\begin{equation}\label{A2}
    A_{2} (s, t, u) = \frac{s - m_{\pi}^{2}}{f_{\pi}^{2}},
\end{equation}
and $A_4^{pol}$, $A_4^{loop}$ are, respectively, the polinomial LECs contribution coming from diagrams b,e and the loop contribution from diagrams c,d,f,g,h,

\begin{eqnarray}
   A_4^{pol} &=& \frac{1}{96 {\color{red} \pi^{2}} f_{\pi}^{4}} ( 2 ( \bar{l}_{1} - 4/3) + ( s - 4 m_{\pi}^{2})  \notag\\
    &+& ( \bar{l}_{2} - 5/6) ( s^{2} - (t - u)^{2}) \notag\\
    &+& 12 m_{\pi}^{2} s ( \bar{l}_{4} - 1) - 3 (\bar{l}_{3} + 4 l_{4} - 5) m_{\pi}^{4}),
    \label{A4pol}
\end{eqnarray}

\begin{eqnarray}
    A_{4}^{loop} &=& \frac{1}{6 f_{\pi}^{4}} ( 3 (s^{2} - m_{\pi}^{4}) \bar J(s)  \notag\\
    &+& (t (t - u) - 2 m_{\pi}^{2}t + 4 m_{\pi}^{2} u - 2 m_{\pi}^{4}) \bar{J}(t)\notag\\
    &+& (u (u - t) - 2 m_{\pi}^{2}u + 4 m_{\pi}^{2} t - 2 m_{\pi}^{4}) \bar{J}(u)).\label{A4loop}
\end{eqnarray}

In the above equations, the $\bar J$ are the loop integrals, defined in dimensional regularization $D=4-\epsilon$ as

\begin{eqnarray}
\notag
J(s)&=&-i\int \frac{d^D q}{(2\pi)^4} 
\frac{1}{q^2-m_\pi^2}\frac{1}{(q-P)^2-m_\pi^2}\\
&=& J(0)+\bar J(s)
\label{Jfun}
\end{eqnarray}
where $J(0)$ contains the divergent part as $\epsilon\rightarrow 0$ and

$$\bar{J}(s) = \frac{1}{16 \pi^{2}} \left(2 + \sigma(s) \log{  \frac{\sigma(s) - 1 }{\sigma(s) + 1}}\right)$$
where 

$$\sigma(s) = (1 - 4 m_{\pi}^{2} / s)^{1/2}$$
is the two-particle phase space for equal masses.





In order to connect to physical quantities, the infinite volume amplitude is projected into partial waves of definite isospin $I$ and angular momentum $J=L$ for pions. The isospin projection of the scattering amplitude follows the standard Clebsch-Gordan decomposition of the two-pion states for $I=0,1,2$, namely,

\begin{eqnarray}
\label{isospininf}
    T^0(s,t,u) &=& 3A(s,t,u)+A(t,s,u)+A(u,t,s) \nonumber\\
    T^1(s,t,u) &=& A(t,s,u)-A(u,t,s) \nonumber \\
    T^2(s,t,u) &=& A(t,s,u)+A(u,t,s),
\end{eqnarray}
whereas its angular momentum projection follows from rotational invariance. Thus, the projected isospin and angular momentum amplitude in the center of mass (CM) frame $\vec{P}=\vec{0}$ and $s=E^2$ with $E$ the total energy, reads,

\begin{equation}
\label{pwinf}
t^{IJ} (s)=\frac{1}{64\pi}\int_{-1}^{1} dx P_J (x) T^I \left[s,t(s,x),u(s,x) \right]
\end{equation}
with $t(s,x)=-\vert \vec{p}-\pp\vert^2=-2p^2(s)(1-x)$, $u(s,x)=-\vert \vec{p}+\pp\vert^2=-2p^2(s)(1+x)$, $p^2(s)\equiv\vert \vec{p} \,\vert^2=s\sigma^2(s)/4$ the CM-momentum squared,  $x=\cos\theta$, being $\theta$ the CM scattering relative angle between $\vec{p}$ and $\pp$ and $P_J$ the $J$-degree Legendre polynomial.   

 Note that the partial wave definition, Eq.~\eqref{pwinf}, is the infinite volume limit of the more general expansion of the amplitude for the scattering of particles without spin and with on-shell momenta in the CM frame, expressed
 in terms of the spherical harmonic funcions, $Y_{lm}$,
\begin{eqnarray}
\label{SHexp}
T^I(E;\hat p, \hat p')=128\pi^2 \sum_{l=0}^\infty \sum_{l'=0}^\infty \sum_{m=-l}^l \sum_{m'=-l'}^{l'} Y_{lm} (\theta_p,\varphi_p)\nonumber\\  t^I_{lm,l'm'} (E)  Y^*_{l'm'} (\theta'_p,\varphi'_p) \notag\ ,\nonumber\\
\end{eqnarray}
where the unitary vectors $\hat p=\vec{p}/p(s)$ and $\hat p'=\pp/p(s)$ are labeled respectively by the angles $(\theta_p,\varphi_p)$, $(\theta'_p,\varphi'_p)$. The inverse relation of \eqref{SHexp} reads

\begin{equation} \label{proj}
    t^{I}_{l m,l'm'} (E) =  \int d\Omega \, d\Omega' \, Y^{*}_{l m} (\Omega') \, T^{I}(E;\Omega, \Omega')  \, Y_{l m} (\Omega)  \ . 
\end{equation}

In the infinite volume, the amplitude $T^I(E;\hat p, \hat p')$ depends only on the scalar product $\hat p \cdot \hat p'=\cos\theta$ and there is no partial wave mixing for different angular momentum, so that rotational symmetry and the Wigner-Eckart theorem imply  $t^{I}_{l m,l'm'}=\delta_{ll'}\delta_{mm'}t^I_{l}$. Then, inserting in Eq. \eqref{SHexp} the relation

\begin{equation}
\sum_{m=-l}^l Y_{lm} (\theta_p,\varphi_p) Y^*_{lm} (\theta'_p,\varphi'_p)=\frac{2l+1}{4\pi} P_l (\cos\theta),
\end{equation}
we recover $t^I_l=t^{IJ}$ in Eq.~\eqref{pwinf}.

Unitarity above  physical  thresholds is a key requirement for the scattering amplitude in order to enlarge its applicability range and generate the lightest  resonances. For the case of a single two-particle scattering channel, and above the physical threshold, unitarity for partial waves reads

\begin{equation}
\label{unitsingle}
\im t^{IJ}(s)=\sigma (s) \vert t^{IJ}(s) \vert^2 \quad \mbox{for $s\geq 4m_\pi^2$}\ .
\end{equation}

The unitarity requirement has led to the construction of several types of unitarized amplitudes in the literature 
\cite{Truong:1988zp,Dobado:1989qm,Dobado:1992ha,Dobado:1996ps,Oller:1997ti,Oller:1998hw,GomezNicola:2001as,GomezNicola:2007qj}. Recall that  the single-channel expression, Eq.~\eqref{unitsingle},  amounts to the condition $\im (1/t)=-\sigma$ for the inverse amplitude in terms of partial waves. Thus, any amplitude of the form, $t=(\re (1/t)-i\sigma)^{-1}$, for $s\geq 4m_\pi^2$ on the real axis above the unitarity cut, is exactly unitary. Thus, unitarized amplitudes satisfy exactly Eq.~\eqref{unitsingle}, while the various approaches differ in the approximation followed for  $\re (1/t)$. Actually, note that the only imaginary part in the physical region comes from the loop integral in the $s$-channel, see Fig. \ref{fig:diag}$f$. This integral is of the type of Eq.~\eqref{Jfun}, which satisfies $\im J(s+i\epsilon)=\sigma (s)/16\pi$ for $s\geq m_\pi^2$ in the CM frame $P=(E,\vec{0})$. 

The Inverse Amplitude Method (IAM) developed in \cite{Truong:1988zp,Dobado:1989qm,Dobado:1992ha,Dobado:1996ps,GomezNicola:2001as} consists in approximating $\re(1/t)$ by its ChPT expansion up to $\Od(p^4)$, i.e, $\re (1/t)=1/t_2 - \re t_4/t_2^2+\Od(p^6)$. Recall that $t_2$ is real for real $s$ and that perturbative unitarity implies that  $\im t_4=\sigma t_2^2$ above the threshold. Thus, the unitarized IAM yields for any partial wave scattering amplitude,

\begin{equation}
\label{IAMsingle}
t_{IAM}=\frac{t_2^2 (s)}{t_2(s)-t_4(s)}
\end{equation}

The IAM amplitude above is not only unitary, but it also has the expected analytical behavior for complex $s$. Thus, although derived formally for real $s\geq 4m_\pi^2$, Eq.~\eqref{IAMsingle} is written explicitly as an analytic function in $s$ with a  unitarity right-hand cut (RHC) in the real axis for $s\geq 4m_\pi^2$, as well as a left-hand-cut (LHC) for $s<0$ coming from the $t,u$-channel diagrams, Fig.~\ref{fig:diag}g, h. Actually, note that for $s<0$, there are $x$ values in Eq.~\eqref{pwinf} such that  $t(s,x)>4m_\pi^2$ and so on for $u(s,x)$, so that both the $\bar J(t), \bar J(u)$ integrals in Eq.~\eqref{A4loop} give rise to the LHC. As mentioned, the imaginary part of the amplitude in the unitarity cut is exact by construction, while in the IAM   its value on the left cut is approximated by $\Od(p^4)$ ChPT.  Actually,
such analytic structure allows to derive formally the IAM amplitude from a dispersion relation applied to $1/t$ \cite{Dobado:1992ha} and when extended to the second Riemann sheet alongside the unitarity cut, it gives rise to poles corresponding to the lightest resonances, which for elastic $\pi\pi$ scattering are the $f_0(500)$ and the $\rho(770)$. 

Moreover,  if one imposes as an additional constraint that the zeroes of the unitarized amplitude (the so called Adler zeroes) lie at the same position that those of the perturbative ChPT one, it can be shown through dispersion relations that the IAM is modified as follows \cite{GomezNicola:2007qj}

\begin{equation}
\label{mIAMsingle}
t_{mIAM}=\frac{t_2^2 (s)}{t_2(s)-t_4(s)+A_z(s)}
\end{equation}
where 

\begin{equation}
\label{Azsingle}
A_z(s)=t_4(s_2)-\frac{(s-s_2)(s_2-s_A)}{s-s_A}\left[t_2'(s_2)-t_4'(s_2)\right],
\end{equation}
and $s_A\simeq s_2+s_4$ is the ChPT expansion of the Adler zero, i.e.,  $t_2(s_2)=t_2(s_2+s_4)+t_4(s_2+s_4)=0$, which lies below threshold.  One can check that $t_{mIAM}(s_A)=0$ with the same linear power as the perturbative amplitude and $t_{mIAM} (s_2)\neq 0,$ while $t_{IAM}$ has a quadratic zero at $s=s_2$. Actually, such mismatch for  the position and power of the Adler zero leads to the appearance of a spurious pole for $t_{IAM}$ below threshold, which disappears for $t_{mIAM}$. The above Adler zero correction is absent in the vector-isovector channel, where $s_A=s_2=4m_\pi^2$, while its effect in the scalar channels is sizable only around the subthreshold region. 

The IAM can be extended to coupled channels for a given partial wave in matrix form as \cite{GomezNicola:2001as}

\begin{equation}
\label{IAMcoupled}
T_{mIAM}=T_2(T_2-T_4+A_Z)^{-1} T_2
\end{equation}
where $T,T_2,T_4$ are matrices whose $nm$ elements 
correspond  to the amplitudes for different scattering processes $n\rightarrow m$ contributing to that partial wave, so that diagonal terms account for the elastic contributions. The elements of the Adler zero  matrix $A_Z$  are defined for every scattering process through \eqref{Azsingle}. The matrix-valued IAM at infinite volume satisfies  the unitarity relation $\im T_{mIAM}=T_{mIAM}\Sigma T_{mIAM}$, which is the extension of the single-channel expression \eqref{unitsingle} and  where $\Sigma$ is a diagonal matrix with elements $\sigma_n$, the two-particle phase space corresponding to the $n$-state above the corresponding threshold. We will actually rely on the matrix formalism in the formulation of the IAM in the finite volume, discussed in detail in section \ref{sec:IAMV}.

As mentioned above, other non-perturbative unitary amplitudes have been used in the literature. A significant case is the Bethe-Salpeter (BS) type approach, also called the Chiral Unitary approach \cite{Oller:1997ti,Oller:1998hw} used also in finite volume analyses in 
\cite{Doring:2011vk,Chen:2012rp,Doring:2012eu,Albaladejo:2012jr,Albaladejo:2013bra}. In general, the BS amplitudes have the same functional form as the IAM ones with replacements $T_2\rightarrow V$, $T_4\rightarrow VGV$ where $V$ is the potential  (tree-level amplitude) and $G$ is the loop function, which in the single-channel case reads  $G=J/16\pi$, according to our normalization, and $J$ being the $s$-channel loop integral given in Eq.~\eqref{Jfun}. Note in particular that the BS does not include the LHC coming from $t,u$-channels.  For total angular momentum $J=0$ one can follow the BS approach just with $V=T_2$ \cite{Oller:1997ti} regularizing $G$ with a cutoff $q_{max}$ which is used as a fit parameter. However, for total angular momentum $J=1$ one needs to include also the polynomial fourth-order amplitude containing the LECs, so that $T_4\rightarrow T_4^{pol}+T_2 G T_2$ \cite{Oller:1998hw} in order to reproduce experimental data and the $\rho (770)$ parameters.

\section{The $\pi\pi$ ChPT scattering amplitude at finite volume}
\label{sec:chptfv}

In this section we will provide the complete result for the finite volume ChPT pion-pion scattering amplitude up to fourth order in the momenta, emphasizing  the main formal differences with respect to the infinite volume one. 

We consider a cubic box of volume $V$ and length size $L$, such that $V=L^3$, where  the scattering amplitude is defined as the amputated four-point Green function, as depicted in Fig. \ref{fig:int}, in the finite volume, for external legs with on-shell momenta (see below). In the finite volume, all spatial momenta, internal (in loops) or external, are discretized as  $\vec{q}=(2\pi/L) \vec{n}$, with $n_i\in\IZ$. In particular,  loop integrals turn into discrete sums for the vector components of internal momenta, i.e,

\begin{equation}
\label{sum}
(-i)\int \frac{d^4 q}{(2\pi)^4}\rightarrow (-i)\frac{1}{L^3}\sum_{\vec{n}}\int \frac{d q_0}{2\pi}\equiv \SumInt
\end{equation}
which we refer to as a loop sum-integral.

In this context, the spatial rotation invariance is lost and, in the CM frame, it is replaced by invariance under the discrete octahedral symmetry group $\mathcal{O}_h$ of the cubic lattice. As a consequence, the usual Lorentz covariant structure of the amplitude has to be modified. In particular, 

\begin{enumerate}

\item The number of independent variables increases with respect to the infinite volume case. With the total three-momenta of the system, $\vec{P}$, fixed (we will consider only the static case here, $\vec{P}=\vec{0}$) one would have as free variables, $P_0$, plus the $2\times 4$ components for the incoming and outgoing four-momenta in the CM frame, $p$ and $p'$, respectively. Setting the four external legs $p_i$ on-shell leaves us with 5 independent variables. In the infinite volume case one is left with only 3 independent variables coming from the only three Lorentz invariants $p\cdot P$, $p'\cdot P$, $p\cdot p'$ which  can be written in terms of the $s,t,u$ Mandelstam variables. One can further reduce the number of independent variables to  two, from $s+t+u=\sum p_i^2=4m_\pi^2$. In the finite volume, the invariants under $\mathcal{O}_h$ extend to the six variables $\vec{p}\cdot\pp$, $\vec{p}\cdot\vec{P}$, $\pp\cdot\vec{P}$, $p_0,p'_0,P_0$ that reduce to five using again $s+t+u=4m_\pi^2$. We will choose here  the variables $s=P^2=E^2-\vert\vec{P}\vert^2$, plus the four angles labelling the directions of $\vec{p},\pp$, which we encode in the unitary vectors $\hat p=\vec{p}/\vert \vec{p} \vert$ and $\hat p'=\pp/\vert \pp \vert$, each one depending on two angles.

\item The usual Veltman-Passarino (VP) relations between loop integrals with powers of momenta $q_\mu q_\nu$ in the numerator \cite{Passarino:1978jh,Gasser:1984gg} (coming from derivative vertices) are based on Lorentz covariance and therefore  do not hold for sum-integrals of the type of Eq.~\eqref{sum} in the finite volume \cite{Bijnens:2013doa}. Consequently, in a discrete momentum space, there are more independent loop sum-integrals of that type. Actually, as we will discuss below, the fourth-order ChPT pion scattering amplitude in the finite volume cannot be just written  in terms of two of those sum-integrals, as it happens in the continuum limit.

\item The loss of rotational invariance implies that the  standard projection of the amplitude into partial waves of well-defined total angular momentum $J$ cannot be implemented. Thus, in general one has to deal with the expansion given in Eq.~\eqref{SHexp}, where partial waves $t_{lm,l'm'}$ can mix \cite{Doring:2012eu}. In fact,  as we will see in detail in section \ref{sec:cubic}, the separate dependence of the  amplitude on the external momenta $\hat p, \hat p'$  will give rise to  nontrivial projections based on the symmetry under the group $\mathcal{O}_h$. Such  projections will allow  for a  better identification of the energy labels corresponding to the irreducible representations of $\mathcal{O}_h$ and, therefore, a better comparison with lattice results.

\end{enumerate}

The Feynman diagrams contributing to the ChPT amplitude at finite volume to fourth order are again those in Fig. \ref{fig:diag}, but one needs to take into account the finite volume modifications described above. 

Regarding the loss of Lorentz covariance, 
we will rely on the analysis of the pion-pion scattering amplitude at finite temperature in \cite{GomezNicola:2002tn}, extended to pion-kaon scattering in \cite{GomezNicola:2023rqi},  where the generalization of the amplitudes to a Lorentz non-covariant situation has been  accomplished.  Thus, following Ref.~\cite{GomezNicola:2002tn}, the generalization of the relations \eqref{crossinginf} and \eqref{isospininf} for the $\pi^a\pi^b\to \pi^c\pi^d$ scattering amplitude, which we denote $\T_{abcd}$ in the finite volume, read

\begin{widetext}
\begin{eqnarray}
\label{crossing}
\T_{abcd}(S,T,U;L)=A(S,T,U;L)\delta_{ab}\delta_{cd} +
A(T,S,U;L)\delta_{ac}\delta_{bd} + A(U,T,S;L)\delta_{ad}\delta_{bc}, 
\end{eqnarray}

\begin{eqnarray}
    \T^0(S,T,U;L) &=& 3A(S,T,U;L)+A(T,S,U;L)  
    +A(U,T,S;L) \nonumber\\
    \T^1(S,T,U;L) &=& A(T,S,U;L)-A(U,T,S;L) \nonumber \\
    \label{isospin}
    \T^2(S,T,U;L) &=& A(T,S,U;L)+A(U,T,S;L)\nonumber \\
\end{eqnarray}
\end{widetext}
where $S^{\mu}=P^{\mu}$, $T^{\mu}=(p-p')^{\mu}$, $U^{\mu}=(p+p'-P)^{\mu}$ are generalized Mandelstam variables and  $A$ is the scattering amplitude for the process $\pi^+ \pi^- \rightarrow \pi^0\pi^0$ in the finite volume.

The next task is to write the $\T^I$ amplitudes above in terms of   the minimum number of independent loop sum-integrals of the form given in Eq.~\eqref{sum}. In order to do that, one needs to generalize the VP relations to a Lorentz non-covariant case \cite{GomezNicola:2002tn}. This is done in App.~\ref{app:sumint}, see Eqs.~\eqref{id1}-\eqref{idn}.  By using the relation in Eq.~\eqref{J1vsJ0} for general external momentum and Eq.~\eqref{J2vsJ0S} for the $s$-channel, one can express the $\pi\pi$ scattering amplitude only in terms of the functions $J_H, J_{s(t,u)}$ and $J_{2t(u)}$ listed in Eqs.~\eqref{JH}-\eqref{J2u}, coming from the tadpole (Fig. \ref{fig:diag} c and d), the $s$-channel diagram (Fig. \ref{fig:diag} f), and the $t$- and $u$-channel diagrams (Fig. \ref{fig:diag} g and h), respectively. In particular, $J_t$, $J_s$ and $J_u$ are just the same function $J_0(Q_0,\vec{Q},L)$ given in Eq. \eqref{J0sum} but evaluated at different values of $Q$. Namely, $J_s$ is evaluated at $Q_0=E$, $\vec{Q}=0$ and $J_t$, $J_u$ are evaluated at $Q_0=0$ and $\vec{Q}=\vec{T}=\vec{p}-\vec{p}\ '$ and $\vec{U}=\vec{p}+\vec{p}\ '$, respectively. In App.~\ref{app:sumint} we have also provided alternative ways to evaluate the above functions in different representations, which will be used extensively  throughout this work.

%

Note that in the continuum limit, Lorentz covariance implies that, on the one hand, the $J_k$ in Eq.~\eqref{intdefs} depend only on one variable $Q^2$, and, on the other hand,  the usual VP relations allow to express all relevant integrals of the type $\int f(q) \Delta (q) \Delta (q-Q)$, with $f$ a  polynomial in $q$, in terms of one integral, usually chosen as $J_0(Q^2)=J(s)$ in Eq.~\eqref{Jfun}. Therefore, the infinite volume $\pi\pi$ scattering amplitude can be expressed just in terms of $J_H$, $J(s)$, $J(t)$ and $J(u)$. Previous works \cite{Albaladejo:2012jr,Albaladejo:2013bra} assumed that the ChPT finite-volume amplitude inherits such continuum structure, just replacing the $J$ functions by their finite-volume versions, i.e,  $J(s)\rightarrow J_s$  $J(t)\rightarrow J_t, J(u)\rightarrow J_u$ which we have just shown that it does not hold due to the modification of the VP relations.

The ChPT scattering amplitude at finite volume for every isospin channel is given by

\begin{eqnarray}\label{eq:tfin}
\T^I(S,T,U;L)=\T^I (s,t,u;L\rightarrow\infty)+ 
\Delta \T^I(S,T,U;L),\nonumber\\
\end{eqnarray}
where the $\pi\pi$ scattering amplitude at infinite volume, $T^I (s,t,u)=\T^I (s,t,u;L\rightarrow\infty)$, is given by Eqs.~(\ref{Ainfgen}-\ref{A4loop}), \eqref{isospininf} and our full results for the finite volume corrections   for the three isospin channels $I=0,1,2$ are  given in App.\ref{app:amp} for zero total three-momentum.  Specifically, in Eqs.~\eqref{dT4I0}-\eqref{dT4I2} we provide $\Delta \T^I=\Delta \T^I_4$ according to the general structure, up to $\Od(p^4)$, 

\begin{eqnarray}
\label{eqx}
    \T^I (E,\hat p, \hat p';L)&=&\T_2^I (E,\ppp)+\T_4^I (E,\hat p, \hat p';L),\nonumber\\
\end{eqnarray}
with $\T_2 (E,\ppp)$ (diagram a in Fig. \ref{fig:diag}) and $\T_4 (E,\hat p, \hat p';L)$ being the $\Od(p^2)$ and $\Od(p^4)$ finite-volume scattering amplitudes, respectively. Note that the $\T_2^I$ amplitudes in the finite volume coincide with those in the infinite volume. Recall also that the UV loop divergences in dimensional regularization as $D\rightarrow 4$ appear only in the $L\rightarrow\infty$ part and are absorbed by the LECs and wave-function renormalizations, as we have explained in section \ref{sec:IAMinf}, so the finite-volume scattering amplitude is finite and scale-independent. 
\footnote{In the following we omit the superscript $I$ for the isospin projection.}

For our subsequent discussion, it will be convenient to write the  $\T_4$ contribution   as

\begin{widetext}
\begin{eqnarray}\label{T4fv}
    \T_4 (E,\hat p, \hat p';L) =\T_{4tree} (E,\ppp) +\T_{4H}(E,\ppp;L)+ \T_{4S} (E,\ppp;L) + \T_{4T} (E,\hat p, \hat p';L) + \T_{4U} (E,\hat p, \hat p';L),
\end{eqnarray}
\end{widetext}
 where $\T_{4tree}$ stands for the tree-level scattering amplitudes (diagram b in Fig. \ref{fig:diag}), while $\T_{4H}$ and $\T_{4S(T,U)}$ contain the dependence on the different loop sum-integrals, which we write as: 

\begin{eqnarray}
\T_{4H} (E,\ppp;L)&=& f_H (E,\ppp) J_H(L)
\label{TgenH}\\
\T_{4S} (E,\ppp;L)&=&
f_s (E,\ppp) J_s(E;L) \label{Tgen4S}\\ 
\T_{4T} (E,\hat p, \hat p';L)&=&
f_t (E,\ppp) J_t (E,\hat p, \hat p';L) + \nonumber\\  
&&f_{2t} (E,\ppp) J_{2t} (E,\hat p, \hat p';L) \label{Tgen4T}\\ 
\T_{4U} (E,\hat p, \hat p';L)&=&
f_u (E,\ppp) J_u (E,\hat p, \hat p';L) + \nonumber\\
&&f_{2u} (E,\ppp) J_{2u} (E,\hat p, \hat p';L)\ .
\label{Tgen4U}
\end{eqnarray}
where the $f_j$ functions can be identified in Eqs. \eqref{dT4I0}-\eqref{dT4I2} as the functions multiplying $J_{H,s,t,u,2t,2u}$. Accordingly, we will use the notation $\Delta T_{4H,4S,4T,4U}$ to denote the finite-volume corrections to the amplitude as given by Eq. \eqref{eq:tfin}, corresponding  to the  different contributions in \eqref{T4fv}.  Note that $t=-\vert \vec{p}-\pp\vert^2$ and $u=-\vert \vec{p}+\pp\vert^2$ introduce a dependence on $\ppp$ even in the tree-level contributions to the amplitude.  In addition, the tadpole contribution $\T_{4H}$  includes  the $L$-dependent renormalization of the external legs depicted in Fig.\ref{fig:diag}d. The same comments about the renormalization of the external legs and asymptotic states at infinite volume apply now, with $m_\pi^2(L)$ and  $Z_\pi(L)$ already included in the $\T_{4H}$ correction at this order and the final result for the amplitude expressed in  terms of the physical $m_\pi,f_\pi$.

The nontrivial momentum dependence in $\hat p, \hat p'$ in the amplitude reflects the loss of rotational invariance at finite volume, which, as mentioned above, implies that in general one  cannot project the amplitude into partial waves of definite angular momentum $J$. Note that, if Eqs. (\ref{TgenH})-(\ref{Tgen4U}) would depend only on $E$, $\ppp = \cos{\theta}$, such partial wave projection would be meaningful. That is the case for previous analyses of pion scattering at finite volume where only the $s$-channel diagrams are included, like those in  \cite{Doring:2011vk,Chen:2012rp}. The additional nontrivial  dependence of $J_{t(u)}, J_{2t(u)}$ on $\hat p$ and $\hat p'$  in Eqs. \eqref{Tgen4T}-\eqref{Tgen4U} comes from the $t,u$ channels,  needed for the full ChPT amplitude from which the IAM one is constructed. In the continuum limit, the sum-integrals in  Eqs. \eqref{Jt}-\eqref{J2u} become three-dimensional integrals in $\vec{q}$ whose $z$ axis can be chosen by rotation invariance as $\vec{T}$ for $J_{t,2t}$, and as $\vec{U}$ for $J_{u,2u}$, being those integrals dependent only on $t,u$ respectively, as it should be.  Therefore, we have to consider  generalized projections of the amplitude suitable to the cubic symmetry, which we explain in detail in the next section. 
In turn, recall that the continuum limit of the $J_{t,2t,u,2u}$ finite-volume sum-integrals contribute to the LHC, as explained above.

In addition, we recall that, from the $J_s$ representation \eqref{Js} one readily gets  that the ChPT amplitude diverges at the free energy levels $E_n=2\sqrt{M^2+(2\pi/L)^2 \vert \vec{n}\vert^2}$ with $\vec{n}\in\IZ^3$. Thus, in the following sections, we construct the counterpart of the IAM at finite volume suitable to reproduce interacting energy levels. The corresponding QC 
will be discussed in section \ref{sec:IAMV} in terms of the cubic symmetry projections of the amplitude. 

\section{Cubic symmetry projections for the scattering amplitude}
\label{sec:cubic}

In order to construct consistently the QC defining interacting energy levels from the ChPT scattering amplitude in the finite volume, we need first to project the amplitude onto the generalization of partial waves $t^{IJ}$ for a system with cubic symmetry.  As explained, such projections are needed due to the terms in the ChPT amplitude depending  on the in-going and out-going momenta, $\vec{p}$ and $\vec{p}\,'$ separately (from $t,u$ channels) and not only through $\vec{p}\cdot\vec{p}\,'$. An immediate consequence is that we cannot longer guarantee that the usual projection of $\T^I$ into partial waves will render diagonal terms, so that partial waves will in general mix. In addition, as we are about to see, the most general cubic projection gives rise to a nontrivial inner space for the amplitude and hence a matrix-like formulation for the IAM in the finite volume (see section \ref{sec:IAMV}) analogously to the multiple channel one discussed in section \ref{sec:IAMinf} and consistently with other approaches  \cite{Doring:2012eu,Doring:2018xxx}. In   \cite{Albaladejo:2012jr,Albaladejo:2013bra} these cubic symmetry projections were not accounted for.

We therefore seek well-defined projections on each of the ten irreducible representations (irreps $\Gamma$)  of the Octahedral group $\mathcal{O}_h$, namely $\Gamma = \{A_{1}^{\pm}, A_{2}^{\pm}, E^{\pm}, T_{1}^{\pm}, T_{2}^{\pm}\}$ whose more relevant properties for this work are reviewed in 
 App.~\ref{app:cubic}. In turn, those projections can be directly compared with lattice QCD (LQCD) analyses, where energy levels are provided for each irrep. For that purpose, we will follow \cite{Doring:2018xxx} and references therein, where that problem was studied within the context of three-body scattering. 
 
 A crucial concept here is the one of shells. We organize discretized three-momenta $\vec{p}=(2\pi/L)\vec{n}$ in the box ($\vec{n} \in \mathbf{Z}^{3}$) in shells defined as follows: two momenta belong to the same shell $r$ if they are connected by a group element $g$, i.e., all momenta in a given shell can be obtained from a certain reference vector $\vec{p}_{0}$ in that shell as

\begin{equation}
     \vec{p} = g \vec{p}_0\in r, \,\ g \in \mathcal{O}_h\ .
 \end{equation}

Since the elements  of the octahedral group are only rotations and inversions, $\det(g) = \pm 1$, so that  $|\vec{p}\,| = |\vec{p_{0}}|$. We can therefore organize the shells by increasing vector modulus in $2\pi/L$ units. Thus, the shell $r=1$ contains only the vector $(0,0,0)$, $r=2$ corresponds to the six vectors obtained from $(1,0,0)$, i.e, $(\pm 1,0,0),(0,\pm 1,0), (0,0,\pm 1)$ and so on for other shells. The multiplicity ${\vartheta}(r)$ of a shell $r$ is the number of vectors it contains. It is important to observe that although all the vectors in a shell have the same modulus,  not all the vectors with the same modulus belong to the same shell. For example, the vector $(3, 0, 0)$ and $(2, 2, 1)$ belong to different shells, $r = 9$ and $r = 10$, respectively.   All possible shells can be categorized into seven types:
\begin{equation}
\label{shelltypes}
    (0,0,0), (0,0,a), (0,a,a), (0,a,b), (a,a,a), (a,a,b),(a,b,c)
\end{equation}
with $a$, $b$, $c$ positive integers with $a \neq b \neq c$.
All shells belonging to the same type have the same multiplicity.

Following \cite{Doring:2018xxx}, we will use two different methods for determining the cubic projections: one based on the matrix representation of irreps and another one using the so called Cubic Harmonics (CH).

\subsection{Irreps matrices projection}
\label{sec:projirrep}

The general projection of a function  $f(\vec{p}\,)$ in a cubic lattice is given by

\begin{equation}
\label{fexp}
f(\vec{p}\,)=f(g\vec{p}_0)=\sum_{\Gamma} \sum_{\alpha\beta} D^\Gamma_{\alpha\beta} (g) f^\Gamma_{\beta\alpha} (\vec{p}_0)
\end{equation}
with $g\in\mathcal{O}_h$ a group element and $\vec{p}_0$ a reference vector of the shell $r$ such that $\vec{p}\in r$. In the above equation,  $\Gamma$ runs over all the group irreps, $D^\Gamma (g)$  is the matrix representation of the element $g$ in the $\Gamma$
irrep  (see table I of \cite{Doring:2018xxx}, and App.~\ref{app:cubic}) and $f^\Gamma_{\beta\alpha}$ are the counterparts of the partial waves in the angular momentum expansion in the infinite volume, i.e. the projection of $f(\vec{p}\,)$ in the irrep $\Gamma$ 
\cite{Doring:2018xxx}: 
\begin{equation}
\label{fGcoeff}
f^\Gamma_{\beta\alpha} (\vec{p_0})=\frac{s_\Gamma}{G} \sum_{g\in \mathcal{O}_h}\left[D^\Gamma_{\alpha\beta}(g)\right]^*f(g\vec{p_0})
\end{equation}
where we have made use of the orthogonality relation of the $D^{\Gamma}(g)$ matrices, 
\begin{equation}
\label{orthorel}
\sum_{g \in \mathcal{O}_h} \left[D^{\Gamma}_{\alpha\beta}(g)\right]^{*} D^{\Gamma'}_{\alpha'\beta'}(g)= \frac{G}{s_{\Gamma}} \delta_{\Gamma \Gamma'} \delta_{\alpha \alpha'}\delta_{\beta \beta'}.
\end{equation}
Here, $G$ represents the number of  elements in ${\cal O}_h$, $G=48$, and $s_\Gamma$ is the dimension of $\Gamma$.
Next, we will apply this expansion to our general expression for the amplitude in Eqs. \eqref{eqx}-\eqref{Tgen4U}:
\begin{widetext}
\begin{eqnarray}
\label{Texpgen}
\small
\T (E,\vec p, \pp;L) =\sum_{\Gamma}  \sum_{\Gamma'}  \sum_{\alpha\beta}  \sum_{\alpha'\beta'} 
D^\Gamma_{\alpha\beta} (g) t_{\beta\alpha, \alpha'\beta'}^{\Gamma\Gamma'}(\vec{p_0},\vec{p_0}') \left[D^{\Gamma'}_{\beta'\alpha'} (g')\right]^{*} 
\end{eqnarray}
\end{widetext}
being $\vec{p}=g \vec{p_0}$, $\vec{p_0}\in r$ and $\pp=g' \vec{p_0}'$ with $\vec{p_0}' \in r'$.  Note that the coefficients $t_{\beta\alpha, \alpha'\beta'}^{\Gamma\Gamma'}$ are, in principle non-diagonal in the irrep space. However, from the functional dependence discussed in section \ref{sec:chptfv} and the inherited group properties of representations, we can obtain a simplified diagonal form in the irrep space, as we show below. From \eqref{fGcoeff} and \eqref{orthorel}:
\begin{equation}
\label{Tcoffgen}
\small
t_{\alpha\beta, \alpha'\beta'}^{\Gamma\Gamma'} = \frac{s_{\Gamma}s_{\Gamma'}}{G^{2}} \sum_{g, g' \in \mathcal{O}_h } \left[D^{\Gamma}_{\beta\alpha}(g)\right]^{*} \T (E,g \vec p_{0}, g' \pp_{0};L) D^{\Gamma'}_{\beta'\alpha'}(g'). 
\end{equation}


First, we consider the contributions to the amplitude coming from  $\T_2,\T_{4tree},\T_{4H},\T_{4S}$ in Eqs. \eqref{eqx}-\eqref{Tgen4S}. Since those contributions depend only  on  $\vec{p}.\pp$,  we have
\begin{widetext}
\begin{eqnarray}
t_{\alpha\beta, \alpha'\beta'}^{\Gamma\Gamma'} 
&=&
\frac{s_{\Gamma}s_{\Gamma'}}{G^{2}} \sum_{g, g' \in \mathcal{O}_h } \left[D^{\Gamma}_{\beta\alpha}(g)\right]^{*} \T (E,(g \vec p_{0}).(g' \vec{p}\,'_{0});L) D^{\Gamma'}_{\beta'\alpha'}(g')\nonumber\\
&=& 
\frac{s_{\Gamma}s_{\Gamma'}}{G^{2}} \sum_{g, g' \in \mathcal{O}_h } \left[D^{\Gamma}_{\beta\alpha}(g)\right]^{*} \T (E,(g'^{-1} g  \vec p_{0}).( \vec{p}\,'_{0});L) D^{\Gamma'}_{\beta'\alpha'}(g')\nonumber\\
&=& 
\frac{s_{\Gamma}s_{\Gamma'}}{G^{2}} \sum_{g', g'' \in \mathcal{O}_h } \left[D^{\Gamma}_{\beta\alpha}(g' g'')\right]^{*} \T (E,(g''  \vec p_{0}).( \vec{p}\,'_{0});L) D^{\Gamma'}_{\beta'\alpha'}(g')\nonumber\\
&=& 
\frac{s_{\Gamma}s_{\Gamma'}}{G^{2}} \sum_{g', g'' \in \mathcal{O}_h } \left[\sum_{\omega} D^{\Gamma}_{\beta\omega}(g')D^{\Gamma}_{\omega\alpha}(g'')\right]^{*} \T (E,(g''  \vec p_{0}).( \vec{p}\,'_{0});L) D^{\Gamma'}_{\beta'\alpha'}(g')\nonumber\\
&=& 
\frac{s_{\Gamma}}{G}  \delta_{\Gamma \Gamma'} \delta_{\beta\beta'} \sum_{g'' \in \cal G} \left[D^{\Gamma}_{\alpha'\alpha}(g'')\right]^{*} \T (E,(g''  \vec p_{0}).( \vec{p}\,'_{0});L) \qquad \\
&&(\T=\T_2,\T_{4tree,4H,4S})\nonumber
\end{eqnarray}
\end{widetext}

Therefore,

\begin{equation}
\label{TirrepcoefnoTU}
t_{\alpha\beta, \alpha'\beta'}^{\Gamma\Gamma'}  \equiv  s_{\Gamma}\delta_{\Gamma \Gamma'} \delta_{\beta\beta'} t^{\Gamma r r'}_{\alpha \alpha'},
\end{equation}
with

\begin{eqnarray}
\label{redTg}
    t^{\Gamma r r'}_{\alpha \beta} &\equiv& \frac{1}{G} \sum_{g \in \cal G } \left[D^{\Gamma}_{\beta\alpha}(g)\right]^{*} \T (E,(g  \vec p_{0})\cdot(\vec{p}\,'_{0});L) \qquad \\\notag
    &&(\T=\T_2,\T_{4tree,4H,4S}). 
\end{eqnarray}
where  $g'' \equiv g'^{-1}g$ and we have used  that $(g\vec{a})\cdot (g\vec{b})=\vec{a}\cdot\vec{b}\Rightarrow (g_a\vec{a})\cdot (g_b\vec{b})=\vec{a}\cdot(g_a^{-1}g_b\vec{b})=(g_b^{-1}g_a\vec{a})\cdot\vec{b}$. Note that in Eq. \eqref{TirrepcoefnoTU} we have made explicit the dependence on the shells $(r,r')$ instead of their reference vectors $(\vec{p_0},\vec{p_0}')$,  following the notation of \cite{Doring:2018xxx}.
The factorization property \eqref{TirrepcoefnoTU} is, hence, a consequence of the amplitude $\T (E, (g'^{-1}g)\vec p_{0},\vec{p}\,'_{0};L)$ depending only on the group element $g'^{-1}g$, and not on $g,g'$ separately.


Consider now the contributions $\T_{4T}$ and $\T_{4U}$ to the amplitude. According to Eqs. \eqref{Tgen4T}-\eqref{Tgen4U} and \eqref{Jt}-\eqref{J2u}, they depend separately on $\vec p$ and $\vec p\ '$ as

\begin{widetext}
\begin{eqnarray}
\T (E,\vec p, \vec p';L) &=&\frac{1}{L^3}\sum_{\vec{n}} I \left[    \modq \, ,\, \qp \, ,\, \qpp \,, \,\vppp \,,\,E   \right]
\nonumber\\
&=& \frac{1}{L^3}\sum_{s}\frac{\vartheta(s)}{G} \sum_{g_q\in \mathcal{O}_h}  
I \left[ \modqzero\, ,  (g_q \vec{q_0} \cdot g\vec{p_0}) \, ,(g_q \vec{q_0}\cdot g'\vec{p_0}') \,, (g\vec{p_0}\cdot g'\vec{p_0}') \,,E\right] 
\nonumber\\
&=& \frac{1}{L^3}\sum_{s}\frac{\vartheta(s)}{G} \sum_{g_q\in \mathcal{O}_h} 
I \left[ \modqzero\, ,  (g_q \vec{q_0} \cdot g\vec{p_0}) \,, \left((g'^{-1} g_q) \vec{q_0}\cdot \vec{p_0}' \right) \,, \left((g'^{-1}  g) \vec{p_0}\cdot \vec{p_0}' \right) \,,E\right] 
\nonumber\\
&=& \frac{1}{L^3}\sum_{s}\frac{\vartheta(s)}{G} \sum_{g_q'\in \mathcal{O}_h} 
I \left[   \modqzero \,, 
\left(g_q' \vec{q_0}\cdot (g'^{-1}g) \vec{p_0}\right)\,, \left(g_q'\vec{q_0} \cdot \vec{p_0}' \right)\,, \left((g'^{-1}g) \vec{p_0}\cdot \vec{p_0}'\right)\,,E \right] 
\qquad (\T=\T_{4T,U}). 
\nonumber\\
\label{TirrepTU}
&\Rightarrow& 
\T (E,g \vec p_{0}, g' \vec{p} \,'_{0};L) = \T (E, g'^{-1} g \vec p_{0}, \vec{p}\,'_{0};L)
\qquad (\T=\T_{4T,4U}). 
\end{eqnarray}
\end{widetext}
with $\vec{q}=(2\pi/L)\vec{n}$, $\vec{q}=g_q\vec{q}_0$, $\modq=\modqzero$, $g'_q=g'^{-1}g_q$ and  $\vec{q}_0\in s$.

Therefore, following the same steps as in \eqref{TirrepcoefnoTU} for these contributions, we also obtain:

\begin{equation}
\label{TirrepcoefTU}
t_{\alpha\beta, \alpha'\beta'}^{\Gamma\Gamma'}  \equiv  s_{\Gamma}\delta_{\Gamma \Gamma'} \delta_{\beta\beta'} t^{\Gamma r r'}_{\alpha \alpha'},
\end{equation}
where 
\begin{eqnarray}
    t^{\Gamma r r'}_{\alpha \beta} &\equiv& \frac{1}{G} \sum_{g \in \cal G } \left[D^{\Gamma}_{\beta\alpha}(g)\right]^{*} \T (E,(g  \vec p_{0}),(\vec{p}\,'_{0});L)
    \label{redTgTU}
    \\ && (\T=\T_{4T,4U}). \nonumber
\end{eqnarray}

Thus, Eqs. \eqref{redTg} and \eqref{redTgTU}
provide the projection of ${\cal T}$ onto a single irrep $\Gamma$, where the corresponding projection coefficients $t_{\alpha\alpha'}^{\Gamma r r'}$ span over the internal shell and group spaces. In App.~\ref{app:cubic} we provide more details about those coefficients and their connection with angular momentum.

\subsection{Cubic Harmonics projection}
\label{sec:CH}

The cubic harmonic functions (CH), $X^{\Gamma \nu \alpha}_l (\hat p)$, are the counterpart in the {\color{red} cubic} box of the spherical harmonics $Y_{lm} (\hat p)$ in the infinite volume where the label  $\Gamma$ denotes a particular irrep, $\alpha$ stands from the basis vector of a particular $\Gamma$, with $\alpha = \{1, ..., s_{\Gamma}\}$ and $l,\nu$ are the angular momentum and degeneracy at that $l$ respectively. Thus, for every shell, there is a finite set of linearly independent CH (unlike the infinite set of $Y_{lm}$) provided e.g. in \cite{Doring:2012eu} for every shell type in Eq.\eqref{shelltypes}.
Note also that the CH are linear combinations of spherical harmonics,
\begin{equation}
\label{XvsY}
X_l^{\Gamma \nu \alpha} =\sum_{m=-l}^{l}c_{lm}^{\Gamma\nu\alpha} Y_{lm}(\hat p),
\end{equation}
with Clebsch-Gordan coefficients $c_{lm}^{\Gamma\nu\alpha}$. 

From those CH's, one can construct the functions 
$\chi_{u}^{\Gamma \alpha r}$~\cite{Doring:2018xxx}, which form an orthonormal basis for every shell $r$ with respect to the scalar product

\begin{equation}
\langle f,g \rangle_r=\frac{4\pi}{\vartheta (r)}\sum_{j=1}^{\vartheta (r)} f(\hat p_j)^* g(\hat p_j),
\end{equation}
where $\hat p_j$ are all possible orientations of the unit vector within the shell $r$. Therefore,

\begin{equation}
    \langle    \chi_{u}^{\Gamma \alpha r},\chi_{u'}^{\Gamma \alpha r} \rangle_r = \delta_{u u'},
\end{equation}
where $u$ labels the different basis vectors and $\Gamma$, $\alpha$ indices do not mix.

Thus, any  given function $f^r(\hat p_j)$, where the superscript $r$ reminds the shell to which $\hat p_j$ belongs, can be expanded in the above basis as

\begin{eqnarray}
    f^{r}(\hat{p}_{j}) &=& \sqrt{4 \pi} \sum_{\Gamma, \alpha} \sum_{u} f^{\Gamma \alpha r}_{u} \chi_{u}^{\Gamma \alpha r} (\hat{p}_{j}),  \notag
\end{eqnarray}    

where the coefficientes of the expansion

\begin{eqnarray}    
    f^{\Gamma \alpha r}_{u} &=& \frac{\sqrt{4 \pi}}{ \vartheta_{r}} \sum_{j = 1}^{\vartheta(r)} f^{r}(\hat{p}_{j}) \chi_{u}^{\Gamma \alpha r} (\hat{p}_{j}) \,\,\ 
    \end{eqnarray}
play the role of the partial waves in the finite volume.

Therefore, any contribution to the scattering amplitude in Eqs. \eqref{eqx}-\eqref{Tgen4U}, can be expanded as follows:
\begin{equation}
\label{TexpCH}
\T^{r r'} (E,\hat p_{j}, \hat p'_{j'};L) =
4\pi \sum_{\Gamma \alpha} \sum_{u u'} \chi^{\Gamma \alpha r}_{u} (\hat p_{j})
t^{\Gamma r r'}_{u u'}
\chi^{\Gamma \alpha r'}_{u'} (\hat p'_{j'}) 
\end{equation}
with
\begin{widetext}
\begin{equation}
\label{CHcoeff}
t^{\Gamma r r'}_{uu'} =\frac{4\pi}{\vartheta(r)\vartheta(r')}\sum_{j = 1}^{\vartheta(r)} \sum_{j' = 1}^{\vartheta(r')}\chi^{\Gamma \alpha r}_{u} (\hat p_j) \T^{r r'} (E,\hat p_j, \hat p'_{j'};L) \chi^{\Gamma \alpha r'}_{u'} (\hat p_{j'}'),
\end{equation}
\end{widetext}
and $\chi_{u}^{\Gamma \alpha s}$ given explicitly in table II of \cite{Doring:2018xxx}.


For completeness, we provide the connection between the irreps and cubic harmonic projections, which can be obtained as follows. On the one hand,  replacing in Eq. \eqref{CHcoeff} the expression of amplitude ${\cal T}$ in Eq. \eqref{Texpgen} and using Eq. \eqref{TirrepcoefTU}, we get:

\begin{widetext}

$$
t^{\Gamma r r'}_{uu'} \equiv \frac{4\pi}{\vartheta(r)\vartheta(r')}\sum_{j = 1}^{\vartheta(r)} \sum_{j' = 1}^{\vartheta(r')}
\sum_{\Gamma'} s_{\Gamma'}   \sum_{\alpha\alpha'\beta}  
\chi^{\Gamma \alpha r}_{u} (\hat p_j) 
D^{\Gamma'}_{\alpha\beta} (g) t_{\beta\alpha'}^{\Gamma' r r'}\left[D^{\Gamma'}_{\alpha\alpha'} (g')\right]^{*}
\chi^{\Gamma \alpha r'}_{u'} (\hat p_{j'}')
$$
\end{widetext}

On the other hand, replacing Eq. \eqref{TexpCH} in Eq. \eqref{redTgTU}, we obtain the inverse relation of the above, i.e.,  

\begin{widetext}

$$
t^{\Gamma r r'}_{\alpha \beta} \equiv 
\frac{4\pi}{G} \sum_{\Gamma' \alpha'} \sum_{u u'}  \sum_{g \in \cal G } \left[D^{\Gamma}_{\beta\alpha}(g)\right]^{*} \chi^{\Gamma' \alpha' r}_{u} (\hat{g p_{0}}) t^{\Gamma' r r'}_{u u'} 
\chi^{\Gamma' \alpha' r'}_{u'} (\hat{p_0}').
$$

\end{widetext}

It is interesting to observe that the relation of Eq. (\ref{XvsY}) allows to connect the above coefficients with the usual partial wave mixing representation with the general form provided in Eq. (\ref{SHexp}), but accounting for the additional labels $r,r'$ coming from the explicit dependence of the amplitude on the external momenta 
$\vec{p}\in r, \pp\in r'$.
Namely,
\begin{eqnarray}
\T^{r r'} (E,\hat p_{j}, \hat p'_{j};L) &=& 4\pi\sum_{lm,l'm'} Y_{lm}({\hat p_{j}}) t^{r r'}_{lm,l'm'} Y^{*}_{l'm'}({\hat p'_{j'}})\notag\\
\end{eqnarray}
with the generalized partial waves given by
\begin{eqnarray}
\label{pwmixshell}
t^{r r'}_{lm,l'm'}&=&\sum_{\Gamma \alpha} \sum_{u u'} \hat c_{lm}^{\Gamma  \alpha r,u} t_{uu'}^{\Gamma rr'} \left[\hat c_{l'm'}^{\Gamma \alpha r', u'} \right]^*
\end{eqnarray}
and $\hat c$ coefficients defined as
\begin{equation}
\label{hatccoeff}
\chi_u^{\Gamma\alpha s} (\hat p_{j}) =\sum_{lm} \hat c_{lm}^{\Gamma  \alpha s,u} Y_{lm} (\hat p_{j})\ ,
\end{equation}
since the cubic harmonic functions in the finite volume, $\chi_{u}^{\Gamma \alpha r}$, are linear combinations of those in the infinite box, $X_{l}^{\Gamma \nu \alpha}$, for different values of the angular momenta $l$. 

Therefore, one ends with a partial wave mixing representation, $t^{r r'}_{lm,l'm'}$, providing for an extension of the one discussed in \cite{Doring:2012eu} by including the shell dependence of the external momenta.\footnote{Actually, in \cite{Doring:2012eu} it was shown  within the finite volume BS approach, that, for generic nonzero total momentum, $\vec{P}$, partial waves with $l\neq l'$ do mix, although for $\vec{P}=\vec{0}$, $t_{lm,l'm'}$ is still diagonal in $l,l'=0,1,2$.}

It is instructive to calculate the $t^{r r'}_{lm,l'm'}$ elements in simple cases:

\begin{itemize}

\item[1.] $r=r'=1$: In this case, $\vec{p}=\vec{p}\,'=(0,0,0)$, $\vartheta(r)=\vartheta(r')=1$, so that there is only one independent function which can be chosen as \cite{Doring:2018xxx} $\chi_1^{A_1^+ 11}=X_0^{A_1^+ 11}=Y_{00}=1/\sqrt{4\pi}$, so that the only nonzero coefficient is $\hat c_{00}^{A_1^+ 11,1}=1$. Also, there is just one nonzero partial wave in this case, which is 
$$t_{00,00}^{11}(E;L)=t_{11}^{A_1^{+}1 1}
=\T^{1 1} (E,0,0;L).$$

\item[2.] $r=1,r'=2$: Here, $\vec{p}=(0,0,0)$ and $\vec{p}\,'=g \,(1,0,0)$ with $g$ any group element. The possible irreducible representations contributing to the $\chi_u$ functions in $r'$ are $\Gamma=A_1^+,E^+,T_1^-$ \cite{Doring:2018xxx} and the $\vartheta(r')=6$. However, since Eq. \eqref{pwmixshell} is diagonal in $\Gamma$, the only surviving contribution in the sum is that corresponding to $A_1^+$, i.e., 
\begin{align} t_{00,00}^{12}(E;L)=t_{00,00}^{21}(E;L)=t_{11}^{A_1^{+} 1 2 } \nonumber \\
=\frac{1}{6}\sum_{j=1}^6 \T (E,0,\hat p_j;L)=\T(E,0,\hat p_1;L)
\nonumber
\end{align}
where $\hat{p}_1=(1,0,0)$. Note that the particular dependence of $\T$ on $\hat p, \hat p'$ discussed in section \ref{sec:chptfv}
implies that the $\T(E,0,\hat p_j;L)$ are equal to each other for all $j=1,\dots 6$. Thus, $t_{00,00}^{12}\neq t_{00,00}^{11}$, i.e., as commented above, partial waves not only can mix but they also differ for the several shell pairs of  external momenta.

\item[3.] $r=r'=2$: In this case,  $\vec{p}=g \, (1,0,0)$, $\pp=g' \,(1,0,0)$ and $\Gamma=A_1^+,E^+,T_1^-$ can enter in the sum of Eq. \eqref{pwmixshell}. The partial waves contributing are $l=0,1,2$, which in principle  could mix. However, for the shells  of type $(0,0,a)$, see Eq. \eqref{shelltypes}, the six $X_l^{\Gamma \nu\alpha}$ contributing  to the $\chi_u^{\Gamma \alpha 2}$ satisfy that different values of $l$ correspond to different values of $\Gamma$ \cite{Doring:2018xxx} and  the coefficients $\hat c_{lm}^{\Gamma \alpha 2,u}$ share that property, so that there is no partial wave mixing in the sum of Eq. \eqref{pwmixshell} for this pair of shells. Furthermore, working out expressions of Eqs. \eqref{pwmixshell} and \eqref{hatccoeff}, with the $\T$ properties discussed in section \ref{sec:chptfv}, we obtain that the only nonzero values of $t_{lm,l'm'}^{22}$ are 
$t_{00,00}^{22}$, $t_{11,11}^{22}=t_{10,10}^{22}=t_{1-1,1-1}^{22}$, $t_{20,20}^{22}$, $t_{22,22}^{22}=t_{22,2-2}^{22}=t_{2-2,22}^{22}=t_{2-2,2-2}^{22}=t_{20,20}^{22}/2$. 
These relations coincide with those obtained  with the BS approach \cite{Doring:2012eu}, both for $l=1$ and $l=2$.

\end{itemize}

Examining other pairs of shells, according to the $X_l^{\Gamma \nu\alpha}$ functions contributing for every $\Gamma$ \cite{Doring:2018xxx}, we conclude that generalized partial waves $t_{lm,l'm'}^{r r'}$ do not mix among themselves for $l,l'=0,1,2$ but they do for higher values of $l$ in the rest frame. For those partial waves, we have that $l=0$ is  attached to $\Gamma=A_1^+$ with contributions  $t_{00,00}^{r r'}\neq 0$  for all $rr'$ pairs. In the case of $l=1$, only  $\Gamma=T_1^-$ contributes, being  $t_{1m,1m'}^{r r'}\neq 0$  for $r\neq 1$ and $r'\neq 1$, while  for $l=2$,   $\Gamma=E^+,T_2^+$ survive, being $t_{2m,2m`}^{rr'}\neq 0$ for  $r\neq 1$ and $r'\neq 1$. In the latter case, only the $E^+$ contributes for a shell of the type $(0,0,a)$, and only $T_2^+$ enters with the $(a,a,a)$ shell type. Both irreducible representations are present in the rest of shells that differ from $(0,0,0)$.

Therefore, Eq. \eqref{CHcoeff} provides the finite volume counterpart of the partial wave decomposition. However, given a finite number of $u \, u'$ combinations for a pair of external momentum shells, the representation $t_{lm,l'm'}^{rr'}$ is more suitable if we want to discuss partial wave mixing. Thus, blocks $[t^{r r'}_{lm,l'm'}]$ of size $N_r\times N_r$  generalize the $t_{lm,l'm'}$ in \cite{Doring:2012eu}, with $N_s$ the highest shell number for a given momentum cutoff $q_{max}$. If we restrict to $l\leq 2$, blocks of different $l$ do not mix, so that the energy levels quantization condition could be formulated in terms of the determinant of those particular blocks.



\section{Inverse Amplitude Method (IAM) at finite volume and quantization condition}
\label{sec:IAMV}

As we have already discussed, the energy levels for the interacting states in the cubic box arise from the Quantization Condition (QC) defined as the condition to have poles in the finite-volume scattering amplitude \cite{Doring:2011vk,Doring:2012eu,Albaladejo:2012jr,Albaladejo:2013bra,Doring:2018xxx}. As it is well-known, and as commented in the Introduction, the energy levels can be used to extract information of the interactions and properties of resonances in the continuum limit.

Here, we will construct the IAM amplitude in the finite volume taking into account the cubic symmetry projections and derive the corresponding  QC. Thus, following our discussion in section \ref{sec:IAMinf},  we will construct the finite-volume IAM amplitude for a given isospin channel as

\begin{equation}\label{eq:iamt}
\T_{IAM}
= \T_{2} \left( \T_{2} - \T_4  + \mathcal{A}_{Z} \right)^{-1} \T_{2},
\end{equation}
where we have omitted the upper index for isospin and $\T_{IAM,2,4},\mathcal{A}_Z$ are matrix-valued functions of $E,L$ in the inner space of cubic symmetry projections discussed in section \ref{sec:cubic}, i.e. the $\T$  matrix elements  are labelled $t^{\Gamma r r'}_{ij}$, where $ij$ stand either for $\alpha\alpha'$ in the irreps projection or for $uu'$ in the CH one, as evaluated from Eqs.~\eqref{redTg}, \eqref{redTgTU} and~(\ref{CHcoeff}), respectively.  In particular, $\T_{2}$ is the isospin-projected amplitude matrix constructed upon the cubic symmetry projections of $\T_2(E,\ppp)$ in Eq.~\eqref{eqx}, which is nothing but the corresponding infinite volume isospin combination, Eq.~\eqref{isospininf} of $A_2(s=E^2,t,u)$ in Eq.~\eqref{A2}, writing $t,u$ in terms of $\ppp$ as explained in section \ref{sec:chptfv}.
On the other hand, the $\T_4$ matrix elements are the cubic projections of 
$\T_4(E,\hat p, \hat p';L)$ in Eq.~\eqref{T4fv}, whose infinite-volume part is given by 
Eqs. (\ref{Ainfgen}-\ref{A4loop}), \eqref{isospininf} and whose finite-volume correction is provided in Eqs. (\ref{dT4I0}-\ref{dT4I2}). 

Finally, $\mathcal{A_{Z}}$ is the Adler zero contribution. Following our discussion in section \ref{sec:IAMinf}, we need to include this contribution in the finite volume, since the perturbative amplitude vanishes for certain channnels below threshold, generating unphysical energy levels due to spurious poles. In practice, for our present analysis, this affects only the $A_1^+$  irrep, corresponding to scalar channels, analogously to the continuum limit \cite{GomezNicola:2008}.  Its matrix elements  $(\mathcal{A_{Z}})^{\Gamma r r'}_{i j}$ are given by Eq.~\eqref{Azsingle} replacing $t_{2,4}\rightarrow$ $(t_{2,4)})^{\Gamma r r'}_{i j}$, $s_2\rightarrow (s_{2})^{\Gamma r r'}_{i j}$, the LO Adler zero such that $(t_{2})^{\Gamma r r'}_{i j}(E;L)$ vanishes at $E^2=(s_{2})^{\Gamma r r'}_{i j}$, and $\left[s_{A} = s_{2} + s_{4}\right]^{\Gamma r r'}_{i j}$, being the NLO Adler zero such that $(t_{2})^{\Gamma r r'}_{i j}(E;L)+(t_{4})^{\Gamma r r'}_{i j}(E;L)$ vanishes at $E^2=(s_{A})^{\Gamma r r'}_{i j}$.

In turn, we note that the IAM amplitude in Eq.~\eqref{eq:iamt} can be seen as the  matrix-valued $[1,1]$ Padé approximant of the perturbative amplitude\ Eq.~\eqref{eqx}, corrected with the Adler zero contribution. 
We remark that, although we are formulating the IAM in \eqref{eq:iamt} in a matrix-valued fashion as a consequence of the finite-volume structure, we are dealing with just one physical channel here, i.e. the $\pi\pi$ one. Our procedure could be generalized to a coupled-channel physical system by considering the additional indices corresponding to scattering processes as discussed at the end of section \ref{sec:IAMinf}.

Since the $\T_2$ matrix is regular, the QC for the IAM can then be read off from \eqref{eq:iamt} as

\begin{equation}
\det\left( \T_{2} - \T_4  + \mathcal{A}_{Z} \right)=0
\end{equation}

However, for practical purposes, it is convenient to write the IAM amplitude and its corresponding QC in a similar fashion 
as the BS approach in \cite{Doring:2012eu}.  For that purpose, we recall that the  $s$-channel contribution $\T_{4S}$ \eqref{Tgen4S} of the perturbative ChPT amplitude can be written as (see App.~\ref{app:amp}):

\begin{widetext}\begin{eqnarray}
\T_{4S}(E,\ppp;L)= \frac{1}{2} \frac{1}{L^3}\sum_{\vec{q}}\T_2 (E,\pq;L)I_q (E;L) \T_2 (E,\qpp;L)
\label{T4SJI02}
\end{eqnarray}
for $I=0,2$ and
\begin{eqnarray}
\T_{4S}(E,\ppp;L)&=& \frac{1}{2} \frac{1}{L^3} \sum_{\vec{q}}\T_2 (E,\pq;L)I_q (E;L) \T_2 (E,\qpp;L)-\frac{1}{6f_\pi^4}(t-u)J_H(L)
\label{T4SJI1}
\end{eqnarray}
for $I=1$.
\end{widetext}

In the above relations, we define

$$
I_q (E;L)=  \frac{1}{\omega_{\vec q}(4 \omega_{\vec q}^{2} - E^{2})}
$$
with $\vec{q}=2\pi\vec{n}/L$, $\vec{n}\in \IZ^3$ and $\omega_q  =\sqrt{ \vert\vec{q}\,\vert^{2} + m_{\pi}^{2}}$, so that

\begin{eqnarray}
J_s(E;L) &=&
\lim_{q_{max}\rightarrow\infty} 
\frac{1}{L^3}\sum_{\vec{q}}^{q_{\max}}I_q(E;L),
 \end{eqnarray}
$q_{max}$ being a momentum cutoff, which we introduce for 
our numerical analysis.

Projecting now both sides of Eqs. \eqref{T4SJI02}  and \eqref{T4SJI1} onto  irreps  using Eq.~\eqref{redTg} or onto CH using Eq.~\eqref{CHcoeff} we obtain for the corresponding matrices in those internal spaces:

\begin{equation}
    \T_{4S}=\frac{1}{2}\T_2 \J \T_2 + \T_{4H1}
\label{T4Smatrixirrep}
\end{equation}
where $\J$ has matrix elements $\J^{rr'}_{ij}=\delta_{ij}\delta^{r r'}I_s^r$ for any irrep $\Gamma$ 
with $I_{s}^{r} (E;L)= (\vartheta_{r}/L^3) I_q(E;L)$ and $\vec{q}\in r$. The $\T_{4H1}$ matrix in Eq.~\eqref{T4Smatrixirrep} is the  projection of the $J_H$ tadpole term in Eq.~\eqref{T4SJI1}, present only for $I=1$.

From Eq.~\eqref{T4Smatrixirrep}, we write the IAM amplitude, Eq.~\eqref{eq:iamt} as

\begin{align}
\label{IAMLSint}
\T_{IAM}=\T_2 \left( \T_2-\tilde \T_4-\frac{1}{2}\T_2\Delta\mathcal{J}\T_2 \right)^{-1} \T_2  \nonumber\\
=\T_2\left[ \I - \frac{1}{2}\left( \T_2-\tilde\T_4 \right)^{-1} \T_2\Delta\mathcal{J}\T_2 \right]^{-1}\left( \T_2-\tilde\T_4 \right)^{-1} \T_2
\end{align}
where

\begin{align}
    \tilde\T_4&=&\T_{4tree}+\T_{4T} + \T_{4U} + \T_{4H} + \T_{4H1} -\mathcal{A_Z}+\frac{1}{2}\T_2\J_\infty \T_2,
\end{align}
$(\Delta \J)_{ij}^{r r'}=\delta_{ij}\delta^{r r'}\Delta I_r$, $(\J_\infty)_{ij}^{r r'}=\delta_{ij}\delta^{r r'}J^{q_{max}}/N_{max}$
with $N_{max}$  the maximum number of shells considered i.e., the largest shell number for which $\modq<q_{max}$ and

\begin{eqnarray}
\Delta I_r(E;L)= I^r_{s} (E;L)-\frac{1}{N_{max}}J^{q_{max}}(E)
\end{eqnarray}

\begin{eqnarray}
    J^{q_{max}}(E) &=& \lim_{L\rightarrow\infty} J_s(E;L)=\int_{\vert q \vert<q_{max}}\frac{d^3\vec{q}}{(2\pi)^3} \frac{1}{\omega_{\vec q}(4 \omega_{\vec q}^{2} - E^{2})}\nonumber\\
    &=&\frac{1}{(4 \pi)^{2}} \left(2 \log{ \left(\frac{q_{max}}{m_{\pi}} \left[ 1 + \sqrt{1 + \frac{m_{\pi}^{2}}{q_{max}^{2}}}\right]\right)} \right.\notag\\
    &-& \left. \sigma \log{\left[\frac{1 + \sqrt{ 1 + \frac{m^{2}_{\pi}}{q^{2}_{max}} }\sigma}{-1 + \sqrt{ 1 + \frac{m^{2}_{\pi}}{q^{2}_{max}} }\sigma}\right]} \right),
\label{cutoffreg}
\end{eqnarray}

Now, from \eqref{IAMLSint}, we write the IAM in a  typical Lippmann-Schwinger form:

\begin{align}
    \T_{IAM}=\left[\I-\frac{1}{2}\V\Delta \J\right]^{-1}\V=\V+\frac{1}{2}\V \Delta \J \T_{IAM}
\label{iamrew}
\end{align}
with the ``potential" term

\begin{eqnarray}
\V&=&\T_2\left[ \T_2-\tilde\T_4 \right]^{-1} \T_2
\end{eqnarray}

The form \eqref{iamrew} resembles the BS-like construction \cite{Doring:2012eu} except that, for a better convergence of our numerical analysis, we have absorbed  $\J_\infty$ (the $L\rightarrow\infty$ part of $\J$) in the $\tilde\T_4$ part of the ``potential" $\V$. The  BS-like limit would correspond to neglecting the $\T_{T,U,H,4H1},\mathcal{A}_Z$ contributions and replacing $\T_2$ by a generic potential. The  BS-like approach used in \cite{Oller:1997ti,Oller:1998hw} for infinite volume, would amount  respectively to replace $\tilde \T_4\rightarrow \J_\infty$ and $\tilde\T_4\rightarrow \tilde \T_{4tree}+ \J_\infty$ although it must be pointed out that in \cite{Oller:1997ti,Oller:1998hw} the analyses are carried out in a coupled-channel $SU(3)$ formalism, i.e., including kaon and eta degrees of freedom. 

Therefore, since the $\V$ matrix is regular, from the IAM form \eqref{iamrew}  we obtain equivalently the QC for the energy levels as

\begin{equation}
\label{QCIAM}
    \det{\left( \delta_{rr'} \delta_{ij} - \frac{1}{2}\V^{\Gamma r r'}_{ij}\Delta {\color{red}I_r}  \right)} = 0.
\end{equation}
where the indices $ij$ stand respectively  for  $\alpha\beta$ or $u u'$ for the irrep and CH representations. 

In the following section, we will obtain our main results for the energy levels using the QC \eqref{QCIAM}.
\section{Results and discussion}
\label{sec:results}

In this section we employ the quantization condition given by Eq.\eqref{QCIAM} to analyze the energy levels obtained with the finite-volume-IAM method presented here. As we have already explained, our approach accounts for the correct volume dependence of the loops in the $s$, $t$ and $u$ channels as well as tadpole contributions, arising from the full ChPT $\pi\pi$ scattering amplitude, which at infinite volume treats adequately the right- and left-hand-cuts in the complex $s$-plane. 

We will show our results for various sets of LECs obtained through different  approaches. It is particularly useful for this analysis to recall that the IAM, being constructed from ChPT, encodes the correct pion mass dependence at low energies, which will allow us to analyze the combined dependence of the energy levels on volume and pion mass. In addition,  we will pay special attention to the comparison of the present  IAM approach with the BS one and we will study  the consistency of our analysis using the two basis, irrep or cubic harmonics, discussed in Secs.~\ref{sec:projirrep} and~\ref{sec:CH}.

Before presenting our results, it is important to highlight the domain of validity of the current approach. On the one hand, the pion mass should be sufficiently small to ensure the convergence of the chiral expansion~\cite{Hanhart:2008mx}, typically requiring $m_{\pi} < 450 \, \text{MeV}$. Note that around this upper limit of the pion mass it could also be important to consider the effect of $K\bar{K}$ and $\eta\eta$ coupled channels, and, consistently,  kaon and eta loops. However, for simplicity, we restrict to the one channel case, $\pi\pi\to\pi\pi$, considering only loops in this channel.

Regarding the box size, the size of the perturbative  corrections within our chiral expansion, as well as the reliability of our numerical analysis, set a typical lower limit $L > 1/m_{\pi}$ for energy levels. Actually, since the relevant dimensionless parameter is $m_\pi L$, as the pion mass increases, smaller box sizes become viable. In the physical limit ($m_{\pi} = m_{\pi}^{\text{exp}}$), $L_{\text{min}} \sim 1/m_{\pi} \sim 1.4\, \text{fm}$. 

Thus, we will show results typically in the range $L=1-4$ \, \text{fm}. Within that range, we will be able to see sizable, albeit small, contributions to the energy levels coming from the exponentially suppressed terms in the amplitude, consistently included within our present approach and responsible in particular of the left cut in the infinite volume limit.
Actually, as we are about to see, our lower $L$ limit still allows to distinguish the differences between our approach and previous ones such as the BS one, hence providing useful predictions for smaller lattices when they become available in LQCD.

For  the purpose of showing some results on the finite volume amplitude and  energy levels, we first take the $l_i^r$ LECs of~\cite{Hanhart:2008mx}, where $l_{1,2}^r$ are fitted to experimental data using the mIAM, while $l_{3,4}^r$ are fixed to the values in \cite{Gasser:1983yg}.

\begin{figure}
\begin{subfigure}{\linewidth}
   \includegraphics[width=\linewidth]{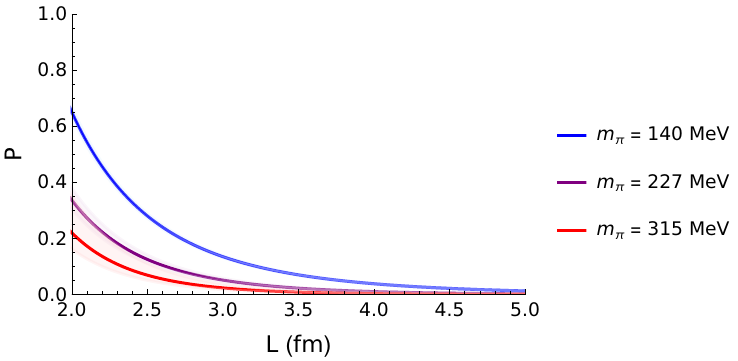}
    \caption{$I = 0$}
\end{subfigure}
\begin{subfigure}{\linewidth}
   \includegraphics[width=\linewidth]{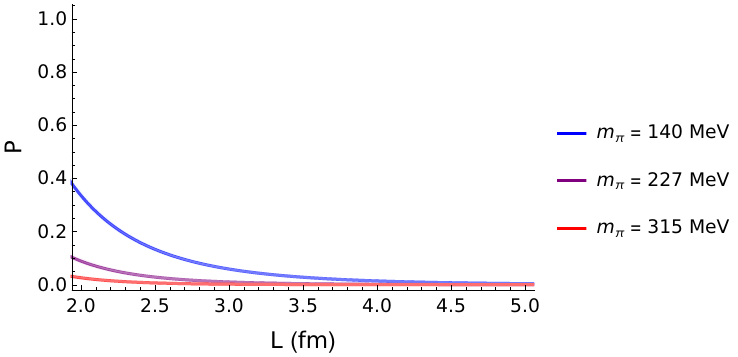}
    \caption{ $I = 2$}
\end{subfigure}
\caption{Finite-volume corrections to the Lüscher approach for the $I=0,2$ energy levels near threshold, using the LECs of Ref.~\cite{Hanhart:2008mx} and the definition of $P$ given in Eq.~\eqref{Pdeftext}. The uncertainty bands correspond to the variation of the renormalized couplings $l_{1-4}^r$.
}
 \label{fig:perc}
 \end{figure}

In order to estimate the size of the different contributions to the amplitude, let us consider first the $I=0,2$ projections of the amplitude at threshold, i.e., for $E=2m_{\pi}$, $\vec{p}=\vec{p}'=\vec{0}$. In that limit, as we show in detail App.~\ref{app:amp}, one can find closed analytic expressions for the finite-volume amplitude and the energy levels, which is particularly useful to compare the $s,t,u$ and tadpole contributions at that point. Thus, in Fig.~\ref{fig:perc} we show for different pion masses 
 (those for which there are lattice data available for energy levels and which will be analyzed  in this section) the  quantity $P$ defined in Eq. \eqref{Pdef} with respect to the amplitude $\T$ at threshold as

\begin{equation}
    P = 32\pi m_{\pi}\,
    \left| a\, \frac{\delta \T_{4}}{\T_{2}^{2}} \right|_{s \to 4m_{\pi}^{2}},
\label{Pdeftext}
\end{equation}
which corrects the energy levels near threshold with respect to the standard Lüscher approach where only the $J_s$ contribution is considered and where $\delta \T_4$ includes the  $t,u$-channel and tadpole finite-volume corrections to the amplitude at threshold, as well as the exponentially suppressed part of the $s$-channel (see App.~\ref{app:amp}), and $a$ is the continuum scattering length.
 Recall that when $P\simeq 1$, those additional contributions at the threshold energy are of the same order as the Lüscher one, as far as the energy levels are concerned.  We see that, at least around threshold, the corrections, although formally suppressed exponentially for large $m_{\pi} L$, become significant already for $m_\pi L\lesssim 2$. The results shown in Fig.~\ref{fig:perc} are calculated using the exact threshold result in Eq.\eqref{deltat4ex}. Retaining only a few terms in the asymptotic  expression in Eq.~\eqref{T4thexp} provides essentially the same curve for $m_\pi=315$ MeV, but  many terms are needed for the other two values of the pion mass for $L< 3$ fm.

We also use the set of LECs in ~\cite{Hanhart:2008mx}  to test  our results with the two methods presented here, the Irreps and CH. The Irrep method is computationally faster and simpler, making it particularly suitable for fitting procedures or quick consistency checks. However, its simplicity comes at the cost of numerical stability. In contrast, the cubic harmonic expansion offers a more accurate and robust projection, albeit at a higher computational cost. This approach is well-suited for detailed physical analyses, as it is more intuitive and closely resembles the partial wave expansion in the continuum. For this reason, throughout this section we will use the CH method when refering to the IAM.

\begin{figure}
\begin{subfigure}{\linewidth}
    \includegraphics[width=\linewidth]{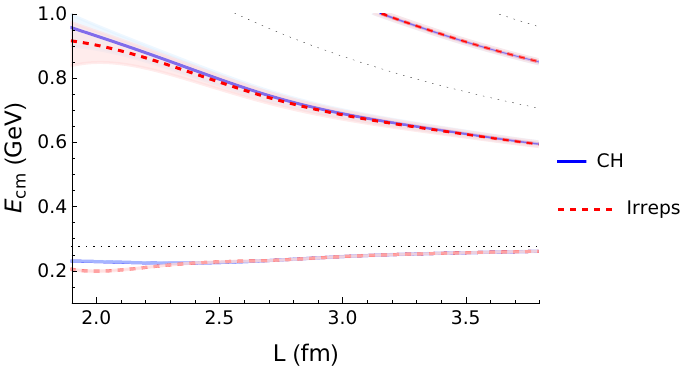}
    \caption{}
\end{subfigure}
\caption{$A_{1}^{+}$ energy levels for \(I = 0\) as a function of the box size, using the LECs given in \cite{Hanhart:2008mx}.}
 \label{fig:compaCHI}
 \end{figure}

In Fig. ~\ref{fig:compaCHI} we show the comparison of both methods. As one can see, these yield essentially identical energy levels, with only minimal numerical differences. The results are depicted for $I=0$ and a similar conclusion is reached for the cases of $I=1,2$, which we omit for simplicity. This is an important consistency check of our present analysis.

\begin{figure}[!h]
\centering
\includegraphics[width=1.0\linewidth]{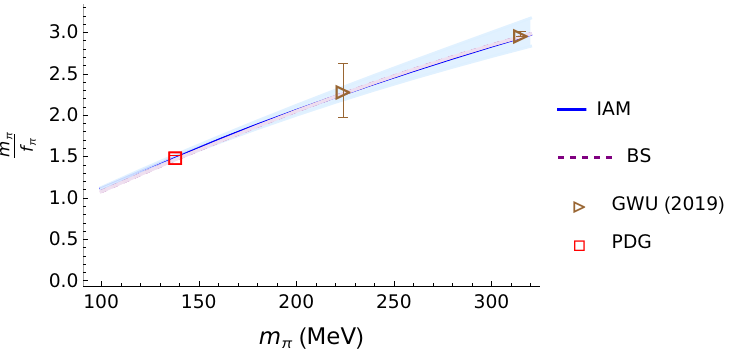}
\caption{Mass dependence of the ratio $m_{\pi}/f_{\pi}$. The data points correspond to lattice results provided by GWU collaboration \cite{Guo:2016zos,Guo:2018zss,cross:prd}. The LECs are given in Tables \ref{tab:iam} and \ref{tab:bs}. }
 \label{fig:fpiovermpilt}
 \end{figure}

Next, we show some results comparing the IAM and BS approaches. In order to produce the energy levels with the BS approach we rely on previous  finite-volume works~\cite{Doring:2011vk,Chen:2012rp}. Here and in what follows, we compare both IAM and BS approach taking the same input for the $m_\pi$ dependence of $f_\pi$, by fitting to some available LQCD data and the experimental point. For this we use the pion mass dependence of the decay constant given in~\cite{Oller:1998hw} in the case of the BS approach and Eq.~(\ref{pionmassdecay}) for IAM. The result of conducting this preliminary fit is given in Fig.~\ref{fig:fpiovermpilt}.
Regarding the $\pi\pi$ scattering amplitude, since we are interested both in scalar and vector channels, we follow the BS approach in \cite{Oller:1998hw} restricted to one channel, $\pi\pi$.

\begin{figure}
\begin{subfigure}{\linewidth}
   \includegraphics[width=\linewidth]{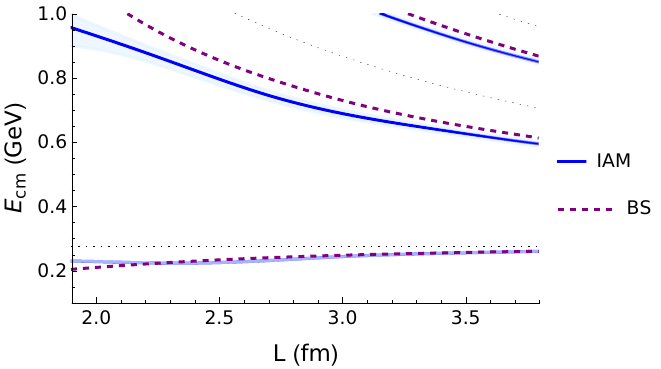}
    \caption{\(A_{1}^{+}\); \(I = 0\)}
\end{subfigure}
\begin{subfigure}{\linewidth}
   \includegraphics[width=\linewidth]{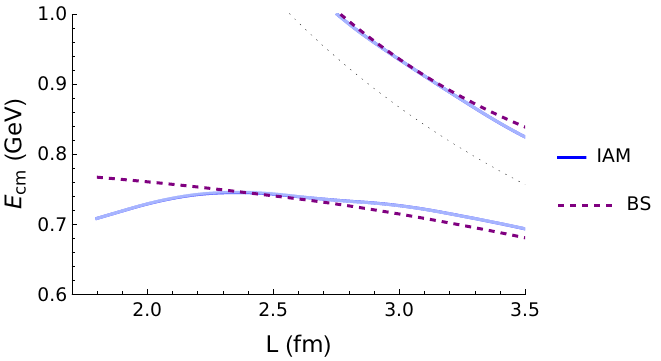}
    \caption{ $T_{1}^{-}: \, I = 1$}
\end{subfigure}
\begin{subfigure}{\linewidth}
    \includegraphics[width=\linewidth]{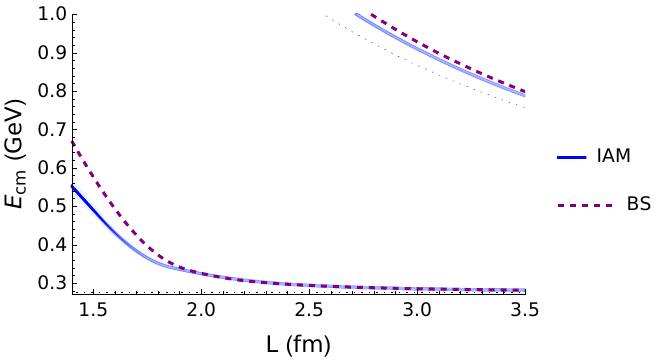}
    \caption{ $A_{1}^{+}: \, I = 2$}
\end{subfigure}
\caption{Energy levels as a function of the box size for the IAM in comparison with the BS, with the LECs given in \cite{Hanhart:2008mx, Oller:1998hw} for the IAM and BS methods respectively,restricted to the $\pi\pi$ channel, and for the physical pion mass.}
 \label{fig:compaBSIH}
 \end{figure}

 In Fig.~\ref{fig:compaBSIH} we compare the energy levels obtained with the IAM method (taking the LECs of~\cite{Hanhart:2008mx}) with those from the BS approach, for which we use the same LECs as~\cite{Oller:1998hw}, which come from a fit to experimental data within the cutoff regularization given in eq.\eqref{cutoffreg} for those loop integrals.
As commented above, we do observe sizable differences for the energy levels between both methods at small volumes, arising from the additional contributions considered in this work in the full ChPT amplitude. At higher volumes, the discrepancies are minimal, coming mostly from differences in the fits to experimental $\pi\pi$ scattering data.

In the following, we will perform our own fits to infinite-volume limit data, where we distinguish two different fitting strategies,
\begin{itemize}
    \item[StA.] We fit the experimental $\pi\pi$ scattering data in the continuum limit. Then, the energy levels with the LECs obtained from this fit are compared with available two-flavour $\pi\pi$ lattice data.
    \item[StB.] A fit of $\pi\pi$ phase shifts data from LQCD is performed~\cite{Guo:2016zos,Guo:2018zss,cross:prd}. Then, we extrapolate to the continuum limit and compare the results with the experimental data.  
\end{itemize}

Let us present our results for those two strategies. We employ a standard $\chi^2$ minimization of experimental phase shifts in the StA case  and of the phase shifts from LQCD~\cite{Guo:2016zos,Guo:2018zss,cross:prd} in the StB case. The different sets of LECs we obtain, once $I=0,1,2$ data are fitted, are the following:

\begin{itemize}
 \item[Set 1.] Corresponds to the mIAM fit to the experimental $\pi\pi$ phase shift data from \cite{estabrooks:pro, Batley:eurp, Froggatt:nucp, oller:prd, schenk:nucphys, janssen:arx, rosselet:prd, estabrooks:nucph, lindenbaum:plb}. The resulting LECs are given in Table \ref{tab:iam}.
 


\item[Set 2.] This is the set that we get from the mIAM fit to phase shifts extracted from the lattice data sets \cite{Guo:2016zos,Guo:2018zss,cross:prd}. The resulting LECs are also given in Table \ref{tab:iam}.

 \item[Set 3.] In this case, the experimental phase shift data from \cite{estabrooks:pro, Batley:eurp, Froggatt:nucp, oller:prd, schenk:nucphys, janssen:arx, rosselet:prd, estabrooks:nucph, lindenbaum:plb} are fitted within the BS approach.
Following \cite{Guo:2018zss}, we use as fit parameters $L_2$ and the following combinations of the $SU(3)$ $L_i$, 
 
 \begin{align} \label{l1L}
&\hat{l}_{1} \equiv 2 L_{4} + L_{5},\\
\label{l2L}
&\hat{l}_{2} \equiv 2 L_{1}- L_{2} + L_{3},\\
&\hat{l}_{3}\equiv 2 L_{6} + L_{8},
\end{align}

The resulting LECs values are provided in Table \ref{tab:bs}.

 \item[Set 4.]This set of LECs is also given in Table \ref{tab:bs} and  stands for the BS fit to the lattice phase shifts \cite{Guo:2016zos,Guo:2018zss,cross:prd}. 
\end{itemize}

\begin{table*}
\centering
\setlength{\tabcolsep}{1.0em}
{\renewcommand{\arraystretch}{1.5}
\begin{tabular}{|P{2.5cm}|P{1.5cm}|P{1.5cm}|P{1cm}|}
\hline
\rowcolor{gray!25} & $l^{r}_{1}$ & $l^{r}_{2}$& $\chi^2_{\mathrm{d.o.f}}$ \\
\hline
Set 1 (Exp.) & $-3.95(0.09)$ & $4.17 (0.27)$ & $0.77$\\
Set 2 (Lat.) & $-4.38(0.14)$ & $5.31 (0.29)$ & $1.55$\\
\hline
\end{tabular}}
\caption{The LECs obtained in this work within the IAM fits to experimental or lattice data, as explained in the text, where the LECs entering in the pion mass and decay constant given by Eq. \eqref{pionmassdecay} have been fixed to the values given in \cite{Gasser:1983yg}, \(l^{r}_{3} = 0.8 \, (3.8) \, \times 10^{-3}\), \(l^{r}_{4} = 6.2 \, (5.7) \, \times 10^{-3}\) and $f_{0} \simeq 87 \, \text{MeV}$, the pion decay constant in the chiral limit}. These $l_i^r(\mu)$ values correspond to the renormalization scale $\mu = 0.77$  GeV.
\label{tab:iam}
\end{table*}
\begin{table*}
\centering
\setlength{\tabcolsep}{0.5em}
{\renewcommand{\arraystretch}{1.6}
\begin{tabular}{|P{2.5cm}|P{2.5cm}|P{2.5cm}|P{1.5cm}|}
\hline
 \rowcolor{gray!25} & $\hat{l}_{2}$ & $L_{2}$ & $\chi^2_{\mathrm{d.o.f}}$ \\
\hline
Set 3 (Exp.)   & $-3.14(0.01)$ & $1.06(0.05)$ & $0.77$ \\
Set 4 (Lat.)  & $-3.574(0.005)$ & $1.16(0.06)$ & $1.27$\\
\hline
\end{tabular}}
\caption{ The LECs that we get in this work by fitting the experimental or lattice data with the  BS approach by using a cutoff  $q_{max} = 1.02 \, \text{GeV}$, $f_0 \simeq 87 \, \text{MeV}$, and $L_{1} \times 10^{3}= 0.5 $, $L_{5} \times 10^{3} = 1.7$ and $L_{8} \times 10^{4}= 2.66$ values given by \cite{Oller:1998hw}, with $\hat{l}_{1} \times 10^{3} = 2.099 (0.003)$ and $\hat{l}_{3} \times 10^{3} = -0.372 (0.003)$. The remaining constants, $L_{4}$ and $L_{6}$, are determined by fitting lattice data for the ratio $m_{\pi}/f_{\pi}$ taken from \cite{Guo:2016zos,Guo:2018zss,cross:prd}.
}
\label{tab:bs}
\end{table*}

In the following two sections, Secs. \ref{sec:exfit} and \ref{sec:ltfit}, we will discuss in more detail the results for the energy levels and phase shifts by taking StA and StB, i.e.,  fitting the experimental or lattice data, respectively. As a first general conclusion, the IAM and BS 
differences arise, as expected, for low values of $m_\pi L$ while they agree  for larger values, which serves as a robust baseline for our finite volume analysis.

\subsection{Experimental fit (StA)}
\label{sec:exfit}

\begin{figure*}[!h]
  \begin{subfigure}[t]{0.48\linewidth}
    \includegraphics[width=\linewidth]{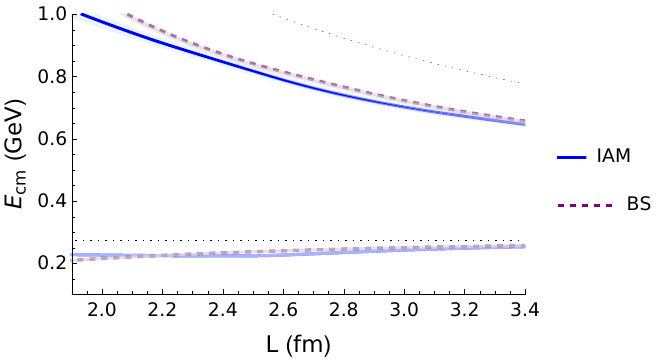}
    \caption{}
\end{subfigure}
  \begin{subfigure}[t]{0.48\linewidth}
    \includegraphics[width=\linewidth]{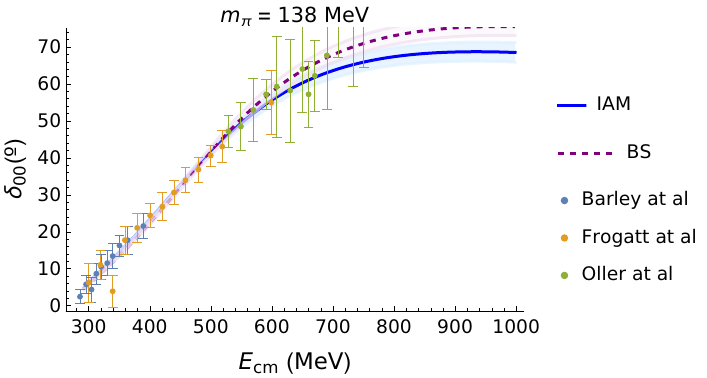}
    \caption{}
\end{subfigure}
  \begin{subfigure}[t]{0.48\linewidth}
    \includegraphics[width=\linewidth]{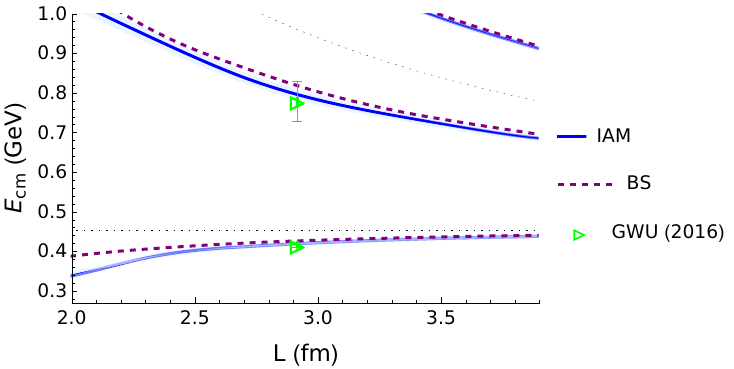}
    \caption{}
\end{subfigure}
  \begin{subfigure}[t]{0.48\linewidth}
    \includegraphics[width=\linewidth]{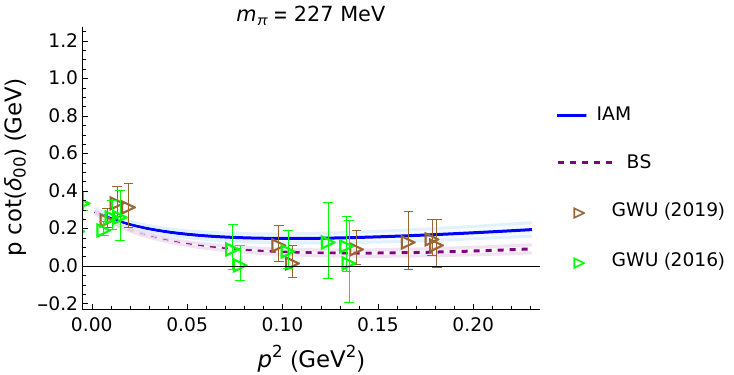}
    \caption{}
\end{subfigure}
\begin{subfigure}[t]{0.48\linewidth}
    \includegraphics[width=\linewidth]{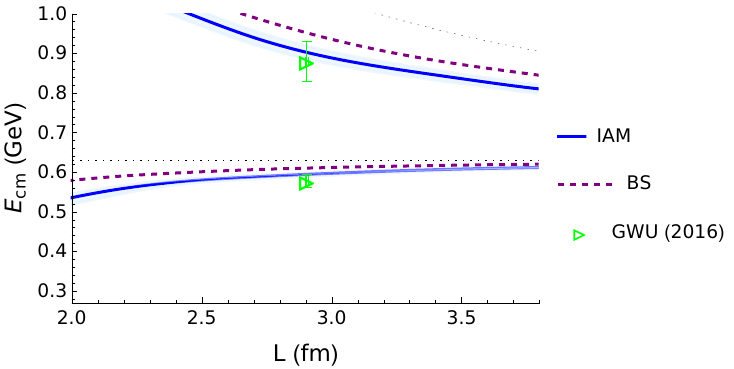}
    \caption{}
\end{subfigure}
  \begin{subfigure}[t]{0.48\linewidth}
    \includegraphics[width=\linewidth]{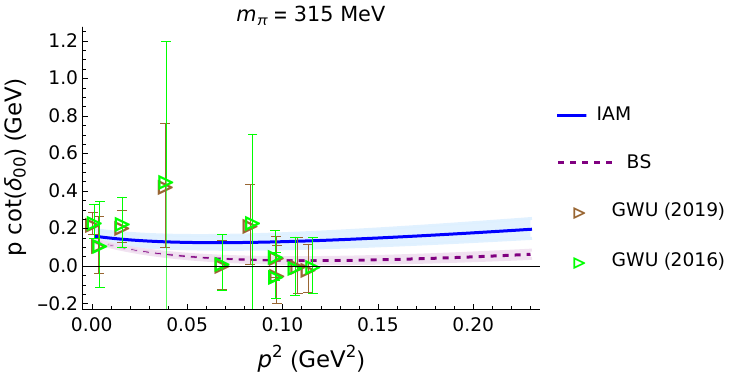}
    \caption{}
\end{subfigure}
 \caption{\(A_{1}^{+}\) energy levels for \(I = 0\) as a function of the box size. The LECs are given in tables. \ref{tab:iam} and \ref{tab:bs}, from the fitting of experimental data \cite{estabrooks:pro, Batley:eurp, Froggatt:nucp, oller:prd, schenk:nucphys, janssen:arx, rosselet:prd, estabrooks:nucph, lindenbaum:plb}.The uncertainty bands correspond to the LECs uncertainties in those tables.}
\label{fig:A10ex}
 \end{figure*}

\begin{figure*}
  \centering
 \begin{subfigure}[t]{0.48\linewidth}
    \centering
    \includegraphics[width=\linewidth]{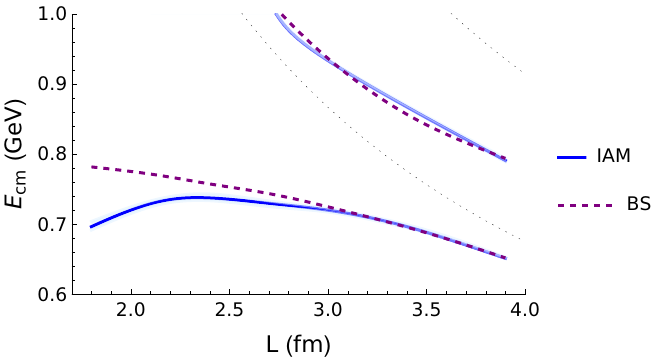}
    \caption{}
  \end{subfigure}
  \hfill
  \begin{subfigure}[t]{0.48\linewidth}
    \centering
    \includegraphics[width=\linewidth]{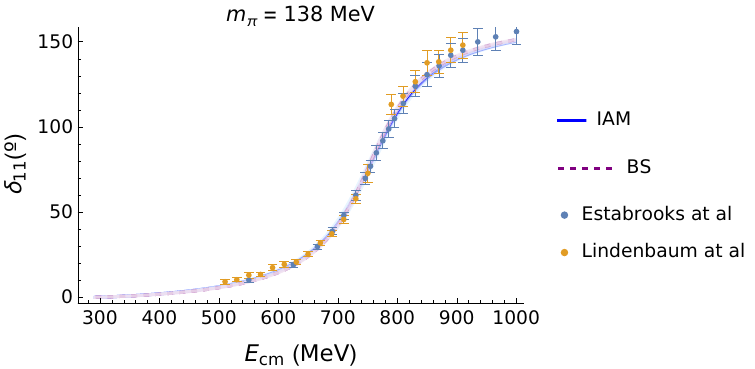}
    \caption{}
  \end{subfigure}
  
  \begin{subfigure}[t]{0.48\linewidth}
    \centering
    \includegraphics[width=\linewidth]{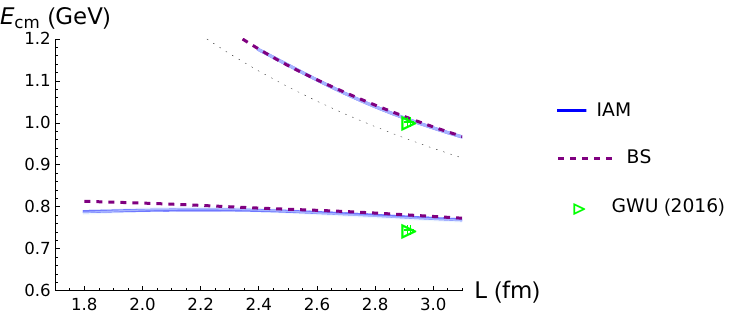}
    \caption{}
  \end{subfigure}
  \hfill
  \begin{subfigure}[t]{0.48\linewidth}
    \centering
    \includegraphics[width=\linewidth]{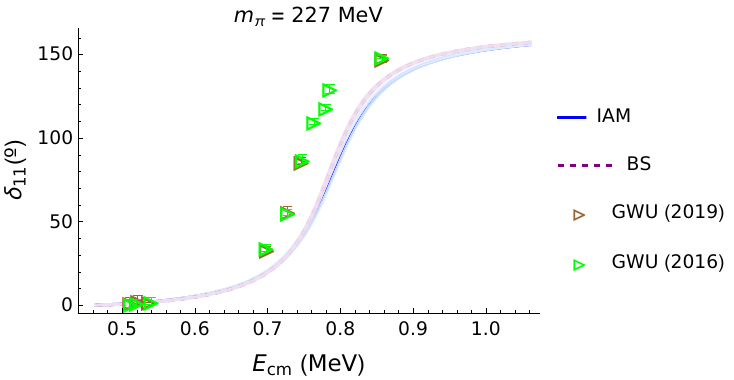}
    \caption{}
  \end{subfigure}

  \begin{subfigure}[t]{0.48\linewidth}
    \centering
    \includegraphics[width=\linewidth]{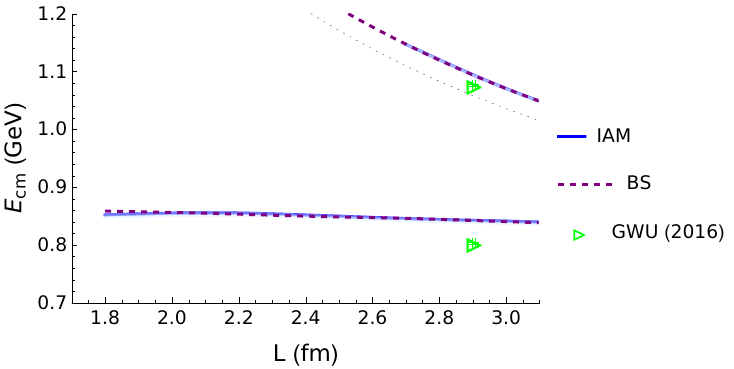}
    \caption{}
  \end{subfigure}
  \hfill
  \begin{subfigure}[t]{0.48\linewidth}
    \centering
    \includegraphics[width=\linewidth]{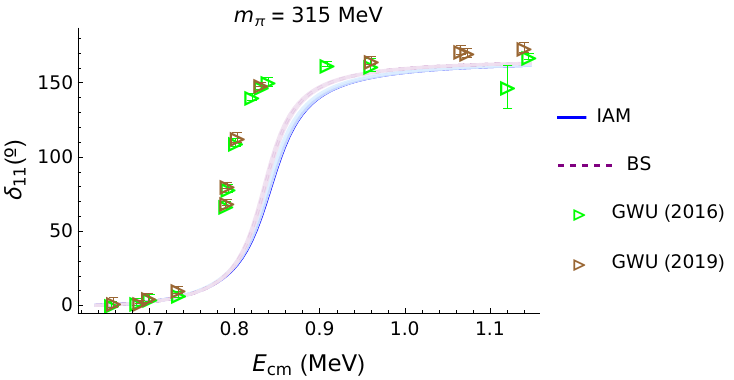}
    \caption{}
  \end{subfigure}

  \caption{\(T_{1}^{-}\) energy levels and phase shifts for \(I = 1\) as functions of the box size and energy, respectively. The LECs are taken from Tables~\ref{tab:iam} and \ref{tab:bs}, corresponding to the fit to experimental data. The uncertainty bands correspond to the LECs uncertainties in those tables.} 
  \label{fig:T11ex}
\end{figure*}

\begin{figure*}[!h]
  \centering
  \begin{subfigure}[t]{0.48\linewidth}
    \centering
    \includegraphics[width=\linewidth]{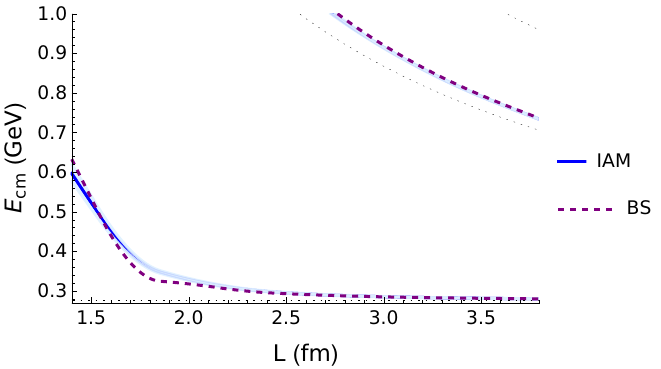}
    \caption{}
  \end{subfigure}
  \hfill
  \begin{subfigure}[t]{0.48\linewidth}
    \centering
    \includegraphics[width=\linewidth]{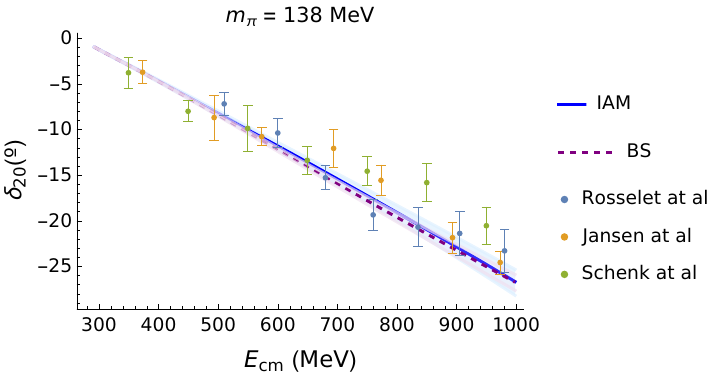}
    \caption{}
  \end{subfigure}
  \begin{subfigure}[t]{0.48\linewidth}
    \centering
    \includegraphics[width=\linewidth]{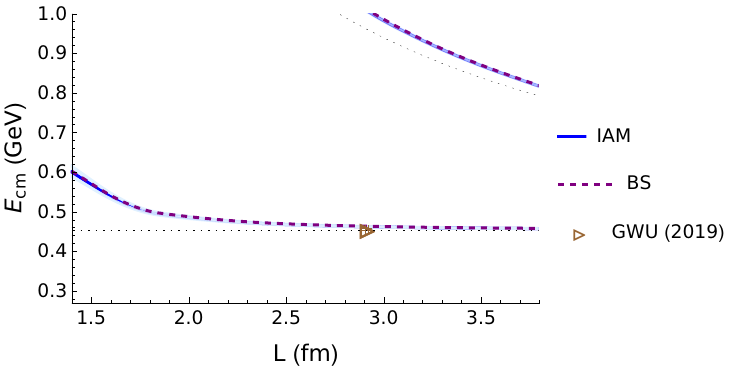}
    \caption{}
  \end{subfigure}
  \hfill
  \begin{subfigure}[t]{0.48\linewidth}
    \centering
    \includegraphics[width=\linewidth]{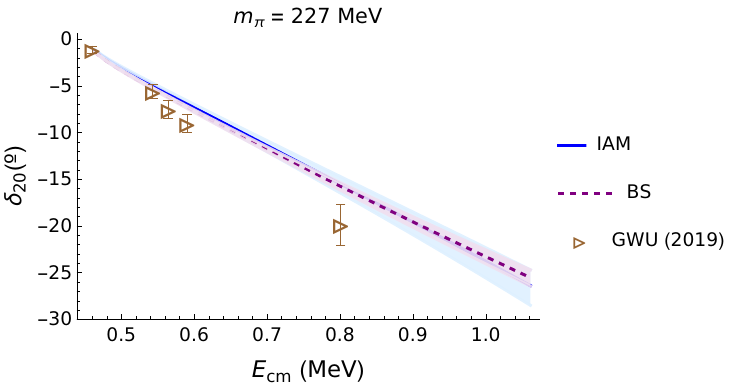}
    \caption{}
  \end{subfigure}
  \begin{subfigure}[t]{0.48\linewidth}
    \centering
    \includegraphics[width=\linewidth]{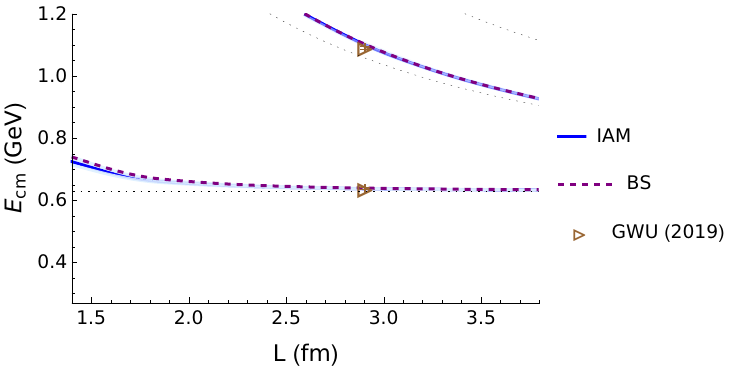}
    \caption{}
  \end{subfigure}
  \hfill
  \begin{subfigure}[t]{0.48\linewidth}
    \centering
    \includegraphics[width=\linewidth]{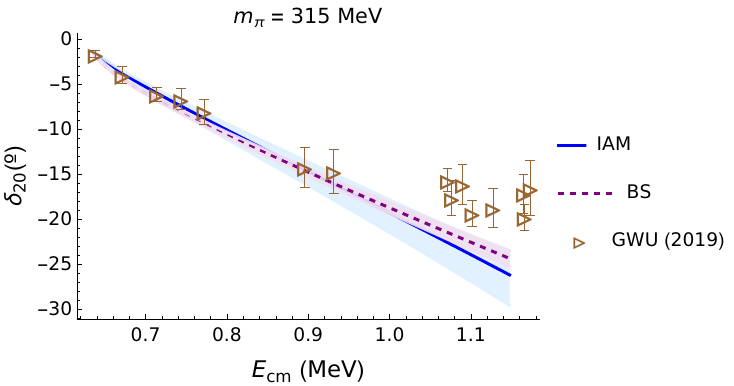}
    \caption{}
  \end{subfigure}
  \caption{\(A_{1}^{+}\) energy levels and phase shifts for \(I = 2\) as functions of the box size and energy, respectively. The LECs are taken from Tables~\ref{tab:iam} and \ref{tab:bs}, corresponding to the fit to experimental data. The uncertainty bands correspond to the LECs uncertainties in those tables.}
  \label{fig:A12ex}
\end{figure*}

\begin{figure}
  \begin{subfigure}{\linewidth}
    \includegraphics[width=\linewidth]{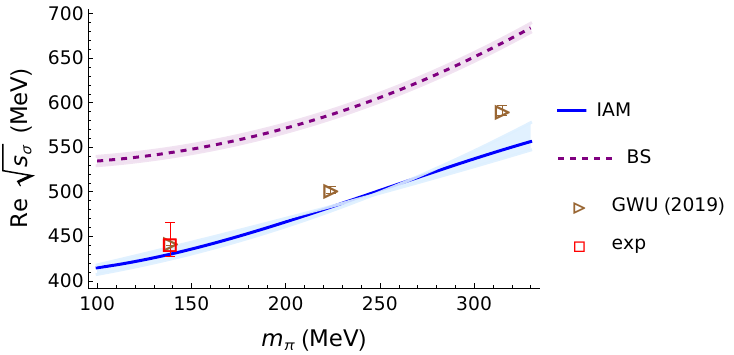}
    \caption{$I=0$}
\end{subfigure}
\begin{subfigure}{\linewidth}
    \includegraphics[width=\linewidth]{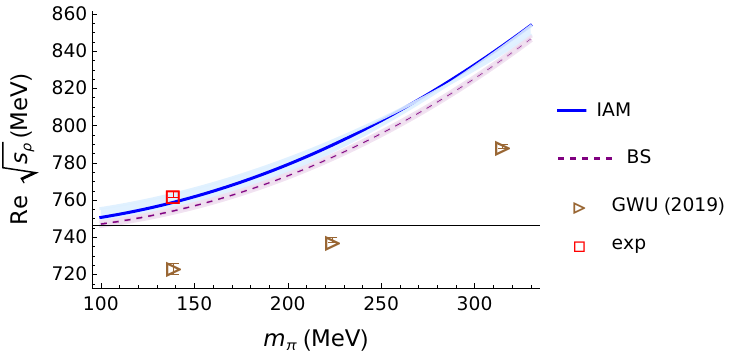}
    \caption{$I=1$}
\end{subfigure}
 \caption{  Mass dependence of the $I=0,1$ poles for  $\pi\pi \rightarrow \pi\pi$ scattering as a function of pion mass. The data points correspond to lattice results provided by GWU collaboration \cite{Guo:2016zos,Guo:2018zss,cross:prd} and the experimental PDG values. The LECs are given in the Tables. \ref{tab:iam} and \ref{tab:bs} for experimental data. The uncertainty bands correspond to the LECs uncertainties in those tables.
 }
 \label{fig:polesex}
 \end{figure}

In this section we discuss the results for the energy levels obtained with the StA. The LECs obtained from fitting the $\pi\pi$ scattering data are given in Tables~\ref{tab:iam} for the IAM and \ref{tab:bs} for the BS. The phase shifts in $(I,J)=(0,0),(1,1),(2,0)$ are shown in Figs. ~\ref{fig:A10ex}, ~\ref{fig:T11ex} and ~\ref{fig:A12ex} from top to bottom, respectively, for the IAM (blue) and BS approach (dashed-purple). For the physical pion mass, both produce a reasonable description of the experimental data. The energy levels obtained from this strategy are depicted in those same three figures for $I=0,1,2$, respectively, both for the physical pion mass and for different uphysical pion masses for which there are LQCD data available~\cite{Guo:2016zos,Guo:2018zss,cross:prd}. Recall that the LECs used here are independent of the pion mass.
Phase shifts are also shown for unphysical masses, compared with lattice data.
In addition, we show in Fig. \ref{fig:polesex} the results  for the $f_0(500)$ and $\rho$ poles for the pion masses considered here and the StA LECs. 

As far as the LQCD data (energy levels and phase shifts) are concerned, for $I=0$ and $2$, both methods (BS and IAM) describe reasonably well the lattice data, the new effects included in our present approach being small for $m_\pi L>2$. There are clear differences between both methods in the phase shifts produced in $I=0$, still, these are compatible with the large errors present in the LQCD simulation in this case. Our results for energy levels are also similar to those obtained previously in Fig. \ref{fig:compaBSIH}, except for $I=2$, where we obtain a better agreement between IAM and BS with the StA LECS.
The differences between both methods for the physical pion mass at short box lenghts can be distinguished, being more sizable for the lowest energy level in the $T_1^-$ irrep ($I=1$) depicted in Fig. \ref{fig:T11ex}. Both methods fail to describe the LQCD lowest energy level for $I=1$ and $m_\pi=227, 315$ MeV.  A possible reason of the poor description of the $I=1$ ground state energy in StA is that, in a two- light-dynamical-quark simulation there are no strange quarks as in the experiment, to which the LECs have been fitted to. This has been noted previously in the analyses of Refs.~\cite{Guo:2016zos,Guo:2018zss}.

Regarding the results in Fig. \ref{fig:polesex}, since only experimental data are fitted here, we expect more differences with lattice points  in the continuum limit for unphysical masses. For the poles, both the IAM and the BS  predict successfully the physical values, and the IAM is not far from lattice points either in the case of $I=0$,  while BS is known to work better for $I=1$ than for $I=0$. As commented previously, for $I=1$, a possible reason for discrepancy with LQCD data is the absence of the strange quark in the simulations of ~\cite{Guo:2016zos,Guo:2018zss}. The presence of $K\bar{K}$ loops can affect the pole position in this case.

In the next subsection, we will provide results from StB, fitting the lattice data. 


\subsection{Lattice data fit (StB)}
\label{sec:ltfit}
\begin{figure*}
  \centering
  \begin{subfigure}[t]{0.48\linewidth}
    \centering
    \includegraphics[width=\linewidth]{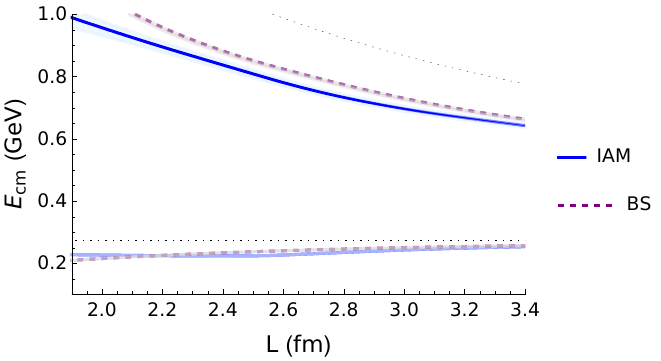}
    \caption{}
  \end{subfigure}
  \hfill
  \begin{subfigure}[t]{0.48\linewidth}
    \centering
    \includegraphics[width=\linewidth]{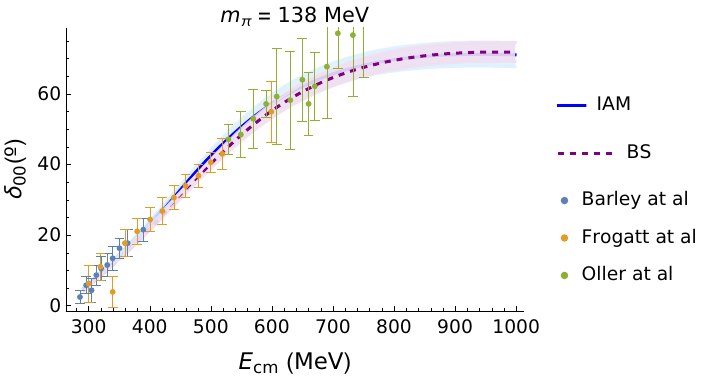}
    \caption{}
  \end{subfigure}
  \begin{subfigure}[t]{0.48\linewidth}
    \centering
    \includegraphics[width=\linewidth]{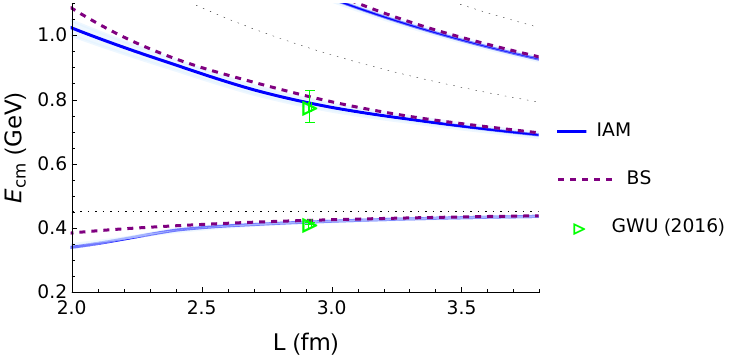}
    \caption{}
  \end{subfigure}
  \hfill
  \begin{subfigure}[t]{0.48\linewidth}
    \centering
    \includegraphics[width=\linewidth]{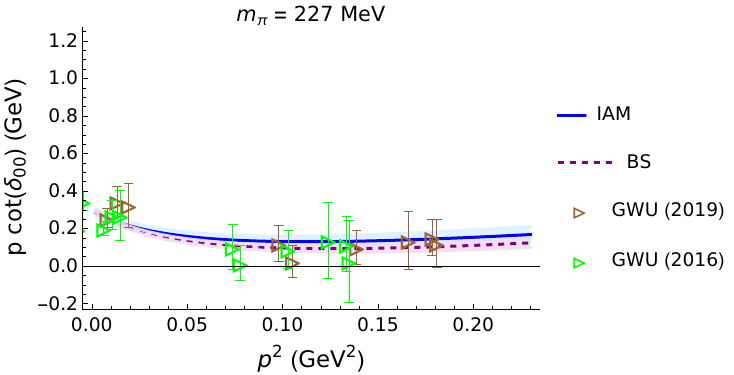}
    \caption{}
  \end{subfigure}
  \begin{subfigure}[t]{0.48\linewidth}
    \centering
    \includegraphics[width=\linewidth]{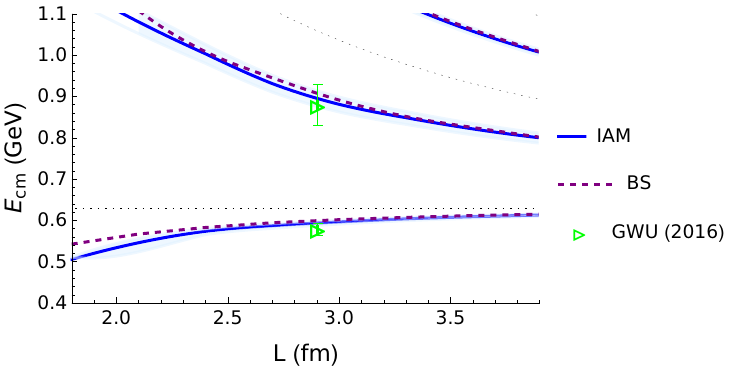}
    \caption{}
  \end{subfigure}
  \hfill
  \begin{subfigure}[t]{0.48\linewidth}
    \centering
    \includegraphics[width=\linewidth]{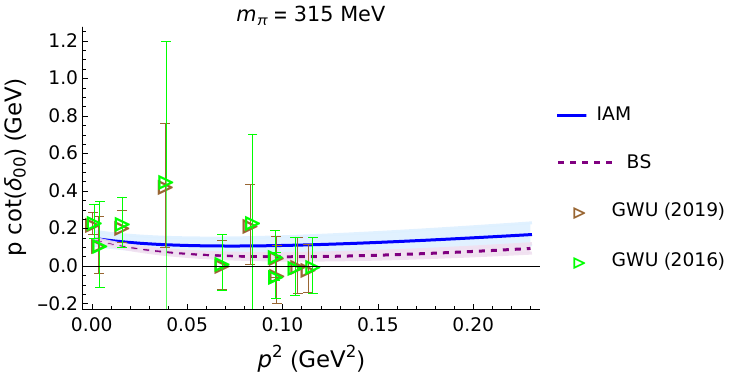}
    \caption{}
  \end{subfigure}
  \caption{\(A_{1}^{+}\) energy levels and phase shifts for \(I = 0\) as functions of the box size and energy, respectively. The LECs are taken from Tables~\ref{tab:iam} and \ref{tab:bs}, corresponding to the fit to lattice data. The uncertainty bands correspond to the LECs uncertainties in those tables.}
  \label{fig:A10lt}
\end{figure*}
\begin{figure*}
  \centering
  \begin{subfigure}[t]{0.48\linewidth}
    \centering
    \includegraphics[width=\linewidth]{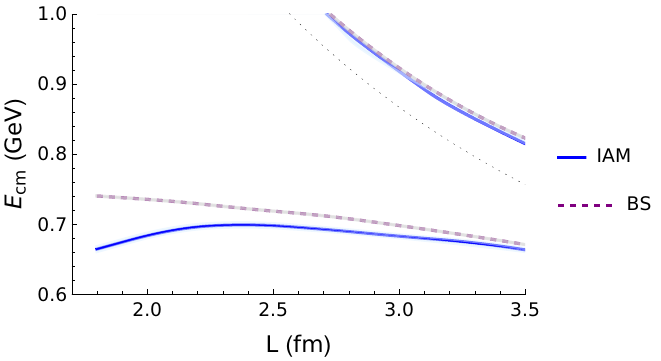}
    \caption{}
  \end{subfigure}
  \hfill
  \begin{subfigure}[t]{0.48\linewidth}
    \centering
    \includegraphics[width=\linewidth]{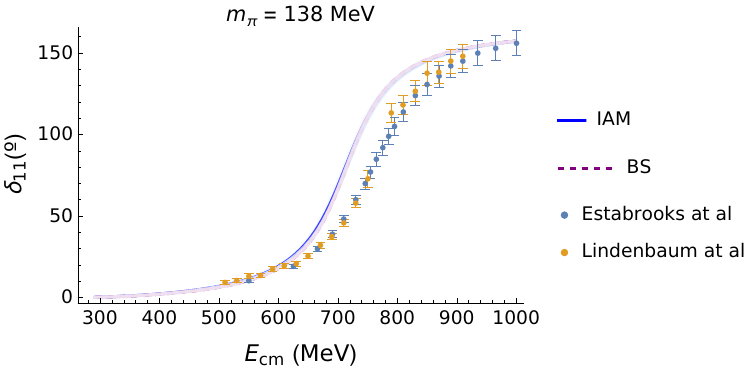}
    \caption{}
  \end{subfigure}
  \begin{subfigure}[t]{0.48\linewidth}
    \centering
    \includegraphics[width=\linewidth]{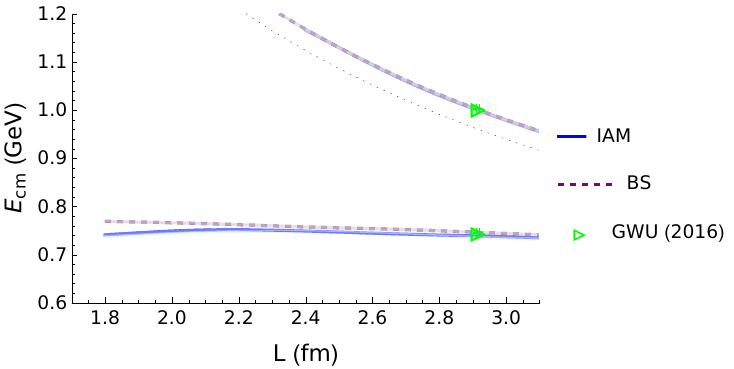}
    \caption{}
  \end{subfigure}
  \hfill
  \begin{subfigure}[t]{0.48\linewidth}
    \centering
    \includegraphics[width=\linewidth]{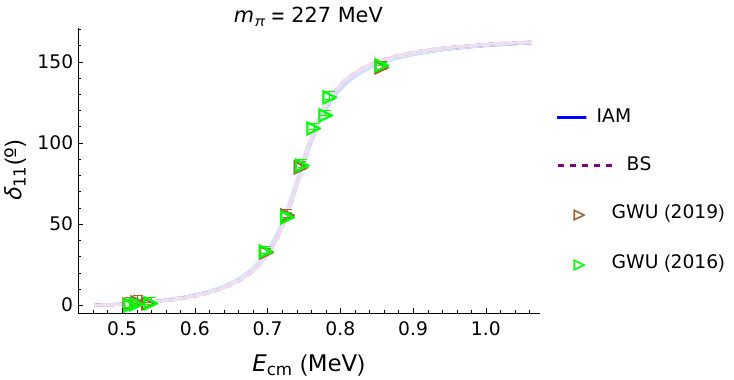}
    \caption{}
  \end{subfigure}
  \begin{subfigure}[t]{0.48\linewidth}
    \centering
    \includegraphics[width=\linewidth]{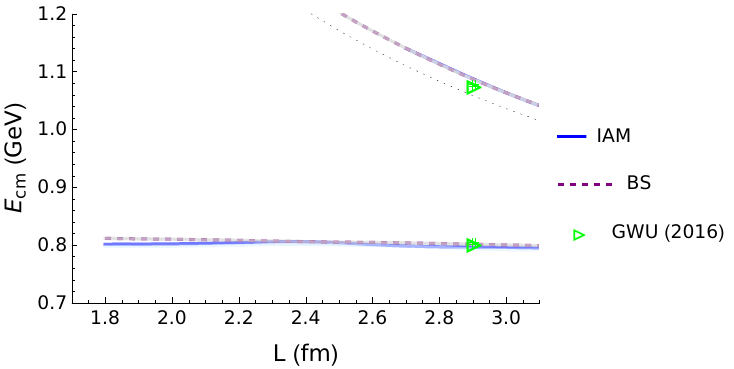}
    \caption{}
  \end{subfigure}
  \hfill
  \begin{subfigure}[t]{0.48\linewidth}
    \centering
    \includegraphics[width=\linewidth]{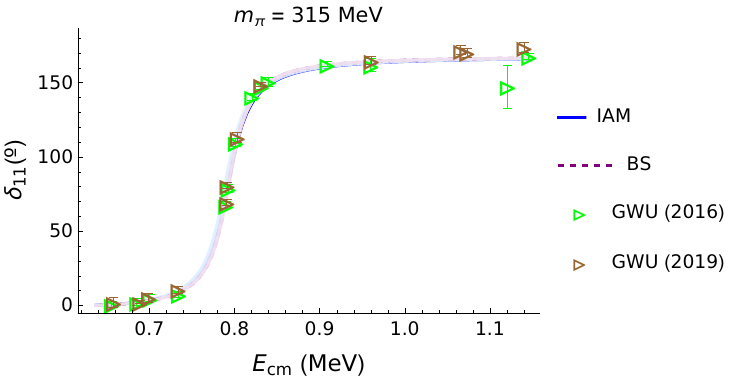}
    \caption{}
  \end{subfigure}
  \caption{\(T_{1}^{-}\) energy levels and phase shifts for \(I = 1\) as functions of the box size and energy, respectively. The LECs are taken from Tables~\ref{tab:iam} and \ref{tab:bs}, corresponding to the fit to lattice data. The uncertainty bands correspond to the LECs uncertainties in those tables.}
  \label{fig:T11lt}
\end{figure*}
\begin{figure*}
  \centering
  \begin{subfigure}[t]{0.48\linewidth}
    \centering
    \includegraphics[width=\linewidth]{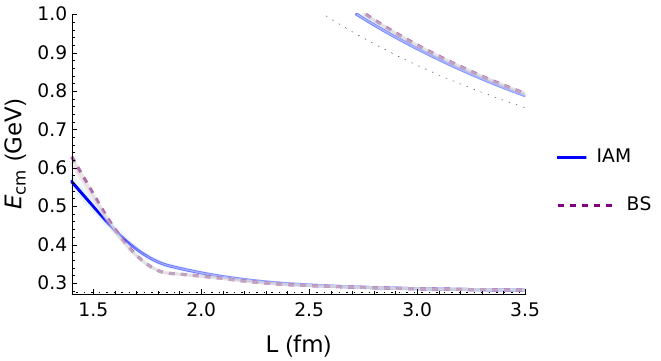}
    \caption{}
  \end{subfigure}
  \hfill
  \begin{subfigure}[t]{0.48\linewidth}
    \centering
    \includegraphics[width=\linewidth]{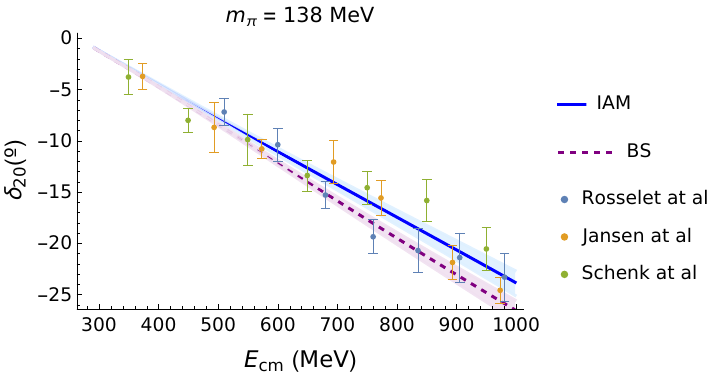}
    \caption{}
  \end{subfigure}
  \begin{subfigure}[t]{0.48\linewidth}
    \centering
    \includegraphics[width=\linewidth]{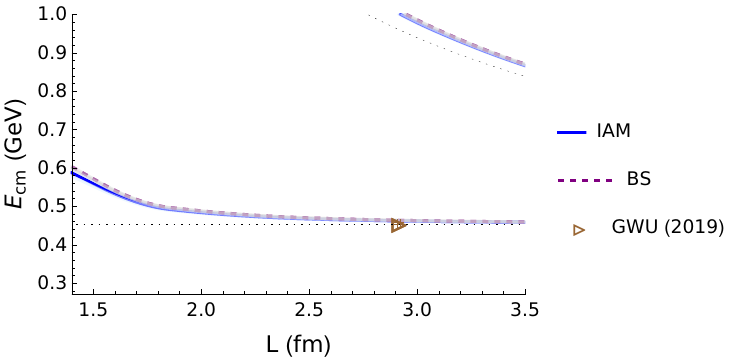}
    \caption{}
  \end{subfigure}
  \hfill
  \begin{subfigure}[t]{0.48\linewidth}
    \centering
    \includegraphics[width=\linewidth]{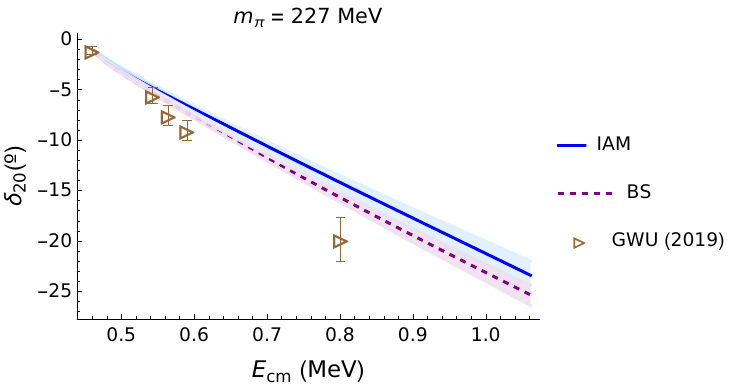}
    \caption{}
  \end{subfigure}

  \begin{subfigure}[t]{0.48\linewidth}
    \centering
    \includegraphics[width=\linewidth]{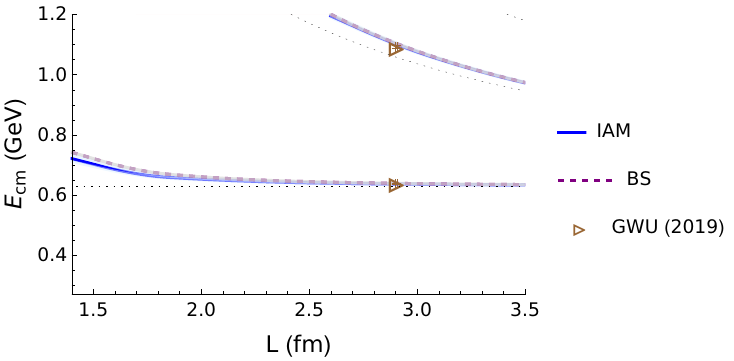}
    \caption{}
  \end{subfigure}
  \hfill
  \begin{subfigure}[t]{0.48\linewidth}
    \centering
    \includegraphics[width=\linewidth]{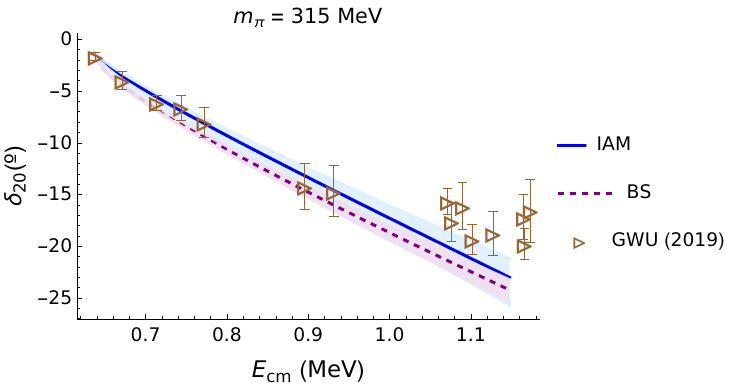}
    \caption{}
  \end{subfigure}
  \caption{\(A_{1}^{+}\) energy levels and phase shifts for \(I = 2\) as functions of the box size and energy, respectively. The LECs are taken from Tables~\ref{tab:iam} and \ref{tab:bs}, corresponding to the fit to lattice data. The uncertainty bands correspond to the LECs uncertainties in those tables.} 
  \label{fig:A12lt}
\end{figure*}

\begin{figure*}
  \centering
  \begin{subfigure}[t]{0.48\linewidth}
    \centering
    \includegraphics[width=\linewidth]{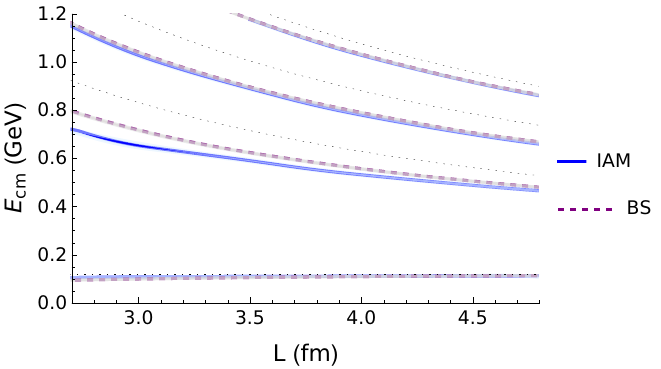}
    \caption{$I = 0$}
  \end{subfigure}
  \hfill
  \begin{subfigure}[t]{0.48\linewidth}
    \centering
    \includegraphics[width=\linewidth]{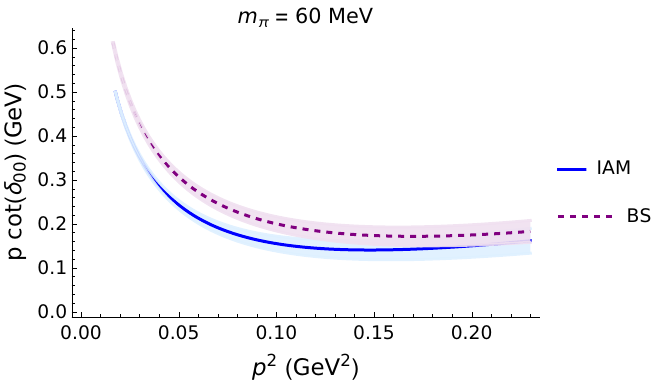}
    \caption{$I = 0$ }
  \end{subfigure}
 \begin{subfigure}[t]{0.48\linewidth}
    \centering
    \includegraphics[width=\linewidth]{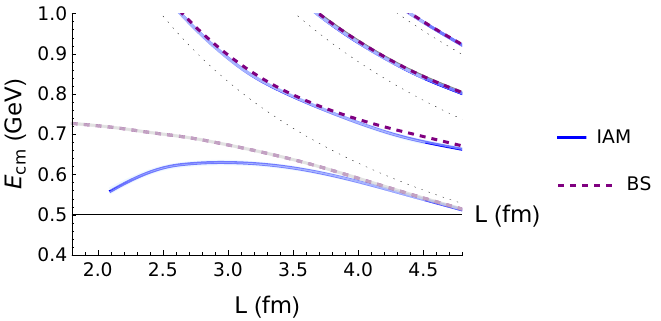}
    \caption{$I=1$}
  \end{subfigure}
  \hfill
  \begin{subfigure}[t]{0.48\linewidth}
    \centering
    \includegraphics[width=\linewidth]{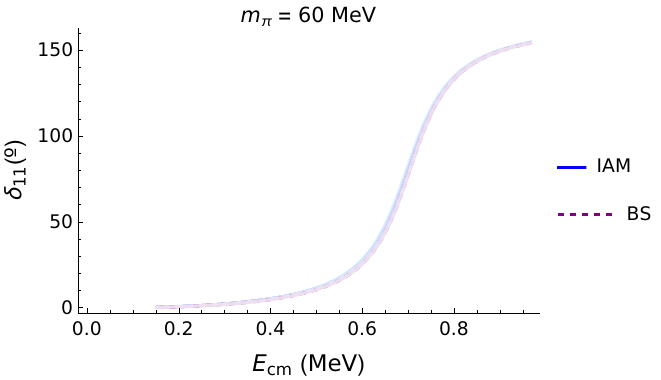}
    \caption{$I=1$}
  \end{subfigure}
  \begin{subfigure}[t]{0.48\linewidth}
    \centering
    \includegraphics[width=\linewidth]{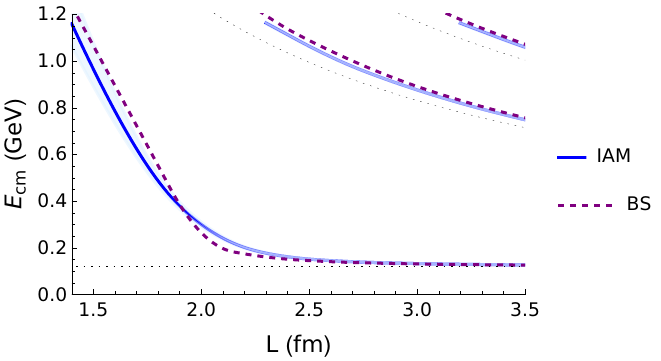}
    \caption{$I=2$}
  \end{subfigure}
  \hfill
  \begin{subfigure}[t]{0.48\linewidth}
    \centering
    \includegraphics[width=\linewidth]{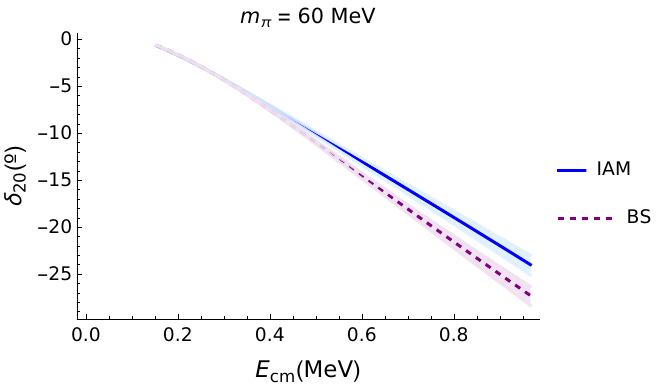}
    \caption{$I=2$}
  \end{subfigure}

  \caption{Energy levels (left) and phase shifts (right) for $I=0,1$ and $2$, from top to bottom as a function of the box size and center-of-mass energy,  for $m_{\pi} = 60 \, \text{MeV}$. The LECs correspond to sets 2 and 4 from Tables~\ref{tab:iam} and \ref{tab:bs}, obtanied from the fit to lattice data with IAM or BS.
  The uncertainty bands correspond to the LECs uncertainties in those tables.}
  \label{fig:mp60lt}
\end{figure*}

In this section we fit the lattice data for $\pi\pi$ phase shifts given in~\cite{Guo:2016zos,Guo:2018zss,cross:prd} within the full IAM framework and the BS method for comparison. The sets of LECs obtained are given in Tables~\ref{tab:iam} and \ref{tab:bs}, that correspond to sets 2 and 4 for the IAM and BS, respectively. The results of these fits for the energy levels and phase shifts are given in Figs.~\ref{fig:A10lt}, \ref{fig:T11lt} and \ref{fig:A12lt} for $I=0,1$ and $2$, respectively.  As it is shown, the IAM is able to reproduce reasonably well the phase shifts and energy levels for different pion masses. The fit with the BS approach provides similar results for those masses. For small lattices, we obtain similar significant differences between the IAM and the BS approaches, highlighting once more the importance of our present analysis.  The quark mass dependence of the pion decay constant, data also included in the fit, is depicted in Fig.~\ref{fig:fpiovermpilt}. Both sets of LECs, 2 and 4, used in the ChPT expression for the pion decay constant reproduce well the lattice data, being consistent with the experimental data point when the extrapolation to the physical point is done.   When such extrapolation is done, we can see that the IAM reproduces very well the experimental data for $I=0$ and $2$, as the BS method also does, with minimal differences compatible with the errors of the experimental data. While for $I=1$, the predicted behavior in the physical limit deviates  around $50$ MeV from experimental data. As discussed in \cite{Hu:2016shf}, this discrepancy stems from the absence of the strange quark in these LQCD simulations.

Once we have shown that the IAM successfully describes the quark mass dependence of the phase shifts in the infinite volume and energy levels in the finite volume, we can predict those for smaller pion masses. In Fig.~\ref{fig:mp60lt}, we show the result for a pion mass of $m_\pi=60$ MeV and $I=0,1$ and $2$ from top to bottom, in comparison with the BS approach. The reduction in $m_\pi L$ enhances the differences between both methods for small $L$ that we have been describing throughout this section, which are more evident for the first excited $I=0$ level and the fundamental $I=1$ one. Noticeably, the latter is enhanced in the presence of a relatively narrow rho meson resonance. The agreement between the two approaches for $I=2$ remains also as a robust conclusion of our present work.

\section*{Conclusions and Outlook}
\label{sec:conclusions}

In this work we have carried out a detailed calculation of the finite-volume pion-pion scattering amplitude in the rest frame, within the Inverse Amplitude Method and   Chiral Perturbation Theory . Our  ChPT calculation takes into account all diagrams contributing up to fourth order, 
including those $t,u$-channel ones responsible for the left cut at infinite volume, as well as tadpole contributions. In addition,  the full ChPT amplitude has by construction the correct pion mass dependence at this order. Those additional contributions are a distinctive novelty of the present approach with respect to previous analyses and they require a proper formalism accounting for the loss of Lorentz covariance in loop sum-integrals as well as a suitable generalization of the partial wave expansion to rotational-breaking projections, accounting for the symmetry of the geometry under consideration, which is the $L^3$ cube in the rest frame case analyzed here. The latter has been carried out within two different approaches, namely the irrep and cubic harmonics projections.

The IAM method has been then applied to the finite-volume ChPT amplitude, constructing a matrix-valued amplitude from which we have generated successfully the interacting energy levels, which we have analyzed numerically for different sets of LECs, including experimental and lattice fits, as well as  different values of the pion mass. We have paid special attention to the comparison of our approach with the Bethe-Salpeter used in previous works. Both approaches are formally equivalent up to corrections exponentially suppressed in $m_\pi L$. Numerically, the differences between them is quite small for $m_\pi L>2$ in the case of the energy levels, although we also find some differences between both methods for these pion masses in $I=0$ for the phase shifts produced. We find  significant differences for smaller volumes in the isospin $I=0,1$ channels which would be relevant when those smaller lattice become available in LQCD. For  $I=1$ the differences arise around the rho meson mass, so that the presence of relatively narrow resonances as $m_\pi L$ decreases might be a qualitatively relevant effect.

In summary, we have developed a comprehensive formalism for computing energy levels that includes exponential contributions, incorporates both left- and right-hand cuts, and offers a refined treatment of momentum discretization and irreducible representations. We have shown that the IAM has an inherent ability to reproduce well the quark mass dependence of physical observables related to the two-hadron scattering due the fact that it is built upon ChPT, and contains all possible sources of quark mass dependence, as it incorporates the correct volume and pion mass dependence  from the loops in the $s$, $t$ and $u$ and tadpole channels. 

Our method can also be extended to moving frames by considering systems with non-zero total momentum. In such cases, it would be necessary to generalize the shell formalism, introducing a new Cubic-like Harmonics basis and recomputing the matrices contributing to the amplitude. This framework is also applicable to other interesting two-hadron processes where exponential contributions may play an even more significant role, including coupled channels and in the presence of left-hand-cuts. For instance, this approach can be applied in the future to an analysis of LQCD data for the $X(3872)$, $Z_c(3900)$ or the $T_{cc}(3875)$, very near threshold resonances studied in LQCD, including one and two-pion exchange loops.

\section*{Acknowledgments}
\label{sec:thanks}

We thank Michael Döring and Raúl A. Briceño for useful discussions.  This work is supported by the Ministerio de Ciencia e Innovación, research contracts PID2022-136510NB-C31 and PID2023-147458NB-C21,
funded by MICIU/AEI/10.13039/501100011033. R. M. also acknowledges support from the ESGENT program with Ref. ESGENT/018/2024 and the PROMETEU program with Ref. CIPROM/2023/59, of the Generalidad Valenciana (GVA), and also from 
the Spanish Ministerio de Economia y Competitividad and European Union (NextGenerationEU/PRTR) by the grant with Ref. CNS2022-136146. R. M. and J. A. S. thank to the Grisolia Program of the Consejería de Innovación, Universidades, ciencia y sociedad digital de la GVA, proyecto con referencia  CIGRIS/2021/089. This work is also partly supported by the Spanish Ministerio de Economia y Competitividad (MINECO) and
European FEDER funds under Contracts No. FIS2017-84038-C2-1-P B, PID2020-112777GB-I00, and by Generalitat Valenciana under contract PROMETEO/2020/023. This project has received funding from the European Union Horizon 2020 research
and innovation program under the program H2020-INFRAIA-2018-1, grant agreement No. 824093 of the STRONG-2020.



 
\appendix
\section{Loop momentum sum-integrals at Finite Volume}
\label{app:sumint}

Here we summarize the main results for the finite-volume, $V=L^3$ sum-integrals of the type \eqref{sum} for  the loops contributing to the pion-pion elastic scattering amplitude. We work in Minkowski space-time with metric $\eta_{\mu\nu} = \text{diag}(+,-,-,-)$ and define
\begin{eqnarray}
J_{H} (L) &=& -\SumInt  G(q,m_{\pi})\label{intdefs}\\
J_{k} (Q_0,\vec{Q};L)&=& (-i)^{k} \SumInt q_{0}^{k} G(q,m_{\pi}) G(q - Q,m_{\pi})\label{eqprop}\notag\\
&&  
\end{eqnarray}
with the propagator
\begin{eqnarray}
\label{deltadef}
 G(q,m_{\pi}) &=& \frac{1}{q^{2} - m_{\pi}^{2} - i \epsilon},
 \end{eqnarray}
the sum-integral symbol $\SumInt$ defined in \eqref{sum} and both $\vec{q}=(2\pi/L) \vec{n}$, $\vec{Q} = (2\pi/L) \vec{N}$ discretized three-momenta. Recall that $J_H$ above corresponds to the tadpole-like   loops in Fig.\ref{fig:diag} c,d, while diagrams f,g,h in that figure give rise to sum-integrals of the type $J_k$ and similar ones with other powers of momenta in the integrand, which we will discuss below. For practical purposes, it is convenient to separate the  $L\rightarrow\infty$ part of the above sum-integrals as
\begin{equation}
\label{Jseparation}
J_{H,k}(Q,L)= J_{H,k}^D(Q^2;L)+\Delta J_{H,k}(Q,L)
\end{equation}
where the superscript $D$ is a reminder that the UV divergent part of the $J$ sum-integrals is contained solely in the $L\rightarrow\infty$ contribution and is regularized as customary in dimensional regularization with $D=4-\epsilon$ \cite{Gasser:1983yg}. 

\subsection{Reduction rules}

First, as mentioned in the main text, we will derive formal relations between different sum-integrals with powers of momenta in the numerator. They  are the generalization of the VP relations in the infinite volume case, arising from Lorentz covariance, to the finite-volume case where Lorentz covariance is lost. In doing so, we will follow the sane steps as in the finite temperature analysis \cite{GomezNicola:2002tn}. The aim is to  write the finite-volume scattering amplitude in terms of the minimum number of independent sum-integrals. Thus, we obtain:
\begin{widetext}
\begin{eqnarray}
\label{id1}
    \SumInt q_{j} G(q,m_{\pi})G(q-Q,m_{\pi}) &=& -\frac{1}{|\vec{Q}|^{2}} \left[ - i Q_{0} J_{1} + \frac{1}{2} J_{0} Q^{2} \right] Q_{j}\\
    Q^{\mu} \SumInt q_{\mu} G(q,m)G(q-Q,m)  &=& \frac{1}{2} J_{0}Q^{2}\\ \label{id3}
    \SumInt q_{i}q_{0} G(q,m_{\pi})G(q-Q,m_{\pi}) &=& \frac{i Q_{0}}{|\vec{Q}|^{2}} \left[ J_{2} - \frac{1}{2} J_{H} + \frac{1}{4} Q^{2} J_{0} \right] Q_{i}\\
    \SumInt q_{i}q_{j} G(q,m_{\pi})G(q-Q,m_{\pi}) &=& I_{2a} Q_{i}Q_{j} - I_{2b} \delta_{ij}\\
    \SumInt q_{0}q^{2} G(q,m_{\pi})G(q-Q,m_{\pi}) &=& i Q_{0}J_{H} + m_{\pi}^{2}J_{1}\\
    \SumInt q_{i}q^{2} G(q,m_{\pi})G(q-Q,m_{\pi}) &=& Q_{i}[ - J_{H} - \frac{m_{\pi}^{2}}{|\vec{Q}|^{2}} ( - i Q_{0} J_{1} + \frac{1}{2} J_{0} Q^{2}) ]\\
     \label{idn}
    \SumInt (q^2)^{2} G(q,m_{\pi})G(q-Q,m_{\pi}) &=& -(Q^{2} + 2 m_{\pi}^{2}) J_{H} + m_{\pi}^{4} J_{0}
\end{eqnarray}
\end{widetext}
with
\begin{widetext}
\begin{eqnarray}
I_{2b} &=& \frac{1}{(D - 2) |\vec{Q}|^{2}} [ - Q^{2} J_{2} - i Q_{0} Q^{2} J_{1} +  \left(\frac{(Q^2)^2}{4} + m^{2} |\vec{Q}|^{2}\right) J_{0} + \frac{Q^{2}}{2} J_{H}]\\
I_{2a} &=& \frac{1}{(D - 2) |\vec{Q}|^{4}} \left[ (D - 1) I_{2b} - (- J_{H} + m^{2} J_{0} + J_{2})\right]
\end{eqnarray}
\end{widetext}
In the following we will provide two alternative representations for the sum-integrals $J_{H,k}$ above, useful for the numerical analysis in this work.

\subsection{$q_{0}$-integration representation}

\begin{figure}[!h]
   \centering
    \includegraphics[width=0.45\textwidth]{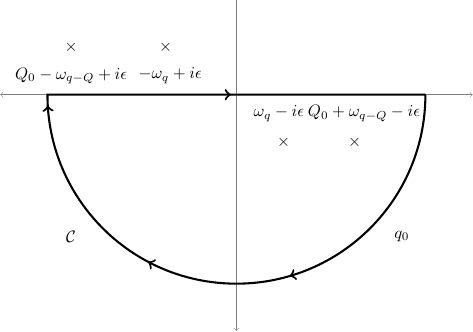}
    \caption{Complex contour in $q_{0}$}
    \label{fig:cc}
\end{figure}

As customarily done in finite volume calculations, we perform the $q_0$-integrals in \eqref{intdefs}  and \eqref{eqprop}  using Cauchy's residue theorem, so that we are just left with frequency sums. Choosing the contour in $q_0$ complex plane depicted in Fig.\ref{fig:cc}, we pick up one of the two poles of $J_H$ at 
$q_{0} =\pm \omega_{q} \mp i \epsilon$ 
with $\omega_{q} = \sqrt{|\;\vec{q}\;|^{2} + m^{2}}$ and obtain
\begin{equation}
\label{tadpolesum}
J_H(L) =\frac{1}{L^3}\sum_{\vec{n}}\frac{1}{2\omega_q}
\end{equation}
Likewise, the same contour picks up two of the poles of $J_k$, which lie at $q_{0} = \{ \pm \omega_{q} \mp i \epsilon , Q_{0} \pm \omega_{q - Q} \mp i \epsilon\}$, with the following result:
\begin{widetext}
\begin{eqnarray}
\label{J0sum}
    J_{0} (Q_0,\vec{Q};L) &=& \frac{1}{L^{3}} \sum_{\vec{n}} \frac{1}{2 \omega_{q}\omega_{q - Q}} \frac{\omega_{q} + \omega_{q - Q}}{(\omega_{q} + \omega_{q - Q})^{2} - Q_{0}^{2}}
    \\
    \label{J1vsJ0}
    J_{1} (Q_0,\vec{Q};L)&=& \frac{- i Q_{0}}{2} J_{0} (Q_0,\vec{Q};L)
    \\
    \label{J2sum}
    J_{2} (Q_0,\vec{Q};L)&=& \frac{1}{2 L^{3}} \sum_{\vec{n}} \frac{\omega_{q} + \omega_{q - Q}}{(\omega_{q} + \omega_{q - Q})^{2} - Q_{0}^{2}} - \frac{Q_{0}^{2}}{2 L^{3}} \sum_{\vec{n}} \frac{1}{\omega_{q - Q}\left[(\omega_{q} + \omega_{q - Q}\right]^{2} - Q_{0}^{2})}
\end{eqnarray}
\end{widetext}
In this representation, the contributions $\Delta J$ in the separation \eqref{Jseparation} can be written as
\begin{widetext}
\begin{equation}
\label{qmaxrep}\\
   \Delta J_{\alpha}(Q; L) = \lim_{q_{max}\rightarrow\infty} \left[  \frac{1}{L^{3}} \sum_{\stackrel{\vec{q} = \frac{2 \pi \vec{n}}{L}}{\vert \vec{q} \vert<q_{max}}} - \int_{\vert \vec{q} \vert<q_{max}} \frac{d^{3}\vec{q}}{(2 \pi)^{3}}\right] f_{\alpha}(Q,\vec{q})
\end{equation}
\end{widetext}
 where $\alpha=H,0,1,2$ and $f_\alpha$ stand for the different functions  inside the $\sum_{\vec{n}}$ sums in \eqref{tadpolesum}-\eqref{J2sum} and $q_{max}$ is a momentum cutoff whose value in practice will be set up to ensure numerical convergence at the range of $L$ considered here. Note that in the case where $\vec{Q} = \vec{0}$, which corresponds to the $s$-channel diagram in Fig.\ref{fig:diag}f in the CM frame, we find 
\begin{equation}
\label{J2vsJ0S}
J_{2}(Q_0,\vec{0};L) = -\frac{Q_{0}^{2}}{4} J_{0} (Q_0,\vec{0};L) + \frac{1}{2} J_{H}
\end{equation}
The relations \eqref{J1vsJ0} and \eqref{J2vsJ0S} allow one to express all $s$-channel contributions to the scattering amplitude solely in terms of $J_0$.  On the other hand, for $t,u$- channel diagrams, $Q_0=T_0=U_0=0$ in the CM frame. Thus, from the analysis in this section, we conclude that the scattering amplitude in the CM can be written solely in terms of the following sum-integrals:
\begin{widetext} 
\begin{eqnarray}
\label{JH}
J_H(L)&=&\frac{1}{L^3}\sum_{\vec{n}}\frac{1}{2\omega_q}=\frac{1}{L^3}\sum_{n\geq 0} x_3(n) \frac{1}{2\omega_q (n)}
\\
\label{Js}
J_s(E;L)&\equiv& J_0(E^2,\vec{0};L) =\frac{1}{L^3}\sum_{\vec{n}}\frac{1}{\omega_q\left(4\omega_q^2 - E^2\right)} =\frac{1}{L^3}\sum_{n\geq 0} x_3(n) \frac{1}{\omega_q (n)\left(4\omega_q(n)^2 - E^2\right)} 
\\
\label{Jt}
J_t (E,\hat p, \hat p';L)&\equiv& J_0(0,\vec{p}-\pp;L)=\frac{1}{L^3}\sum_{\vec{n}} \frac{1}{2\omega_q \omega_{p-p'-q}\left(\omega_q+\omega_{p-p'-q}\right)}
\\
\label{Ju}
J_u (E,\hat p, \hat p';L)&\equiv& J_0(0,\vec{p}+\pp;L) =\frac{1}{L^3}\sum_{\vec{n}}
\frac{1}{2\omega_q \omega_{p+p'-q}\left(\omega_q+\omega_{p+p'-q}\right)}
\\
J_{2t} (E,\hat p, \hat p';L)
&\equiv& J_2(0,\vec{p}-\pp;L) = \frac{1}{L^3}\sum_{\vec{n}}
\frac{1}{2\left(\omega_q+\omega_{p-p'-q}\right)}
\label{J2t}
\\
J_{2u} (E,\hat p, \hat p';L)
&\equiv& J_2(0,\vec{p}+\pp;L) = \frac{1}{L^3}\sum_{\vec{n}}
\frac{1}{2\left(\omega_q+\omega_{p+p'-q}\right)}
\label{J2u}
\end{eqnarray}
\end{widetext}
which we have expressed in terms of the representation \eqref{tadpolesum}-\eqref{J2sum} and where $n=\vert \vec{n} \vert^2$ and $x_3(n)$ are the number of possible choices of $(n_1,n_2,n_3)$ that satisfy $n_1^2+n_2^2+n_3^2=n$ with fixed $n$ \cite{Bijnens:2013doa}. Recall that $J_s$ in \eqref{Js} diverges for $E_n=2\sqrt{m^2+(2\pi/L)^2 n}$ with the $n\geq 0$ integer, which are precisely the free energy levels.

\subsection{Poisson summation formula representation}

An alternative representation for the sum-integrals comes from the use of the Poisson summation formula:
\begin{eqnarray}
\frac{1}{L^{3}}\sum_{\vec{q} = \frac{2 \pi \vec{n}}{L}} f(\vec{q})=\sum_{\vec{q}} \int \frac{dp^{3}}{(2\pi)^{3}} e^{i L \, \vec{n}.\vec{p}} f(\vec{p})\notag \\ 
\label{poisson}
=\int \frac{d^{3}p}{(2\pi)^{3}} f(\vec{p})+\sum_{\vec{n}\neq 0}\int \frac{d^{3}p}{(2\pi)^{3}} e^{iL \, \vec{n}.\vec{p}} f(\vec{q})
\end{eqnarray}
where the first term on the r.h.s. is the $L\rightarrow\infty$ contribution. 
Using \eqref{poisson}, one can cast the above relevant sum-integrals as follows
\cite{Bijnens:2013doa}:
\begin{widetext}
\begin{eqnarray}
\Delta J_H(L)&=&\frac{m}{4\pi^2 L} \sum_{\vec{k}\neq \vec{0}}\frac{K_1 \left(L m \vert \vec{k} \vert \right) }{\vert \vec{k} \vert}=\frac{m}{4\pi^2 L} \sum_{k>0} x_3 (k) \frac{K_1 \left(L m \sqrt{k} \right) }{\sqrt{k}}\nonumber\\
\label{JHPSF}
&=&\frac{m^2}{16\pi^2}\int_0^\infty \frac{d\lambda}{\lambda^2}e^{-\lambda}\left[\theta_3\left(0;e^{-\frac{L^2m^2}{4\lambda}}\right)^3-1\right]
\\
 \Delta J_s(E;L)&=&
\frac{1}{8\pi^2}\sum_{\vec{k}\neq \vec{0}} \int_0^1 dx K_0 \left( L \tilde M_s (x;E) \vert \vec{k} \vert \right)=\frac{1}{8\pi^2}\sum_{k>0} x_3 (k) K_0 \left( L \tilde M_s (x;E) \sqrt{k} \right)\nonumber\\&=&
\frac{1}{16\pi^2} \int_0^1 dx \int_0^\infty \frac{d\lambda}{\lambda}e^{-\lambda} \left[\theta_3\left(0;e^{-\frac{L^2\tilde M^2(x,E)^2}{4\lambda}}\right)^3-1\right]
\label{JsPSF}\\
\Delta J_{t,u}(E,\vec{Q};L)&=&\frac{1}{8\pi^2}\sum_{\vec{k}\neq \vec{0}} \int_0^1 dx \ e^{iL (1-x) \vec{Q}\cdot\vec{k}} 
K_0 \left( L \tilde m_Q (x;Q) \vert \vec{k} \vert \right)\nonumber\\
\label{JtuPSF}
&=& \frac{1}{16\pi^2} \int_0^1 dx \int_0^\infty \frac{d\lambda}{\lambda}e^{-\lambda} \nonumber\\&&
\left[\prod_{i=1}^3 \theta_3\left(\frac{L Q_i (1-x)}{2};e^{-\frac{L^2\tilde m^2(x,E)^2}{4\lambda}}\right)-1\right]
\\
\Delta J_{2t,2u}(E,\vec{Q};L)&=&
\frac{1}{8\pi^2}\sum_{\vec{k}\neq \vec{0}} \int_0^1 dx \ e^{iL (1-x) \vec{Q}\cdot\vec{k}}  \ \frac{\tilde m_Q (x;Q)}{L \vert \vec{k} \vert }
K_1 \left( L \tilde m_Q (x;Q) \vert \vec{k} \vert \right)\nonumber\\
\label{J2tuPSF}
&=& \frac{1}{32\pi^2} \int_0^1 dx \tilde m^2(x;Q) \int_0^\infty \frac{d\lambda}{\lambda^2}e^{-\lambda} \nonumber\\&&
\left[\prod_{i=1}^3 \theta_3\left(\frac{L Q_i (1-x)}{2};e^{-\frac{L^2\tilde m^2(x,E)^2}{4\lambda}}\right)-1\right]
\end{eqnarray}
\end{widetext}
with $\vec{k}\in\IZ^3$, $k=\vert \vec{k} \vert^2$, $K_n(z)$  modified Bessel functions and $\theta_3(u;q)$ the third Jacobi theta function,  $\tilde m_Q^2(x;Q)=m^2+\vert\vec{Q}\vert^2 x(1-x)$, $\vec{Q}=\vec{T}=\vec{p}-\pp$ for $J_{t,2t}$ and $\vec{Q}=\vec{U}=\vec{p}+\pp$ for $J_{u,2u}$.
\section{ChPT pion scattering amplitude at finite volume}
\label{app:amp}

The finite volume part for the ChPT amplitude $\T_4$ is given in the CM frame by
\begin{widetext}
\begin{eqnarray}
    \Delta \T_4^{I = 0} &=& \frac{1}{4 f_{\pi}^{4}} \left[ 
    2 (m_{\pi}^{2} - 2 s)^{2} \Delta J_{s}(E;L) \right. \nonumber\\&+& 
    (30 m_{\pi}^{4} - 4 m_{\pi}^{2} s - 24 m_{\pi}^{2} t + s t + 7 t^{2} ) \Delta J_{t}(E,\hat p, \hat p';L) 
    \nonumber\\
   &+&( 46 m_{\pi}^{4} - 8 m^{2}_{\pi} s - 4 m^{2}_{\pi} t - 32 m_{\pi}^{2} u \nonumber\\ 
   &+& 2 s u +t u + 8 u^{2}  ) \Delta J_{u}(E,\hat p, \hat p';L)
   \nonumber\\
    &+&
    (16 m_{\pi}^{2} + 4 s - 4 t) \Delta J_{2t} (E,\hat p, \hat p';L)
     \nonumber\\
     &+& (32 m_{\pi}^{2} - 4 t - 8 u ) \Delta J_{2u}(E,\hat p, \hat p';L) 
    \nonumber\\ 
     \label{dT4I0}
    &-& \left.  (74 m_{\pi}^{2} - 22 s - 16 t - 16 u) \Delta J_{H}(L) \right]
    \\ 
    \Delta \T_4^{I = 1} &=& \frac{1}{24 f_{\pi}^{4}} \left[ 
    4 ( 4 m_{\pi}^{2} - s) ( 4 m_{\pi}^{2} - s - 2 t ) \Delta J_{s}(E; L)  \right. \nonumber\\
    &+& (12 m_{\pi}^{4} - 12 m_{\pi}^{2} s + 3 s t + 9 t^{2}) \Delta J_{t}(E,\hat p, \hat p';L)
    \nonumber\\
    &+& ( 84 m_{\pi}^{4} - 12 m^{2}_{\pi} s - 24 m^{2}_{\pi} t - 48 m_{\pi}^{2} u \nonumber\\
    &+& 3 s u + 6 t u - 3 u^{2}  ) \Delta J_{u}(E,\hat p, \hat p';L) 
\nonumber\\
    &+&
    (48 m_{\pi}^{2} + 12 s - 12 t) \Delta J_{2t} (E,\hat p, \hat p';L) \nonumber\\
    &+&  (48 m_{\pi}^{2} - 36 s - 24 t - 12 u ) \Delta J_{2u}(E,\hat p, \hat p';L) 
    \nonumber\\
    \label{dT4I1}
    &-& \left. (80 m_{\pi}^{2} - 20 s - 46 t + 6 u) \Delta J_{H}(L) \right]
    \\
    \Delta \T_4^{I = 2} &=& -\frac{1}{8 f_{\pi}^{4}} \left[ 
    -4 ( - 2 m_{\pi}^{2} + s)^{2} \Delta J_{s}(E; L)     \right.\nonumber\\
    &+& (12 m_{\pi}^{4} - 4 m^{2}_{\pi} s + s t - 5 t^{2}) \Delta J_{t}(E,\hat p, \hat p';L)
    \nonumber\\
    &+& (-20 m_{\pi}^{4} + 4 m_{\pi}^{2} s + 8 m_{\pi}^{2} t + 16 m_{\pi}^{2} u \nonumber\\ 
    &-& s u - 2 t u - 7 u^{2}) \Delta J_{u}(E,\hat p, \hat p';L)
    \nonumber\\
    &+&
    (16 m_{\pi}^{2} + 4 s - 4 t) \Delta J_{2t} (E,\hat p, \hat p';L)  \nonumber\\
    &+& (- 16 m_{\pi}^{2} + 12 s + 8 t+4u) \Delta J_{2u}(E,\hat p, \hat p';L) \nonumber\\
     \label{dT4I2}
    &-&\left.   ( -40 m_{\pi}^{2} + 8 s + 14 t + 14 u) \Delta J_{H}(L)\right]
\end{eqnarray}
\end{widetext}
where $s=E^2$, $t=-\vert \vec{p}-\pp\vert^2$, $u=-\vert \vec{p}+\pp\vert^2$, and $\Delta \T_{4} = \delta \T_{4} + f_{s} \Delta J_{s}$. Let us now prove results Eqs. \eqref{T4SJI02} and \eqref{T4SJI1} for the part of the finite-volume amplitude proportional to $J_s(E;L)$, i.e. $\T_{4S}$ according to our notation in section \ref{sec:chptfv}. First, note that 
\begin{widetext}
\begin{equation}
\T_{2}(E,\ppp) = 
\begin{cases}
\dfrac{(2s - m_{\pi}^{2})}{f_{\pi}^{2}}, & \text{for } I = 0, \\[1em]
4\,\dfrac{\vec{p}\cdot\vec{p}'}{f_{\pi}^{2}}=\dfrac{t - u}{f_{\pi}^{2}}  , & \text{for } I = 1, \\[1em]
\dfrac{(2 m_{\pi}^{2} - s)}{f_{\pi}^{2}}, & \text{for } I = 2.
\end{cases}
\label{t2cases}
\end{equation}
\begin{equation}
\T_{4S}(E,\ppp;L) = 
\begin{cases}
\dfrac{(2E^2 - m_{\pi}^{2})^{2}}{2 f_{\pi}^{4}}\, J_s(E;L), & \text{for } I = 0, \\[1em]
\dfrac{2 (E^2 - 4 m_{\pi}^{2})(\vec{p}\cdot\vec{p}')}{3 f_{\pi}^{4}}\, J_s(E;L), & \text{for } I = 1, \\[1em]
\dfrac{(2 m_{\pi}^{2} - E^2)^{2}}{2 f_{\pi}^{4}}\, J_s(E;L), & \text{for } I = 2.
\end{cases}
\label{t4scases}
\end{equation}
\end{widetext}
Therefore, for the $I = 0,2$ cases, we readily arrive to Eq. \eqref{T4SJI02} from Eqs. \eqref{t2cases} and \eqref{t4scases} since in those cases both $\T_{4S}$ and $\T_2$ are functions of $E$ but not of $\ppp$.
In the $I=1$ case, to prove Eq. \eqref{T4SJI1} let us use that in the CM frame
\begin{align}
\frac{1}{L^3}\sum_{\vec{q}} q_i q_j I_q(E;L)=\frac{1}{3}\delta_{ij} \frac{1}{L^3}\sum_{\vec{q}} \vert\vec{q}\vert^2 I_q(E;L)\nonumber\\
=\frac{1}{12}\delta_{ij} \left[ J_H (L)+\left(E^2-m_\pi^2\right)J_s(E;L)\right]
\end{align}
so that
\begin{widetext}

\begin{align}
    \T_{4S}(E,\ppp;L) = \dfrac{2}{3 f_{\pi}^{4}} (\vec{p}\cdot\vec{p}') \left[ \frac{4}{L^3} \sum_{\vec{q}} \vert\vec{q}\vert^2 I_q (E;L)-J_H(L)\right]=
    \dfrac{1}{f_{\pi}^{4}} \left[ \frac{1}{L^3} \sum_{\vec{q}} p_i q^i p'_j q^j I_q (E;L) - \frac{1}{6}(t-u) J_H(L)\right]
\nonumber\\
=\frac{1}{2} \frac{1}{L^3} \sum_{\vec{q}} \T_{2}(E,\pq)I_q \T_{2}(E,\qpp)  - \frac{1}{6f_{\pi}^{4}}(t-u) J_H(L).
\end{align}

\end{widetext}

On the one hand, for $E=2m_\pi$, $\vec{p}=\vec{p}'=\vec{0}$, the  loop sum-integrals coming from the $t,u$- channels  of the finite-volume amplitude, and given in Eqs. \eqref{Jt}-\eqref{J2u} can be written in terms of the tadpole sum-integral Eq. \eqref{JH} as
\begin{eqnarray}
\label{Jtuthreshold}
J_t(2m,0,0,;L)&=&J_u(2m,0,0,;L)=\frac{d}{d m^2} J_H(L)\nonumber\\
J_{2t}(2m,0,0,;L)&=&J_{2u}(2m,0,0,;L)=\frac{1}{2}J_H(L)
\nonumber\\
\end{eqnarray}
On the other hand, in terms of the cubic symmetry projections explained in section \ref{sec:cubic}, it is not difficult to show that for $r=r'=1$ only the irrep $\Gamma=A_1^+$ contributes, the only nonzero projection being simply $t_{11}^{A_1^+ 11}=\T(E,0,0;L)$ both for the irrep and CH cases.
 Therefore, the threshold contribution of the amplitude is only present for scalar channels and therefore for even isospin, since $G$-parity conservation implies that $I+l$ should be even.

Therefore, for energy levels close enough to threshold, we can write a Lüscher-like version of the QC only for the $r=r'=1$ component of the finite-volume scattering amplitude   as \cite{Bedaque:2006yi}
\begin{widetext}
\begin{equation} \label{eq:luscher-like}
    p(s) \cot{\delta (s)} 
    - 16\pi \sqrt{s}\, \frac{\delta \T_{4} (s;L)}{\T_{2}^2(s)} 
    = \frac{1}{\pi L}\mathcal{S}\left(\frac{p^2(s) L^2}{4\pi^2}\right).
\end{equation}
\end{widetext}
where $\delta (s)$ is the infinite-volume phase shift,  related to the full infinite-volume amplitude as customary, i.e., 
\begin{equation}
    T^{\mathrm{IAM}}(s) = \frac{\sqrt{s}}{2}\, (p(s) \cot{\delta (s)} - i p (s) )^{-1},
\end{equation}
and $\mathcal{S}$ is the Lüscher function;
\begin{equation}
    \mathcal{S}\left(\frac{p^2(s) L^2}{4\pi^2}\right)=4\pi^2 L\left[  \frac{1}{L^{3}} \sum_{\vec{q} = \frac{2 \pi \vec{n}}{L}} - \int \frac{d^{3}\vec{q}}{(2 \pi)^{3}}\right] 
    \frac{1}{\vert\vec{q}\vert^2-p^2(s)}
\label{SLuscher}
\end{equation}

Note that, $\Delta J_{s} = - \mathcal{S}(p^{2}L^{2}/4 \pi^{2})/8 \pi^{2}L \sqrt{s} \,+ \,\Delta J_{s}^{exp}$, where $\Delta J_{s}^{exp}$ is formally exponentially suppressed, i.e.,  it contributes at the same order as the $\Delta \T_{4}$ contributions in $t,u$ and tadpole channels. Thus, in Eq.\eqref{eq:luscher-like}, $\delta \T_4(s;L)$ accounts for all fourth order volume-dependent contributions at threshold other than  the Lüscher function  contribution, i.e., $\delta \T_4=\Delta \T_{4H}+\Delta \T_{4T}+\Delta \T_{4U} +f_s\Delta J_s^{exp
}$ in our notation. Note that the extra term $-16\pi \sqrt{s}\, (\delta \T_{4}/\T_{2}^2)$ in \eqref{eq:luscher-like} parametrizes the corrections with respect to the standard Lüscher formula and encodes all the new contributions to the ChPT amplitude studied in the present work, coming in particular from the left cut contribution at infinite volume.

Performing now an expansion around threshold, i.e., $E\simeq 2m_\pi$ and $p \cot{\delta} \simeq 1/a$ with $a$ the scattering length (which includes the LECs contribution) one can find from Eq.~\eqref{eq:luscher-like}  the finite-volume energy shift $\Delta E = E_{0} - 2 m_{\pi}$, with $E_0$ the lowest energy level. The additional near-threshold contribution from $\delta \T_4$ is given by

\begin{equation}
    P = 32\pi m_{\pi}\,
    \left| a\, \frac{\delta \T_{4}}{\T_{2}^{2}} \right|_{s \to 4m_{\pi}^{2}} .
\label{Pdef}
\end{equation}
The Lüscher limit corresponds to $P=0$  while $P\simeq 1$ corresponds to the limit where $\delta \T_4$ corrections are of the same order as the Lüscher one.  

The finite-volume expansion of $\delta \T_{4}/\T_{2}^{2}$ reads
\begin{equation} \label{deltat4ex}
  \frac{\delta \T_{4}}{\T_{2}^{2}} =
   \begin{cases}
     \frac{23 \Delta J_{H}}{98 m_{\pi}^{2}} + \frac{1}{7} \Delta J_{u} + \frac{1}{2} \Delta J_{s}^{exp},
      & I = 0,\\[10pt]
      - \frac{5 \Delta J_{H} - m_{\pi}^{2} (2 \Delta J_{s}^{exp} + \Delta J_{u})}{4 m_{\pi}^{2}},
      & I = 2.
   \end{cases}
\end{equation}
At the large $m_{\pi}L$ limit, we get

\begin{equation}
\label{T4thexp}
    \frac{\delta \T_{4}}{\T_{2}^{2}} 
    = \frac{1}{2^{5/2}\pi^{3/2}} 
    \sum_{n = 1}^{\infty} \sum_{k=0}^{\infty} 
    \frac{\vartheta_{n}}{(n m_{\pi} L)^{k + 1/2}} 
    e^{- n m_{\pi} L} 
    c_{k, n}^{I},
\end{equation}
with coefficients
\begin{equation}
   c^{I}_{k,n} =
   \begin{cases}
      \dfrac{a_{0, k} (-7 + 2 n^{2} m_{\pi}^{2} L^{2})}{28 n^{2} m_{\pi}^{2} L^{2} } + \dfrac{a_{1,k}(23 - 392 \pi)}{98 n m_{\pi} L}\,, & I = 0,\\[6pt]
      \dfrac{a_{0,k} (-2 + n^{2} m_{\pi}^{2} L^{2})}{8 n^{2} m_{\pi}^{2} L^{2}} - \dfrac{(5 + 16 \pi) a_{1,k}}{4 n m_{\pi} L}, & I = 2,
   \end{cases}
\end{equation}
which comes from Eq.\eqref{Jtuthreshold}  and the Bessel functions representation of the $\Delta J_{H,s}$ functions given in Eqns. \eqref{JHPSF}, \eqref{JsPSF} (see also \cite{Bedaque:2006yi}), using  the series expansion  $K_{\nu}(z) = \left( \frac{\pi}{2z} \right)^{1/2} e^{-z} \sum_{k = 0}^{\infty} \frac{a_{\nu,k}}{z^{k}}$ and 
\begin{equation}
   a_{\nu, k} = 
   \frac{ \left(\frac{1}{2} - \nu\right)_{k} 
          \left(\frac{1}{2} + \nu\right)_{k} }
        { (-2)^{k} k! },
\end{equation}
where $(n)_{k}$ denotes the Pochhammer symbol. 
 Note that the expression Eq. \eqref{T4thexp} is particularly useful, since the exponentially suppressed contributions for large $m_\pi L$ are explicitly shown.

Finally, the scattering length $a$ is given by
\begin{equation}\label{asclength}
   16 \pi m_{\pi} a = 
   \begin{cases}
      \dfrac{1}{
      \dfrac{2 f_{\pi}^2}{7 m_{\pi}^2}
      -\dfrac{40 \bar{l}_{1}+80 \bar{l}_{2}-15 \bar{l}_{3}+84 \bar{l}_{4}+105}{2352 \pi ^2}},
      & I = 0,\\
      \dfrac{1}{
      -\dfrac{f_{\pi}^2}{m_{\pi}^2}
      -\dfrac{8 \bar{l}_{1}+16 \bar{l}_{2}-3 \bar{l}_{3}-12 \bar{l}_{4}+3}{96 \pi ^2}},
      & I = 2.
   \end{cases}
\end{equation}

\section{Projections on cubic symmetry}
\label{app:cubic}

The Octahedral group $\mathcal{O}_h$ is the group corresponding to a cubic system within the Crystallographic Point Group, which consists of $24$ possible rotations, $R_{a}$, and the inversion operation, $I = \text{diag}_{3}(-1)$. The symmetry group $\mathcal{O}_h$ corresponds in particular to the proper rotations of a cube. The rotations with $a = 1, ..., 24$ can be parameterized by unitary vectors $n^{(a)}$ along the rotational axis, and the angle $\omega^{(a)}$ or equivalently by the Euler angles, $\alpha^{(a)}$, $\beta^{(a)}$ and $\gamma^{(a)}$, given in  Table \ref{table:1} \cite{Bernard:2008ax}. 
A general element of $\mathcal{O}_h$ is given by $g_{a} = R_{a}I$, where $I$ as the inversion matrix, which commutes with $R_{a}$.  Therefore, the total number of the elements of $\mathcal{O}_h$ is equal to $48$.

\begin{table*}[ht]
\resizebox{\columnwidth}{!}{%
\renewcommand{\arraystretch}{0.5}
\begin{tabular}{ |P{1.5cm}|P{1.5cm}|P{1.5cm}|P{1cm}|P{1cm}|P{1cm}|P{1cm}|  }
\hline
\rowcolor{lightgray} \multicolumn{7}{|c|}{Octahedral group parametrization} \\
\hline
\rowcolor{gray!25} Class & $a$ & $\vec{n}^{(a)}$ & $\omega^{(a)}$ & $\alpha^{(a)}$ & $\beta^{(a)}$ & $\gamma^{(a)}$\\
\hline
\rowcolor{yellow!25}
$E$ & $1$ & $(0,0,0)$ & $0$ & $0$ & $0$ & $0$ \\
\hline
\rowcolor{blue!25!white!40}
$8C_{3}$ & $2$ & $(1,1,1)$ & $-2\pi/3$ & $-\pi/2$ & $-\pi/2$ & $0$ \\
\rowcolor{blue!25!white!40}
 & $3$ & $(1,1,1)$ & $2\pi/3$ & $0$ & $-\pi/2$ & $\pi/2$ \\
\rowcolor{blue!25!white!40}
 & $4$ & $(-1,1,1)$ & $-2\pi/3$ & $0$ & $\pi/2$ & $-\pi/2$ \\
 \rowcolor{blue!25!white!40}
 & $5$ & $(-1,1,1)$ & $2\pi/3$ & $\pi/2$ & $-\pi/2$ & $0$ \\
 \rowcolor{blue!25!white!40}
 & $6$ & $(-1,-1,1)$ & $-2\pi/3$ & $-\pi/2$ & $\pi/2$ & $0$ \\
 \rowcolor{blue!25!white!40}
 & $7$ & $(-1,-1,1)$ & $2\pi/3$ & $0$ & $-\pi/2$ & $\pi/2$ \\
 \rowcolor{blue!25!white!40}
 & $8$ & $(1,-1,1)$ & $-2\pi/3$ & $0$ & $\pi/2$ & $-\pi/2$ \\
 \rowcolor{blue!25!white!40}
 & $9$ & $(1,-1,1)$ & $2\pi/3$ & $\pi/2$ & $-\pi/2$ & $0$ \\
\hline
\rowcolor{green!25!white!40}
$6C_{4}$ & $10$ & $(1,0,0)$ & $-\pi/2$ & $-\pi/2$ & $-\pi/2$  & $\pi/2$ \\
\rowcolor{green!25!white!40}
         & $11$ & $(1,0,0)$ & $\pi/2$  & $\pi/2$  & $-\pi/2$  & $-\pi/2$ \\
\rowcolor{green!25!white!40}
         & $12$ & $(0,1,0)$ & $-\pi/2$ & $0$      & $-\pi/2$  & $0$ \\
         \rowcolor{green!25!white!40}
         & $13$ & $(0,1,0)$ & $\pi/2$  & $0$      & $\pi/2$   & $0$ \\
         \rowcolor{green!25!white!40}
         & $14$ & $(0,0,1)$ & $-\pi/2$ & $-\pi/2$ & $0$       & $0$ \\
         \rowcolor{green!25!white!40}
         & $15$ & $(0,0,1)$ & $\pi/2$  & $\pi/2$  & $0$       & $0$ \\
\hline
 \rowcolor{red!25!white!40}
$6C'_{2}$ & $16$ & $(0,1,1)$ & $\pi$ & $-\pi/2$ & $-\pi/2$  & $-\pi/2$ \\
          \rowcolor{red!25!white!40}
          & $17$ & $(0,-1,1)$ & $\pi$ & $-\pi/2$ & $\pi/2$  & $-\pi/2$ \\
           \rowcolor{red!25!white!40}
          & $18$ & $(1,1,0)$ & $\pi$ & $-\pi/2$ & $-\pi$  & $0$ \\
           \rowcolor{red!25!white!40}
          & $19$ & $(1,-1,0)$ & $\pi$ & $0$ & $\pi$  & $-\pi/2$ \\
           \rowcolor{red!25!white!40}
          & $20$ & $(1,0,1)$ & $\pi$ & $0$ & $\pi/2$  & $-\pi$ \\
           \rowcolor{red!25!white!40}
          & $21$ & $(-1,0,1)$ & $\pi$ & $0$ & $-\pi/2$  & $-\pi$ \\
\hline
\rowcolor{orange!25!white!40}
$3C_{2}$ & $22$ & $(1,0,0)$ & $\pi$ & $\pi$ & $\pi$  & $0$ \\
         \rowcolor{orange!25!white!40}
         & $23$ & $(0,1,0)$ & $\pi$ & $0$ & $-\pi$  & $0$ \\
         \rowcolor{orange!25!white!40}
         & $24$ & $(0,0,1)$ & $\pi$ & $0$ & $0$  & $-\pi$ \\
\hline
\end{tabular}}
\caption{Parametrization of the all possible rotations in the Octahedral group.}
\label{table:1}
\end{table*}


All possible symmetry operations in $\mathcal{O}_h$ can be represented as a combination of rotations of a definite angle around some axis and a reflection in some plane. Thus, the symmetry operations can be classified as:

\begin{itemize}
    \item $E$: Identity.
    \item $8C_{3}$: rotations of $\pm 2\pi/3$ around the four body diagonals.
    \item $6C_{4}$: rotations of $\pm\pi/2$ around the coordinate axis.
    \item  $6C_{2}'$: rotations of $\pi$ around the six axis bisecting  opposite sides.
    \item $3C_{2}$: rotations of $\pi$ around the coordinate axis.
\end{itemize}

The irreducible representations of the Octahedral group $\mathcal{O}_h$ are denoted by $\Gamma = \{A_{1}, A_{2}, E, T_{1}, T_{2}\}$, with dimensions $s_{\Gamma} = \{1,1,2,3,3\}$ \cite{Doring:2018xxx} and with matrices $D^\Gamma_{\alpha\beta}$ given by:
\begin{widetext}
\begin{itemize}
    \item $A_{1}$: trivial $1-$dimensional rotation, $D^{A_{1}}(g_{a}) = 1$.
    \item $A_{2}$: $D^{A_{2}}(g_{a})=-1$, for the conjugacy classes $6C_{4}$ and $6C_{2}'$, $D^{A_{2}}(g_{a}) = 1$ otherwise.
    \item $E$: the matrices in this representation are two-dimensional and real:
    \begin{eqnarray}
        D^{E}(g_{a}) &=& \mathbb{1} \,\ \text{for} \,\ a=1,22,23,24 \notag\\
         &=& \sigma_{3} \,\ \text{for} \,\ a=14,15,18,19\notag\\
        &=& - \cos{\frac{\pi}{3}} \mathbb{1} + i \sin{\frac{\pi}{3}} \sigma_{2} \,\ \text{for} \,\ a=2,5,6,9\notag\\
        &=& - \cos{\frac{\pi}{3}} \mathbb{1} - i \sin{\frac{\pi}{3}} \sigma_{2} \,\ \text{for} \,\ a=3,4,7,8\notag\\
        &=& - \cos{\frac{\pi}{3}} \sigma_{3} - \sin{\frac{\pi}{3}} \sigma_{1} \,\ \text{for} \,\ a=10,11,16,17\notag\\
        \label{eq:mt}
        &=& - \cos{\frac{\pi}{3}} \sigma_{3} + \sin{\frac{\pi}{3}} \sigma_{1} \,\ \text{for} \,\ a=12,13,20,21 
    \end{eqnarray}
    \item $T_{1}$: $(D^{T_{1}}(g_{a}))_{\alpha\beta} = \exp{\left(- i \omega^{(a)} \vec{n}^{(a)} \cdot \vec{J}  \right)}_{\alpha\beta} = \cos{\omega^{(a)}} \delta_{\alpha\beta} + (1 - \cos{\omega^{(a)}} n_{\alpha}^{(a)} n_{\beta}^{(a)}) - \sin{\omega^{(a)}} \epsilon_{\alpha\beta\gamma} n_{\gamma}^{(a)}$, where $(J_{\gamma})_{\alpha\beta} = - i \epsilon_{\alpha\beta\gamma}$ denote the group generator.
    \item $T_{2}$: The matrices are the same as $T_{1}$, but with a change of sigh for the conjugacy classes $6C_{4}$ and $6C'_{2}$.
\end{itemize}
\end{widetext}
 If we add inversions, each of this representations will be duplicated ($A_{1} \rightarrow A^{\pm}_{1}$), and the elements corresponding to $R$ and $R I$ are the same, with opposite sign \cite{Doring:2018xxx}.

Since Lorentz Symmetry is broken in the cube, spatial angular momentum $L$ is not a a good quantum number and  partial wave mixing may occur. The elements of the basis of the irreducible representation $\Gamma$ can be  expressed in terms of  spherical harmonics and the $D^\Gamma$ matrices above as \cite{Doring:2018xxx}
\begin{equation}
    \xi^{l \Gamma }_{\alpha \beta m} (\hat{p}_{0}) = \frac{\sqrt{4 \pi} s_{\Gamma}}{G} \sum_{g \in \mathcal{O}_h} (D^{\Gamma}_{\beta \alpha} (g))^{*} \sum_{m'} D^{l}_{m m'} (g) Y_{l m'}(\hat{p}_{0})
\end{equation}
where $D^{l}_{m m'} (g_{a})$ is the Wigner's matrix,  $G = 48$ and  $\hat{p}_{0}$ is the unitary vector of reference momentum.

The trace of the matrices of Eq. \ref{eq:mt}, given in  table \ref{table:2}, that we call $\chi_{k}^{\Gamma}$, characterise the irreducible representation, being $k$ the symmetry operation index ($ k = \{I, 8C_{3}, 6C_{4}, 6C'_{2}, 3C_{2}\}$), with $s_{k}$ members ($s_{k} = \{1, 8, 6, 6, 3\}$, see Table \ref{table:1}). The multiplicity of an Octahedral irrep, $\vartheta^{\Gamma}_{l}$, counts the number of Irreps  in a given angular momentum $l$ and is given by
\begin{equation}
    \vartheta_{l}^{\Gamma} = \frac{1}{G} \sum_{k} s_{k} \, \chi_{k}^{\Gamma} \, \chi_{k}^{l}
\end{equation}
where the trace of $SU(2)$ matrices for the angular momentum $l$  and $\theta_{k}$ angle, $\chi^{l}_{k}$, is written as follows. 
\begin{equation}\label{trazasu2}
    \chi_{k}^{l} = \frac{\sin{(l + \frac{1}{2}}) \theta_{k}}{\sin{\frac{\theta_{k}}{2}}}
\end{equation}
The corresponding irreducible characters are given in Table \ref{table:2}, with the specific value of $\theta_{k}$ for each symmetry operation, $k$. Thus, we can establish a correspondence between the octahedral group and $SU(2)$ (reduction)  associating to every angular momentum and parity a subgroup of $SU(2)$. For example, from the table \ref{table:3} , we can see that, in $A_{1}^{\pm}$ in $\mathcal{O}_h$ can be associated to $l^{P} = 0^{+}, 4^{+}, 6^{+}, 9^{-}, ... $, $A_{2}^{\pm}$ to $l^{P} = 3^{-}, 6^{+}, 7^{+}, 9^{-}, ...$, $E^{\pm}$ to $l^{P} = 2^{+}, 4^{+}, 5^{-}, 6^{+}, ... $, $T_{1}^{\pm}$ to $l^{P} = 1^{-}, 3^{-}, 4^{+}, 5^{-}, ... $, and $T_{2}^{\pm}$ to $l^{P} = 2^{+}, 3^{-}, 4^{+}, 5^{-}, ... $. This means in particular that a physical system with $l^{P} = 0^{+}$ in the cube, is invariant under $A_{1}^{+}$, while one with $l^{P} = 1^{-}$ is invariant under $T_{1}^{-}$. However, for a $d$-wave system,  both $E^{+}$ and $T_{2}^{+}$ contribute \cite{Johnson:prl} (see also our discussion in section \ref{sec:CH}).

\begin{table*}[ht]
\centering
\begin{center}
{\renewcommand{\arraystretch}{1.5}
\begin{tabular}{ |P{1.5cm}|P{1.5cm}|P{1.5cm}|P{1cm}|P{1cm}|P{1cm}| }
\hline
\rowcolor{lightgray} \multicolumn{6}{|c|}{Irreducible characters of $\mathcal{O}_h$} \\
\hline
\rowcolor{gray!25} Irreps & $I$ & $8C_{3}$ & $6C_{4}$ & $6C'_{2}$ & $3C_{2}$\\
\hline
$A_{1}$ & $1$ & $1$ & $1$ & $1$ & $1$ \\

$A_{2}$ & $1$ & $1$ & $-1$ & $-1$ & $1$ \\

$E$ & $2$ & $-1$ & $0$ & $0$ & $2$ \\

$T_{1}$ & $3$ & $0$ & $1$ & $-1$ & $-1$ \\

$T_{2}$ & $3$ & $0$ & $-1$ & $1$ & $-1$ \\
\hline
$\theta_{k}$ & $2\pi$ & $2\pi/3$ & $\pi/2$ & $\pi$ & $\pi$ \\
\hline
\end{tabular}}
\end{center}
\caption{Irreducible characters for each Irrep, and corresponding  angle of rotation $\theta$ per class.}
\label{table:2}
\end{table*}

\begin{table*}[ht]
\centering
\begin{center}
{\renewcommand{\arraystretch}{}
\begin{tabular}{ |P{1.5cm} P{5cm}|}
\hline
\rowcolor{lightgray} \multicolumn{2}{|c|}{Reduction} \\
\hline
\rowcolor{gray!25} $l^{P}$ &  \\
\hline
 $0^{+}$ & $A_{1}^{+}$  \\
 $1^{-}$ & $T_{1}^{-}$  \\
 $2^{+}$ & $E^{+} \oplus T_{2}^{+}$ \\
 $3^{-}$ & $A_{2}^{-} \oplus T_{1}^{-} \oplus T_{2}^{-}$ \\
 $4^{+}$ & $A_{1}^{+} \oplus E^{+} \oplus T_{1}^{+} \oplus T_{2}^{+}$ \\
 $5^{-}$ & $E^{-}  \oplus   2 T_{1}^{-} \oplus T_{2}^{-}$ \\
 $6^{+}$ & $A_{1}^{+} \oplus A_{2}^{+} \oplus E^{+}  \oplus T_{1}^{+} \oplus 2 T_{2}^{+}$ \\
 $7^{-}$ & $A_{2}^{-} \oplus E^{-}  \oplus 2 T_{1}^{-} \oplus 2 T_{2}^{-}$ \\
 $8^{+}$ & $A_{1}^{+} \oplus 2 E^{+}  \oplus 2 T_{1}^{+} \oplus 2 T_{2}^{+}$ \\
 $9^{-}$ & $A_{1}^{-} \oplus A_{2}^{-} \oplus E^{-}  \oplus 3 T_{1}^{-} \oplus 2 T_{2}^{-}$ \\
 \hline
\end{tabular}}
\end{center}
\caption{Reduction of the $SU_{2}$ to $\mathcal{O}_h$, up to $l=9$ and parity $P = (-1)^{l}$.}
\label{table:3}
\end{table*}


\end{document}